\documentclass{jfm}
\usepackage{graphicx}
\usepackage{epstopdf, epsfig}
\usepackage{xcolor}

\shorttitle{Heat transfer in laminar Couette flow laden with particles}
\shortauthor{M. Niazi Ardekani, O. Abouali, F. Picano and L. Brandt}

\title{Heat transfer in laminar Couette flow laden with rigid spherical particles}

 \author{M. Niazi Ardekani\aff{1},
  O. Abouali\aff{1,2}   \corresp{\email{abouali@shirazu.ac.ir}}, 
  F. Picano\aff{3}
 \and L.  Brandt\aff{1}}

\affiliation{\aff{1} Linn\'e Flow Centre and SeRC (Swedish e-Science Research Centre), KTH Mechanics, \\ S-100 44 Stockholm, Sweden
\aff{2} School of Mechanical Engineering, Shiraz University, \\ Mollasadra Street, Shiraz 71348-51154, Iran
\aff{3} Department of Industrial Engineering, University of Padova, \\ Via Venezia 1, 35131 Padua, Italy}

\begin{document}

\maketitle

\begin{abstract}

We study heat transfer in plane Couette flow laden with rigid spherical particles by means of direct numerical simulations using a direct-forcing immersed boundary method to account for the dispersed phase. A volume of fluid approach is employed to solve the temperature field inside and outside of the particles. 
We focus on the variation of the heat transfer with the particle Reynolds number, total volume fraction (number of particles) and the ratio between the particle and fluid thermal diffusivity, quantified in terms of an effective suspension diffusivity. 
We show that, when inertia at the particle scale is negligible, the heat transfer increases with respect to the unladen case following an empirical correlation recently proposed.
In addition, an average composite diffusivity can be used to predict the effective diffusivity of the suspension the inertialess regime when varying the molecular diffusion in the two phases.
At finite particle inertia, however, the heat transfer increase is significantly larger,  saturating at higher volume fractions. 
By phase-ensemble averaging we identify the different mechanisms contributing to the total heat transfer and show that the increase of the effective conductivity observed at finite inertia is due to the increase of the transport associated to fluid and particle velocity. 
We also show that the heat conduction in the solid phase reduces when increasing the particle Reynolds number and so that  particles of low thermal diffusivity
weakly alter the total heat flux in the suspension at finite particle Reynolds numbers.
On the other hand, a higher particle thermal diffusivity significantly increase the total heat transfer. 

\end{abstract}


\section{Introduction}
\label{introduction}

Problems involving  particle-fluid interactions  are widely encountered in different applications such as sediment transport, air pollution, pharmaceutical industry, blood flow, petrochemical and mineral processing plants.  
Controlling heat transfer in particulate suspensions has also many important applications such as packed and fluidised bed reactors and industrial dryers. In these cases, in addition to momentum/mechanical interactions between particles and fluids, the flow is also characterised by heat transfer between the two phases. Due to the complex phenomena related to diffusion and convection of heat in the particulate flows, a better understanding of the heat transfer between the two phases in the presence of a flow is essential to improve the current models. The aim of this study is to validate the proposed numerical algorithm against existing experiments and numerical studies and then examine, in particular, the effect of finite particle inertia and difference in fluid and solid phase diffusivity  on the heat transfer across a sheared particle suspensions. Ensemble averaged equations are derived to disentangle the energy transfer in the fluid and solid phases.

Simplified  theoretical approaches and experiments have been used in earlier times to study these challenging physical phenomena. 
The experiments of \cite{Ahuja1975a} on sheared suspensions of polystyrene particles at finite particle Reynolds numbers ($Re_p > 1$) revealed a significant enhancement of heat transfer. The author attributed this enhancement to a mechanism based on the inertial effects, in which the fluid around the particle is centrifuged by the particle rotation.  
\cite{Sohn1981} investigated the eddy transport, associated with microscopic flow fields in shearing two-phase flows with volume fraction of spherical particles up to $30\%$. At low Reynolds numbers and high Peclet numbers $Pe$, they found an increase in the heat transfer, which approaches a power law relationship with $Pe$. The Peclet number $Pe$ defines the ratio between fluid viscosity and heat diffusivity in the fluid. \cite{Chung1982} measured experimentally the effective thermal conductivity of a sheared suspension of rigid spherical particles. These author compared their results to the theoretical prediction of \cite{Leal1973} for a sheared dilute suspension at low particle Peclet number, $Pe$, reporting good agreement. In addition, they investigated moderate concentrations (volume fraction $\phi < 25 \% $) and higher Peclet numbers compared to the study by \cite{Leal1973}.  It was later suggested by \cite{Zydney1988} that the increase in solute transport, previously observed for particle suspensions, is caused by shear-induced particle diffusion \citep{Madanshetty1996,Breedveld2002} and the resulting dispersive fluid motion. These authors proposed a model, based on existing experimental results, concluding that the augmented solute transport is expected to vary linearly with the Peclet number. \cite{Shin2000} experimentally studied the heat transfer of suspensions with low volume fractions ($\phi < 10\%$) for different shear rates and particle sizes. They found that the heat transfer increases with shear rate and particle size, however it saturates at large shear rates.


Recently, the rapid development of computer resources and efficient numerical algorithms has directed more attention to approaches based on the Direct Numerical Simulations (DNS) of heat transfer in particle suspensions. Numerical algorithms coupling the heat and mass transfer are complex and require considerable computational resources, which limits the number of available direct simulations.
Hence, in the first attempts, researchers used DNS only for the hydraulic characteristics of the flow and modelled the energy or mass transport equation. Among more recent studies, we
consider here \cite{Wang2009} who presented experimental, theoretical, and numerical investigations of the transport of fluid tracers between the walls bounding a sheared suspension of neutrally buoyant solid particles. In their simulation, these authors used a lattice-Boltzmann method \citep{Ladd1994,Ladd19942} to determine the fluid velocity and solid particle motion and a Brownian tracer algorithm to determine the chemical  mass transfer. They reported that the chaotic fluid velocity disturbances, caused by the motion of the suspended particles, lead to enhanced hydrodynamic diffusion across the suspension. In addition, it was found that for moderate values of the Peclet numbers the Sherwood number, quantifying the ratio of the total rate of mass transfer to the rate of diffusive mass transport alone, changes linearly with $Pe$. At higher Peclet numbers, however, the Sherwood number grows more slowly due to the increase in the mass transport resistance by a molecular-diffusion boundary layer near the solid walls. Further, these authors report that the fluid inertia enhances the rate of mass transfer in suspensions with particle Reynolds numbers in the range between $0.5$ to $7$.

The effect of shear-induced particle diffusion on the transport of the heat across the suspension was investigated more recently by  \cite{Metzger2013} through a combination of experiments and simulations. In this study, the influence of particle size, particle volume fraction and applied shear are investigated. 
Using index matching and laser-induced fluorescence imaging, these authors measured individual particle trajectories and calculated their diffusion coefficients. They also performed numerical simulations using a lattice Boltzmann method for the flow field and a passive Brownian tracer algorithm to model the heat transfer. Their numerical results are in good agreement with experiments and show that fluid fluctuations due to the particles movement can lead to more than $200 \%$  increase in the heat transfer through the suspension. A correlation is presented in this study for the effective thermal diffusivity: this is found to be a linear function of both the Peclet number and the solid volume fraction. \cite{Souzy2015} investigated the mass transport in a cylindrical Couette cell of a sheared suspension of non-Brownian spherical particles. They found that a rolling-coating mechanism (particle rotation convects the dye layer around the particles) transports convectively the dye directly from the wall towards the bulk. 

Including buoyancy forces,  \cite{Feng2008} used direct numerical simulations  to study the dynamics of non-isothermal cylindrical particles in particulate flows. These authors resolve the momentum and energy equations to compute the effect of heat transfer on the sedimentation of particles. They found that the drag force on non-isothermal particles, strongly depends on the Reynolds number and the Grashof number, reporting that the drag coefficient is higher for the hottest particles at relatively low Reynolds numbers. Grashof number quantifies the ratio of the buoyancy to viscous force acting on a fluid.
The same numerical method is used also to study a pair of hot particles settling in a container at different Grashof numbers. The simulations demonstrated that the well-known drafting-kissing-tumbling $DKT$ motion observed  for isothermal particles \citep{Feng2008} disappears in the case of particles hotter than the fluid. 
 \cite{Feng2009} extended these earlier works to $3-D$ cases 
using a finite difference method in combination with the Immersed Boundary (IB) method for treating the particulate phase. They used an energy density function to represent thermal interaction between particle and the fluid similar to that of a force density to represent the momentum interaction without solving the differential energy equation inside the solid particles.
\cite{Dan2010} suggested a Distributed Lagrange Multiplier/Fictitious Domain ($DLM/FD$) method to compute the temperature distribution and the heat exchange between the fluid and solid phases. The Bousinesq approximation was used to model density variations in the fluid. 
These authors employed a Finite Element Method ($FEM$) to solve the mass, momentum and energy conservation equations and used a Discrete Element Method ($DEM$) to compute the motion of particles. Distributed Lagrange multipliers for both the velocity and temperature fields are introduced to treat the fluid/solid interaction. \cite{Tavassoli2013} extended the Immersed Boundary (IB) method proposed by  \cite{Uhlmann2005} to systems with interphase heat transport.  Their numerical method treats the particulate phase by introducing momentum and heat source terms at the boundary of the solid particle, which represent the momentum and thermal interactions between the fluid and the particle. The method is used to investigate the non-isothermal flows past stationary random arrays of spheres.  
 \cite{Hashemi2014} studied numerically the heat transfer from spheres settling under gravity in a box filled with liquid. The simulations in this study employ a three-dimensional Lattice Boltzmann Method to simulate fluid-particle interactions and investigate the effects of Reynolds, Prandtl and Grashof numbers ($Re$, $Pr$, $Gr$) for the case of a settling particle at fixed/varying temperature. These authors also studied the hydraulic and the heat transfer interactions of $30$ hot spherical particles settling in an enclosure. Recently, \cite{Sun2016} investigated and modelled the pseudo-turbulent heat flux in a suspension. These authors report results for a wide range of mean slip Reynolds number and solid volume fraction using particle-resolved direct numerical simulations ($PR-DNS$) of steady flow through a random assembly of fixed isothermal mono-disperse spherical particles. They revealed that the transport term in the average fluid temperature equation, corresponding to the pseudo-turbulent heat flux, is significant when compared to the average gas-solid heat transfer.  
 
In the present work, the numerical code developed by \cite{Breugem2012}, previously used to study  suspensions in laminar and turbulent flows \citep{Lashgari2014,Picano2015,Fornari2015,Lashgari2016}, is extended to resolve the temperature field in the fluid and solid phase of a suspension with the possibility to examine different particle and fluid thermal diffusivities. 
 In combination to the accurate IBM method to resolve the solid-fluid interaction, this enables us to investigate the different heat transport mechanism at work.
We quantify the heat flux across a plane Couette flow  when varying the 
particle volume fraction, the particle Reynolds number, thus including inertial effects, 
 and the ratio between the fluid and solid diffusivity, aiming to identify the conditions for  enhancement/reduction of heat transfer.


\section{Governing equations and numerical method}
\label{Governing equations and numerical method}

\subsection{Governing equations}

The equations describing the flow field in the Eulerian framework  are the incompressible Navier-Stokes equations: 
\begin{eqnarray}
\label{eq:NS1}  
\rho_f (\frac{\partial \textbf{u}}{\partial t} + \textbf{u} \cdot \nabla \textbf{u} ) &=&  -\nabla p + \mu_{f} \nabla^2 \textbf{u} + \rho_f \textbf{f} \, , \\ [8pt]
\nabla \cdot \textbf{u} &=& \, 0 \, .
\label{eq:NS2} 
\end{eqnarray}
with $\textbf{u}$ the fluid velocity, $p$ the pressure with $\rho_f$ and $ \mu_{f}$ the density and dynamic viscosity of the fluid. The additional term $\textbf{f}$ is added on the right-hand-side of equation (\ref{eq:NS1}) to account for the presence of particles. This force is active in the immediate vicinity of a particle surface to impose the no-slip and no-penetration boundary condition indirectly, see description of the numerical algorithm below.  

The motion of rigid spherical particles is described by the Newton-Euler Lagrangian equations,
\begin{eqnarray}
\label{eq:NewtonEuler1}  
\rho_p V_p \frac{ \mathrm{d} \textbf{U}_{p}}{\mathrm{d} t} &=& \oint_{\partial {S}_p}  \pmb{\tau} \cdot  \textbf{n} \mathrm{d}A - V_p \nabla p_e + \left( \rho_p - \rho_f \right)V_p\textbf{g} + \textbf{F}_c , \, \\ [8pt] 
\textbf{I}_p  \frac{ \mathrm{d} \left( \pmb{\omega}_{p} \right) }{\mathrm{d} t} &=& \, \oint_{\partial {S}_p} \textbf{r} \times \left( \pmb{\tau} \cdot  \textbf{n} \right) \mathrm{d}A + \textbf{T}_c  \, , 
\label{eq:NewtonEuler2}
\end{eqnarray} 
with $\textbf{U}_p$ and  $\pmb{\omega}_{p}$ the translational and the angular velocity of the particle. $\rho_p$, $V_p$ and $\textbf{I}_p$ are the particle mass density, volume and moment-of-inertia. The outward unit normal vector at the particle surface is denoted by $\textbf{n}$ and $\textbf{r}$ is the position vector from the center of the particle.
The stress tensor, $\tau = -p\textbf{I} + \mu_f \left( \nabla \textbf{u} + \nabla \textbf{u}^T \right)$, integrated on the surface of particles and the force terms $- \rho_f V_p\textbf{g}\, \,$, $V_p \nabla p_e$ and $\textbf{g}$ account for the fluid-solid interaction, the hydrostatic pressure, any external constant pressure gradient and the gravitational acceleration. $\textbf{F}_c$ and $\textbf{T}_c$ are the force and torque acting on the particles due to particle-particle (particle-wall) collisions.  

The energy equation is resolved on every grid point in the computational domain, including both the fluid and the solid phase to obtain the temperature field,   
\begin{eqnarray}
\frac{\partial T}{\partial t}  \,+\,   \textbf{u} \cdot \nabla T  \,=\,  \nabla \cdot \left( \alpha \, \nabla T \right) \, 
\label{eq:Temperature} 
\end{eqnarray}
where $\alpha$ is the thermal diffusivity equal to $k/(\rho C_p)$, with $C_p$ and $k$ the specific heat capacity and thermal conductivity, respectively. It should be noted here that in this study $\rho C_p$ we assume to be identical for both phases: ${(\rho C_p )}_p = {(\rho C_p )}_f $; $\alpha$ is  however different in the two phases.

\subsection{Numerical scheme}

The IBM code initially developed by \cite{Breugem2012} is extended here to be able to study the heat transfer in a suspension of rigid particles.
More details on the algorithm used for the solution of the isothermal problem and several validations can be found in \cite{Lambert2013,Picano2015,Lashgari2016}. This is therefore just shortly described in the first two sections below.

\subsubsection{Solution of the flow field}

The \textit{pressure-correction} scheme described in \cite{Breugem2012}, is employed here to resolve the flow field on a uniform ($ \Delta x = \Delta y = \Delta z$), staggered, Cartesian Eulerian grid. Equations (\ref{eq:NS1}) and (\ref{eq:NS2}) are integrated in time using an explicit low-storage Runge-Kutta method. Particles are represented by a set of Lagrangian points, uniformly distributed on the surface of each particle. The number of Lagrangian grid points $N_L$ is defined such that the Lagrangian grid volume $\Delta V_l$ is as close as possible to the volume of the Eulerian mesh ${\Delta x}^3$.  $\Delta V_l$ is obtained by assuming the particle as a thin shell with same thickness as the Eulerian grid size.

The IBM point force $F_l$ is calculated at each Lagrangian point based on the difference between the particle surface velocity ($\textbf{U}_p + \pmb{\omega}_p \times \textbf{r}$) and the interpolated first prediction velocity at the same point. A first prediction velocity, obtained by advancing equation (\ref{eq:NS1}) in time without considering the force field $\textbf{f}$ and neglecting pressure correction, is used to approximate the IBM force, and a second one to compute the correction pressure and update the pressure field. The forces, $\textbf{F}_l$, integrate to the force field $\textbf{f}$  using the regularized Dirac delta function $\delta_d$ introduced in \cite{Roma1999}: 
 \begin{equation}
\label{eq:Fl}  
\textbf{f}_{\, ijk} = \sum\limits_{l=1}^{N_L} \textbf{F}_l \delta_d \left( \textbf{x}_{ijk} - \textbf{X}_l \right) \Delta V_l \,  \, 
\end{equation}
with $\textbf{x}_{ijk}$ and $\textbf{X}_l$ referring to an Eulerian and a Lagrangian grid cell. This delta function smoothens the sharp interface over a porous shell of width  $3\Delta x$, while preserving the total force and torque on the particle provided that the Eulerian grid is uniform.
 
\subsubsection{Solution of the particle motion}
Taking into account the motion of rigid spherical particles and the mass of the fictitious fluid phase inside the particle volumes, \cite{Breugem2012} showed that  equations (\ref{eq:NewtonEuler1}) and (\ref{eq:NewtonEuler2}) can be rewritten as:
\begin{eqnarray}
\label{eq:NewtonEulerWim1}  
\rho_p V_p \frac{ \mathrm{d} \textbf{U}_{p}}{\mathrm{d} t} &\approx&  -\rho_f \sum\limits_{l=1}^{N_L} \textbf{F}_l \Delta V_l + \rho_f  \frac{ \mathrm{d}}{\mathrm{d} t} \left( \int_{V_p} \textbf{u} \mathrm{d} V  \right)  + \left( \rho_p - \rho_f \right)V_p\textbf{g} + \textbf{F}_c , \, \\ [8pt] 
\frac{ \mathrm{d} \left( \textbf{I}_p \, \pmb{\omega}_{p} \right) }{\mathrm{d} t} &\approx& -\rho_f \sum\limits_{l=1}^{N_L} \textbf{r}_l \times \textbf{F}_l \Delta V_l + \rho_f  \frac{ \mathrm{d}}{\mathrm{d} t} \left( \int_{V_p} \textbf{r} \times \textbf{u} \mathrm{d} V  \right) + \textbf{T}_c  \, .
\label{eq:NewtonEulerWim2}  
\end{eqnarray}
The equations above are integrated in time using the same explicit low-storage Runge-Kutta method used for the flow.

The interaction force $\textbf{F}_c$ and torque $\textbf{T}_c$ are activated when the gap width between the two particles (or between one particle and the wall) is less than one Eulerian grid size. Indeed, when the gap width reduces to less than one Eulerian mesh, the lubrication force is under-predicted by the IBM. To avoid computationally expensive grid refinements, a lubrication model based on the asymptotic analytical expression for the normal lubrication force between two equal spheres \citep{Brenner1961} is used here for particle/particle interactions whereas the solution for two unequal spheres, one with infinite radius, is employed for particle/wall interactions. A soft-sphere collision model with Coulomb friction takes over the interaction when the particles touch.  The restitution coefficients used for normal and tangential collisions are $0.97$ and $0.1$, with Coulomb friction coefficient set to $0.15$. More details about the short-range models and corresponding validations can be found in \cite{Ardekani2016,Costa2015}.  

\subsubsection{Solution of the temperature field}

\begin{figure}
 \centering
  \includegraphics[width=0.75\textwidth]{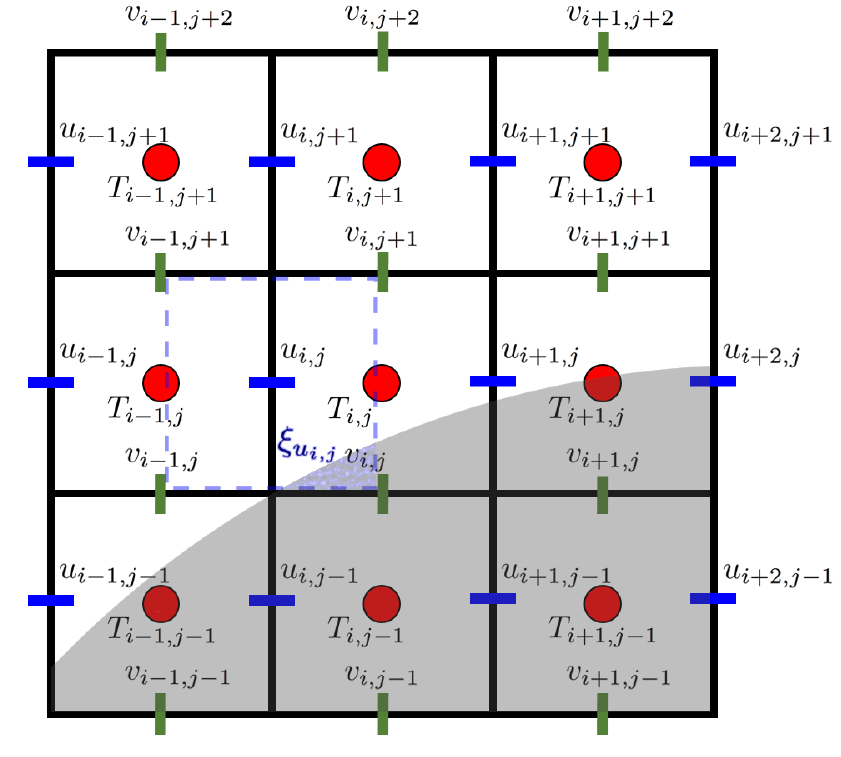}
  \caption{The staggered Eulerian grid and the phase indicator $\xi$ around the velocity point $u_{i,j}$.}
\label{fig:xi}
\end{figure}

A phase indicator, $\xi$, is used to distinguish the solid and the fluid phase within the computational domain. $\xi$ is computed at the velocity (cell faces) and the pressure points (cell center) throughout the staggered eulerian grid. This value varies between $0$ and $1$ based on the solid volume fraction of a cell of size ${\Delta x}$ around the desired point. As we know the exact location of the fluid/solid interface for rigid spheres, a level-set function $\zeta$ given by the signed distance to the particle surface $\mathcal{S}$ is employed to determine $\xi$ at each point. With $\zeta < 0$ inside and $\zeta > 0$ outside the particle, the solid volume fraction is calculated similar to \cite{Kempe2012JCP}:
\begin{eqnarray}
\xi =  \frac{\sum_{n=1}^{8} - \zeta_n \mathcal{H}(-\zeta_n)}{\sum_{n=1}^{8}  |\zeta_n|} \, 
\label{eq:Temperature2} 
\end{eqnarray}   
where the sum is over all $8$ corners of a box of size ${\Delta x}$ around the target point and $\mathcal{H}$ is the Heaviside step function. Figure~\ref{fig:xi} indicates the staggered Eulerian grid and the phase indicator $\xi$ around the velocity point $u_{i,j}$.

Using a volume of fluid (VoF) approach, the velocity and the thermal diffusivity of the combined phase are defined at each point in the domain as
\begin{eqnarray}
\label{eq:Ucp}  
\textbf{u}_{cp} &=&  \left( 1 - \xi \right) \textbf{u}_f \, + \, \xi \textbf{u}_p , \, \\ [8pt] 
\alpha_{cp}  &=&  \left( 1 - \xi \right) \alpha_f \, + \, \xi \alpha_p   \, , 
\label{eq:ALPHAcp}
\end{eqnarray} 
where $\textbf{u}_f$ is the fluid velocity and $\textbf{u}_p$ the solid phase velocity, obtained by the rigid body motion of the particle at the desired point; $\alpha_f$ and $\alpha_p$ denote the thermal diffusivity of the fluid and the solid phase. 

Substituting  $\textbf{u}_{cp}$ and $\alpha_{cp}$ in equation (\ref{eq:Temperature}) results in 
\begin{eqnarray}
\frac{\partial T}{\partial t} \,+\,  \nabla \cdot \left( \textbf{u}_{cp} \, T \right)  \,=\,  \nabla \cdot \left( \alpha_{cp} \, \nabla T \right) \, 
\label{eq:Temperature2} 
\end{eqnarray}
which is discretized  around the Eulerian cell centers (pressure and temperature points on the Eulerian staggered grid) and integrated in time, using the same explicit low-storage Runge-Kutta method used for marching the flow and particle equations. 

\subsection{Flow configuration}

\begin{table}
  \begin{center}
\def~{\hphantom{0}}
  \begin{tabular}{lcccccc}
       \,\,\,\,\,\,\,\,\,\,\,  \,$Re_p$ & $0.5$ & $1$  & $4$ & $8$  & $16$  \\[3pt]
       \,\,\,\,\,\,\,\,\,\,\,  \,$Re_b$ & $18$ & $36$  & $144$ & $288$  & $576$  \\
\hline
       $\, L_x \times L_y \times L_z$  &  \,\,\,\,\,\,\, &  \,\,\,\,\,\,\, &  $12D \times 6D \times 6D$\\[3pt]
       $N_x \times N_y \times N_z$  &  \,\,\,\,\,\,\, &  \,\,\,\,\,\,\, &  $288 \times 144 \times 144$\\
       \hline
       $ \,\,\,\,\,\,\,\,\,\,\,\, \phi \, (\%)$  & $0 $ & $3$ & $10$ & $20$ & $30$ \\[3pt]
       $ \,\,\,\,\,\,\,\,\,\,\,\, \, \,  \,N_p$  & $0 $ & $28$ & $84$ & $168$ & $252$ \\  
       \hline
       $\,\,\,\,\,\,\,\,\,\,\,\, \, \,\,\,   \Gamma$ &  & $0 .1$ & $1$ &  $10$ & & \\[3pt]                            
  \end{tabular}
  \caption{Summary of the different simulations performed. From the top: particle and flow Reynolds number considered and corresponding box size and resolution.
Suspension volume fraction $\phi$ and corresponding number of particles $N_p$. $D$ is the particle diameter and $\Gamma$ is the ratio between the thermal diffusivity of the particles and of the fluid.}
 \label{tab:cases}
 \end{center}
\end{table}

In the present work, the Couette flow between two infinite walls with distance  $2h$ in the wall-normal direction, $y$, is investigated. The size of the computational domain  is $L_x= 4h$, $L_y=2h$ and $L_z=2h$ in the streamwise, wall-normal and spanwise directions. Periodic boundary conditions for velocity and temperature are imposed in both streamwise and spanwise directions ($x$ and $z$), while the upper and lower walls are moving with velocity $0.5 U_b$ and $-0.5 U_b$. $U_b$ is the reference velocity, used to define the flow Reynolds number $Re_b =U_b 2h/ \nu$, with $\nu$ the kinematic viscosity of the fluid phase. 
The diameter of the particles, considered in this study is equal to one sixth of the distance between the planes ($2h/D = 6$) and the particle Reynolds number is defined as $Re_p = \dot{\gamma} D^2/\nu$, where $\dot{\gamma}=U_b/2h$ is the shear rate. 
The temperature of the upper and the lower wall is fixed at $T=0.5$ and $-0.5$. 

Simulations are performed at different particle Reynolds number $Re_p$, volume fraction $\phi$ and thermal diffusivity ratio $\Gamma \equiv \alpha_p/\alpha_f$.
We investigate the effect of each parameter on the heat transfer between the two walls and quantify the results in terms of $\alpha_r \equiv \alpha_e/\alpha_f$, with $\alpha_e$ denoting the effective thermal diffusivity of the suspension. This is the diffusivity that would correspond to the heat transfer extracted from the numerical data for the same  temperature gradient. The different parameters, the flow geometry and numerical resolution used are reported  in table~\ref{tab:cases}.

\section{Validation}
\label{Validation}

The IBM used in this study to resolve the fluid-solid interaction has been fully validated by \cite{Breugem2012} and \cite{Picano2013,Picano2015}. The validation of temperature solver with the mentioned VoF approach for the heat transfer between the phases is presented here, considering first a single sphere.

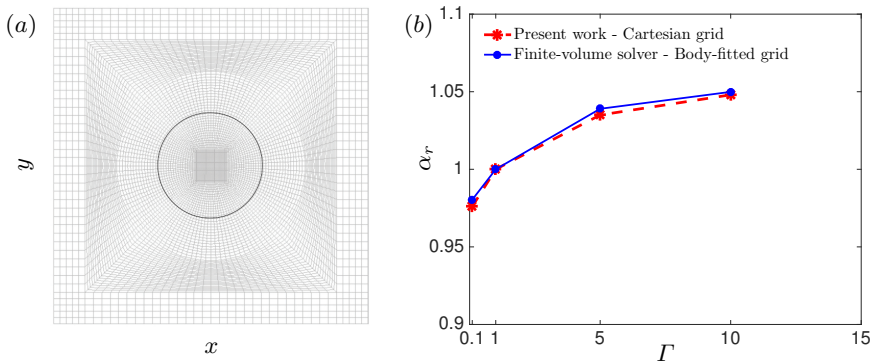
\begin{figure}
 \centering
  \includegraphics[width=0.36\textwidth]{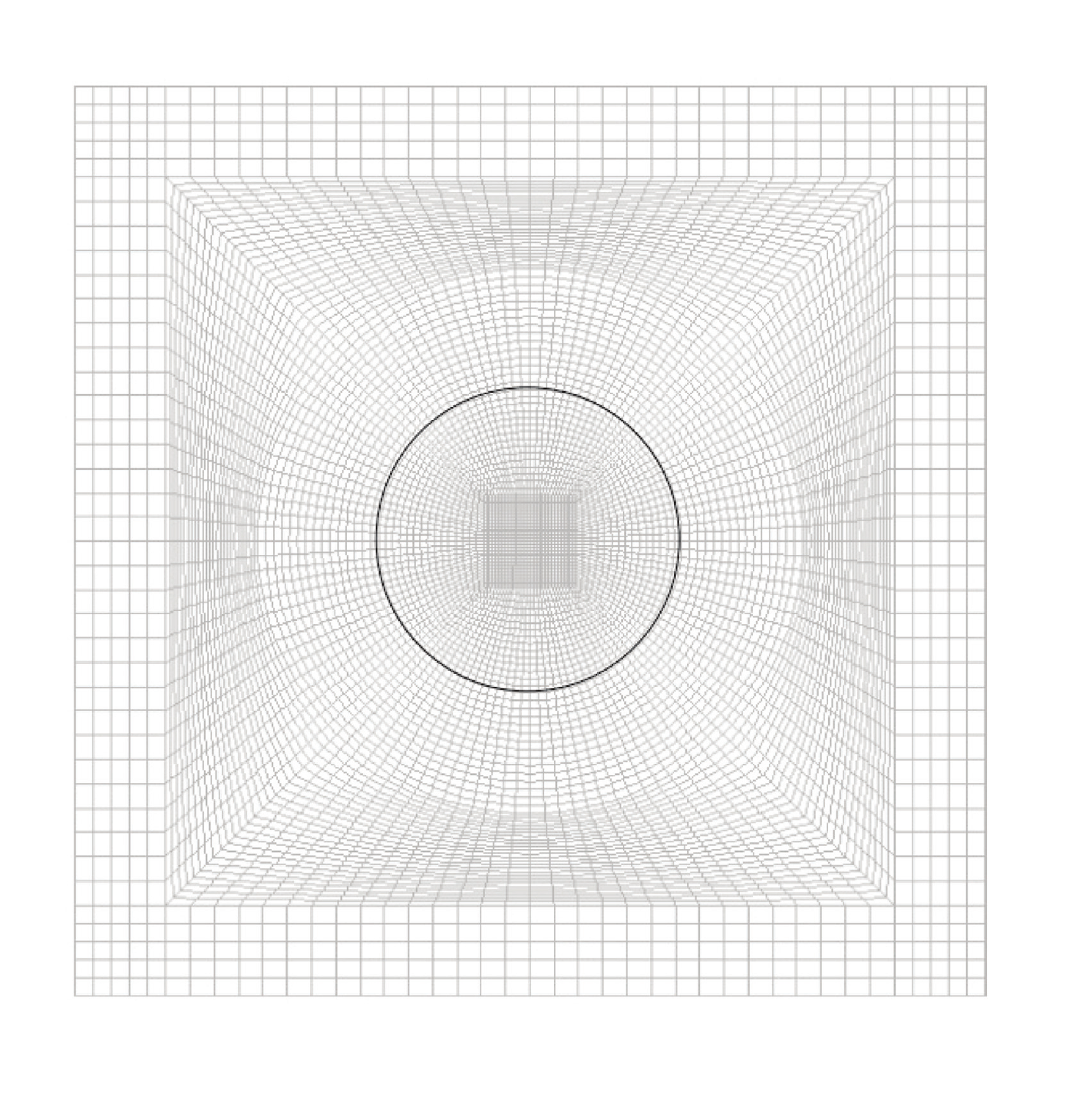}
   \includegraphics[width=0.495\textwidth]{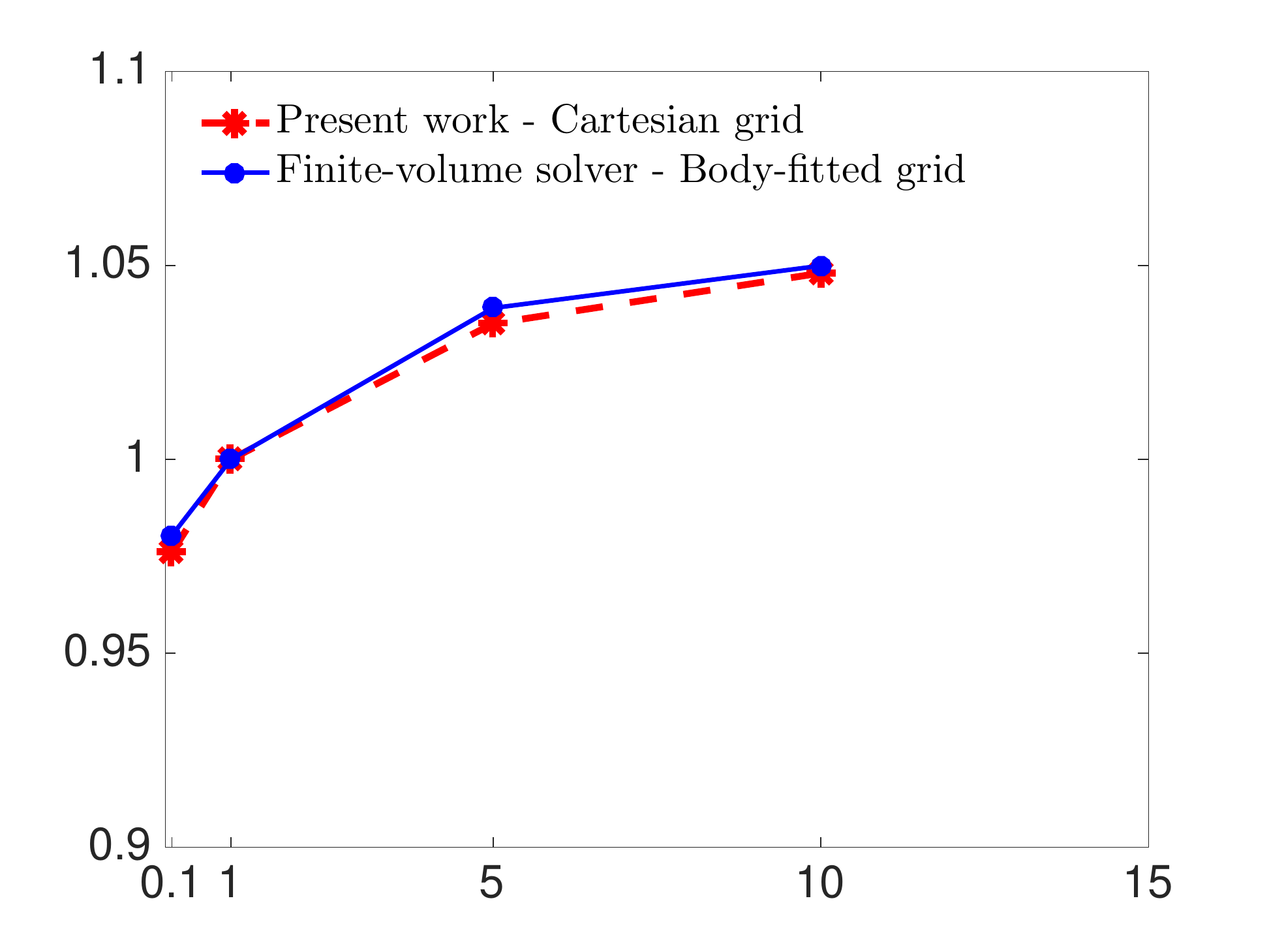}
   \put(-340,125){$(a)$}
   \put(-190,125){$(b)$}   
   \put(-95,2){$\Gamma$} 
   \put(-184,73){\rotatebox{90}{$\alpha_r$}} 
   \put(-265,5){$x$} 
   \put(-335,73){\rotatebox{90}{$y$}} \\   
  \caption{$(a)$ $2d$ section of the boundary-fitted orthogonal grid used in the finite volume solver. $(b)$ Effective thermal diffusivity $\alpha_e$, normalized by the thermal diffusivity of the fluid ($\alpha_r \equiv \alpha_e/\alpha_f$) for the different thermal diffusivity ratio $\Gamma$. }
\label{fig:valid2}
\end{figure}

The numerical code developed for this study enjoys the ability to resolve the temperature field across the domain with different thermal diffusivity for the particles and the fluid. When there is a significant difference between the thermal diffusivity of the solid and the fluid, the coefficient in the diffusion term of the temperature equation experiences a jump across the interface. This jump is smoothened  around the interface in the present numerical scheme. To evaluate how the present numerical model perform for these situations, a simpler validation case is chosen. 

We consider a cubic box of size $3D$ with a sphere in the center and vary the ratio of thermal diffusivity of the sphere and of the surrounding fluid, $\Gamma$. The temperature at the upper and the bottom wall of the domain is set to $1$ and $0$ respectively, while  periodic boundary conditions are imposed on the other four faces of the cube. The fluid is at rest and we thus solve only the temperature equation, in particular the diffusive terms. 

The case just described is simulated with the present numerical code over a uniform cartesian grid with a resolution of $24$ grid per diameter of the sphere. The results of this simulation are compared with a finite volume solver (ANSYS Fluent commercial software), using an orthogonal body-fitted grid around the sphere and the box surfaces, in which the body-fitted mesh allows the solver to capture the sharp temperature gradients at the interface accurately.  Figure~\ref{fig:valid2}a shows a two-dimensional section of the grid across the middle of the computational domain.

As displayed in figure~\ref{fig:valid2}b, the effective thermal diffusivity, normalized by the thermal diffusivity of the fluid phase, computed by the present numerical code very closely matches that obtained by the finite volume solver over the boundary fitted grid: the difference is less than $1\%$. This comparison shows that the current numerical method is very well able to capture the jump in the thermal diffusivity in the range of thermal diffusivity ratio $\Gamma$ investigated in this study.

\begin{figure}
 \centering
 \includegraphics[width=0.6\textwidth]{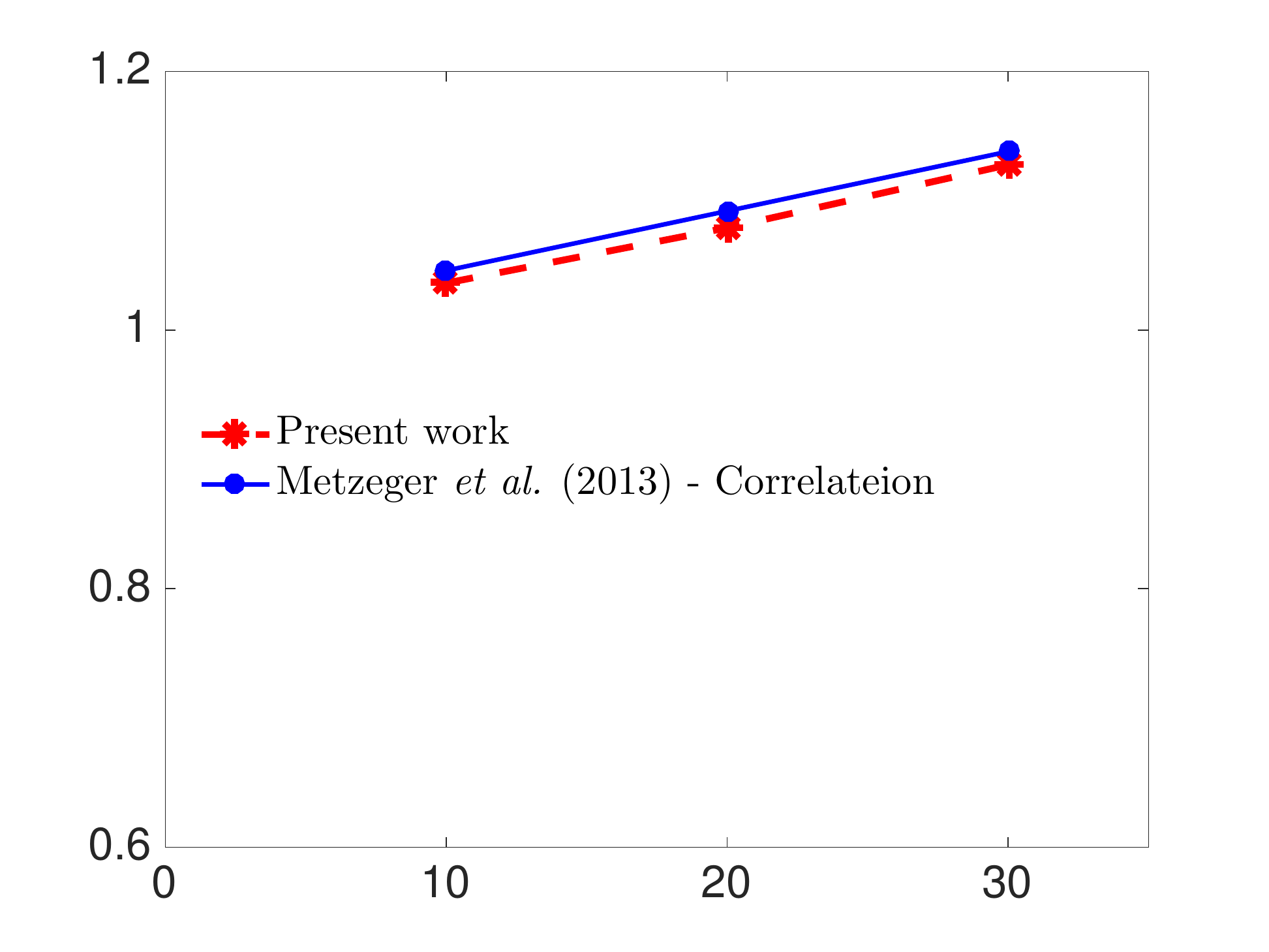}
   \put(-128,0){$\phi \, (\%)$} 
   \put(-220,85){\rotatebox{90}{$\alpha_r$}}  
 \caption{Effective thermal diffusivity, normalized by the thermal diffusivity of the single phase flow, $\alpha_r$, for the different volume fractions under investigation. The parameters for these simulations are set to $Pr=20$, $2h/D=6$, $Re_p=0.5$ and $Pe = Re_p Pr = 10$.}
\label{fig:valid1}
\end{figure}

As further validation, directly relevant to this work, we recall that \cite{Metzger2013} have proposed a correlation ($\alpha_e/\alpha_f=1+0.046 \phi Pe$) based on  experimental and numerical data of spherical particles in a Couette flow in the inertialess Stokes regime, i.e.\ valid when the particle Reynolds number is less than $0.5$. As  validation of our numerical code, simulations are therefore performed at $Re_p =0.5$ for different volume fractions $10\%$, $20\%$ and $30\%$ with $\Gamma=1$. 
The Prandtl number is set to $20$ and the results compared to the  correlation in \cite{Metzger2013}.  
Figure~\ref{fig:valid1} depicts the comparison between the effective thermal diffusivity of the suspension, normalized by the thermal diffusivity of the single phase flow, $\alpha_r$, obtained in this work and the suggested correlation. It is observed that the present numerical results are in good agreement with the empirical fit in literature, as further shown when discussing the results.

\section{Heat transfer in a particle suspension subject to uniform shear}
\label{Results}

\subsection{Effect of the particle inertia on the heat transfer}

\begin{figure}
  \centering
   \includegraphics[width=0.495\textwidth]{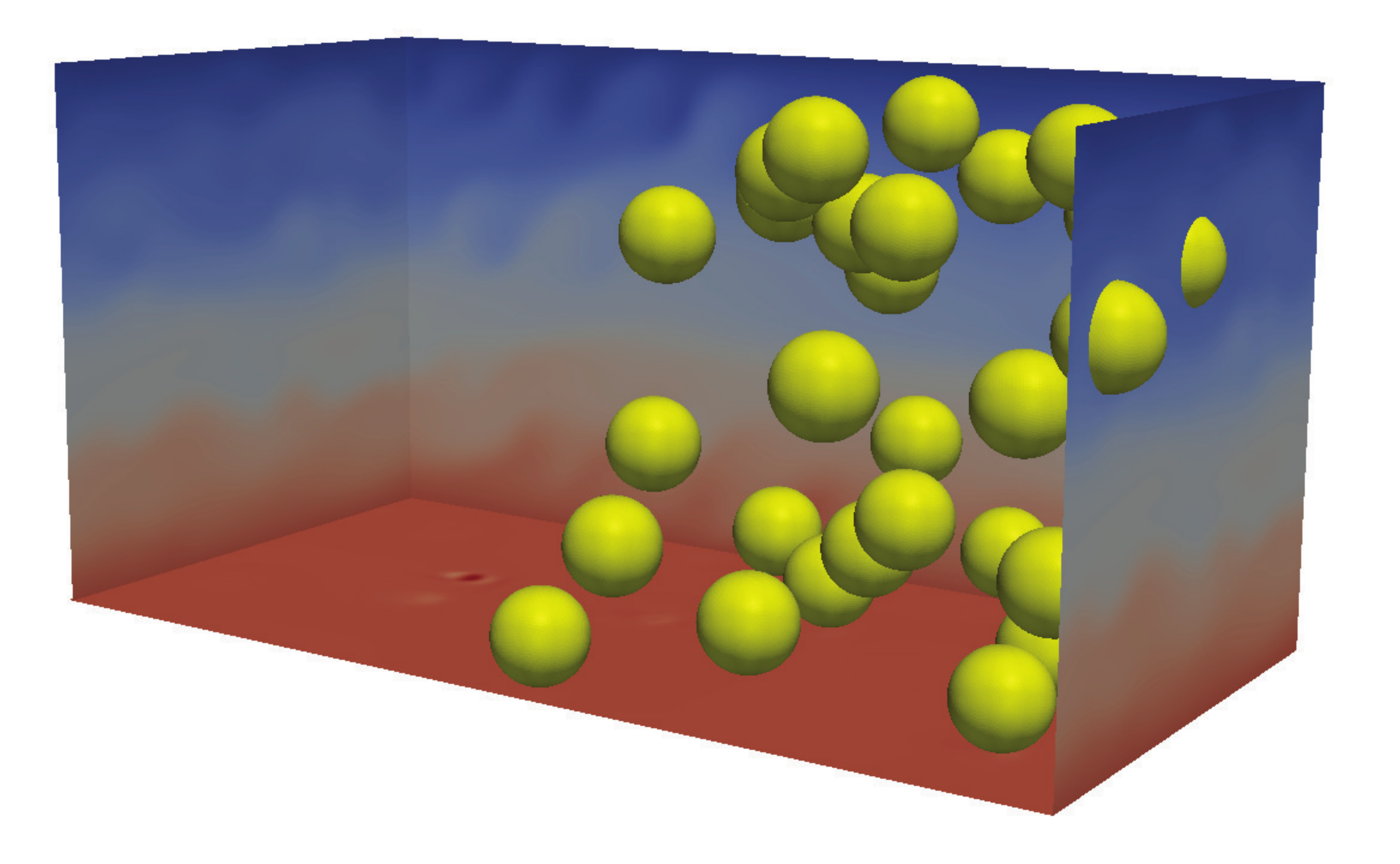}
   \includegraphics[width=0.495\textwidth]{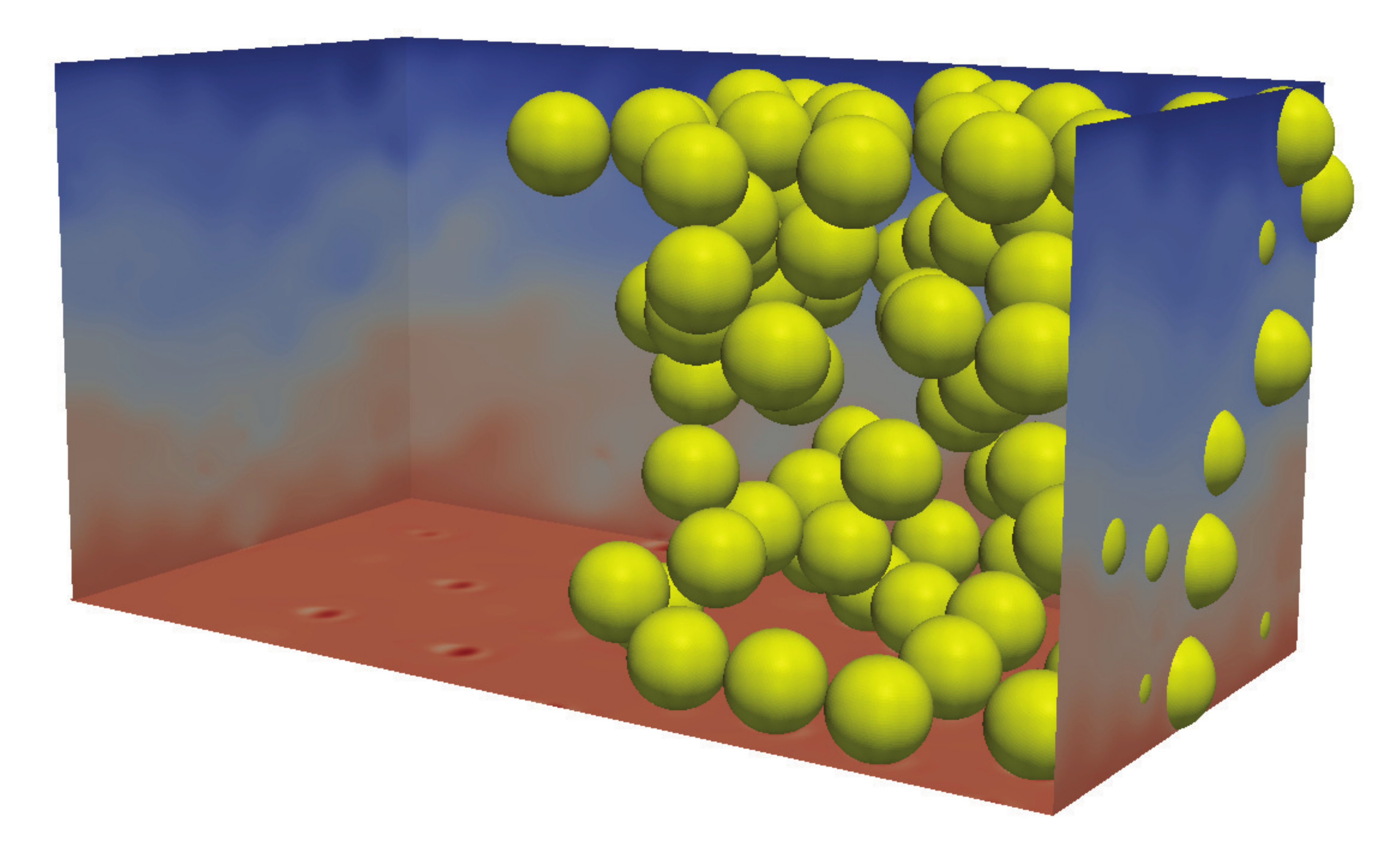}
 \put(-392,102){$(a)$}
\put(-198,102){$(b)$} \\    
   \includegraphics[width=0.495\textwidth]{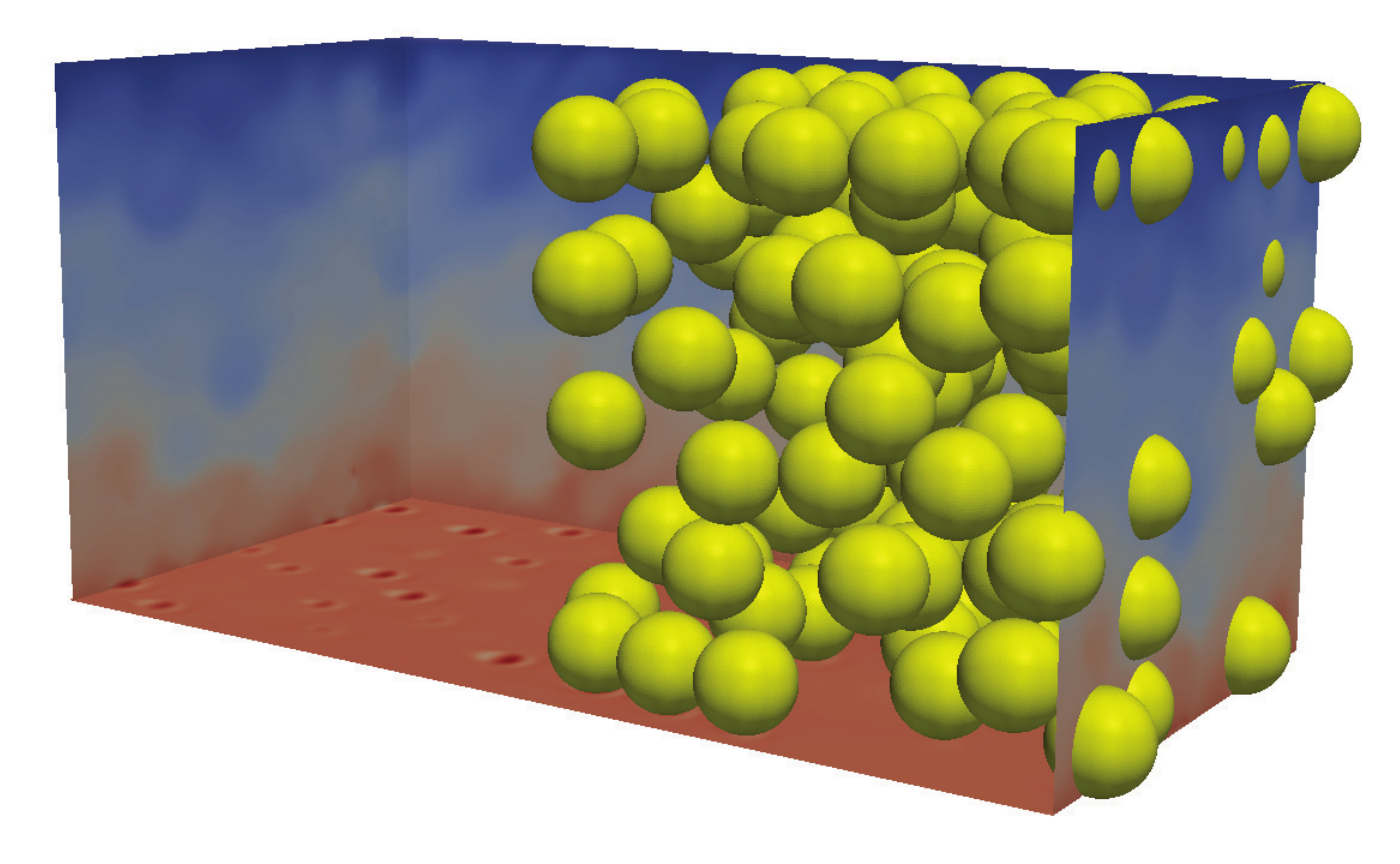} 
\put(-198,102){$(c)$}      
   \vspace{5pt}
  \caption{Snapshots of the temperature in the suspension flow for the volume fractions $\phi = 10, \,20$ and $30\%$ at $Re_p=16$.}
\label{fig:tempSnap}
\end{figure}

Having shown that the current implementation can reproduce the correlations obtained experimentally and numerically in the limit of vanishing inertia, we shall focus first
 on the effect of Reynolds number, finite inertia, on the heat transfer. We will consider dilute and semi-dilute suspensions at volume fractions $\phi = 3, 10, 20$ and $30\%$ with the Prandtl number $Pr=7$ and particles of same diffusivity as the fluid ($\Gamma = \alpha_p/\alpha_f =1$).
The simulations are performed with 4 values of the particle Reynolds number, $Re_p = 1, 4, 8$ and $16$.


Snapshots of the temperature in the suspension flow are shown in figure \ref{fig:tempSnap} for the volume fractions $\phi = 10, \,20$ and $30\%$ at $Re_p=16$. The instantaneous temperature is represented on different orthogonal planes with the bottom plane located near the bottom wall. Finite-size particles are displayed only on one half of the domain to give a visual feeling on how dense the solid phase is. The layering of the particles and movement between the layers can be observed for $\phi=30\%$. There is a noisy pattern in the temperature distribution due to the particles motion and fluctuations. At higher volume fractions these noisy patterns are stronger and more observable, as we will quantify in the following.

\begin{figure}
  \centering
   \includegraphics[width=0.495\textwidth]{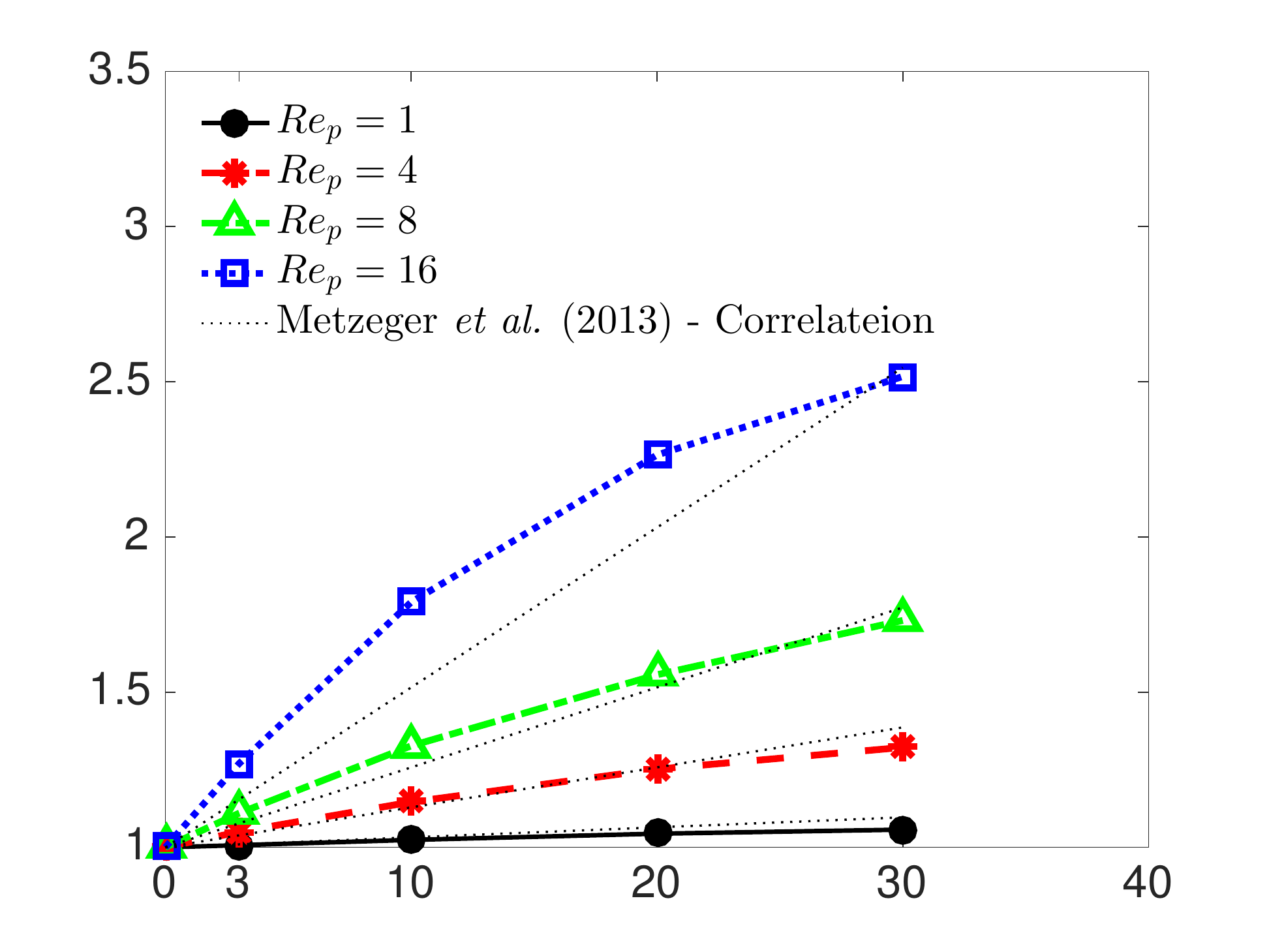}
    \includegraphics[width=0.495\textwidth]{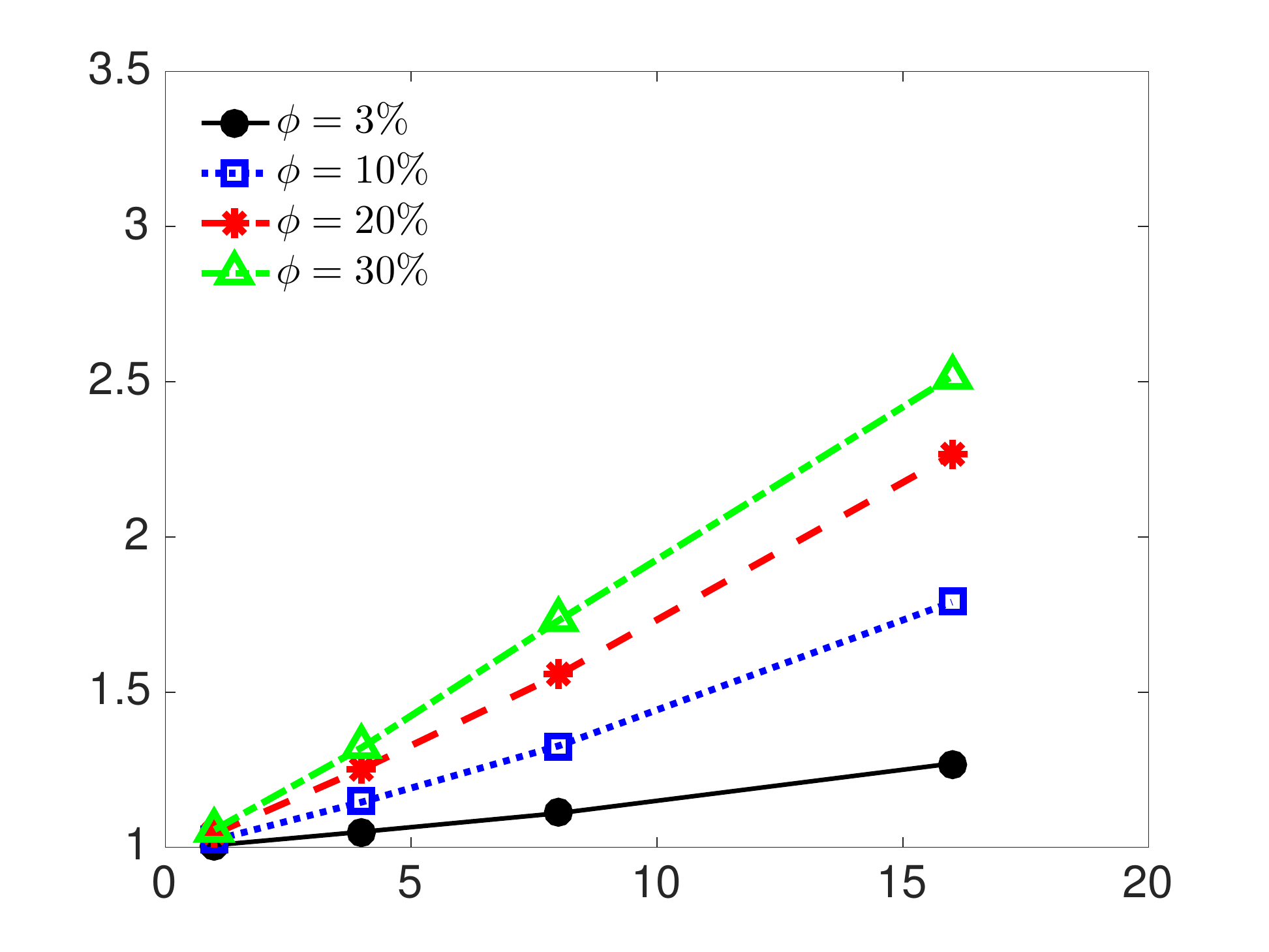} 
 \put(-391,122){$(a)$}
\put(-197,122){$(b)$} 
   \put(-294,-3){$\phi \, (\%)$} 
   \put(-99,-3){$Re_p$} 
   \put(-185,70){\rotatebox{90}{$\alpha_r$}}   
   \put(-378,70){\rotatebox{90}{$\alpha_r$}}  \\         
   \includegraphics[width=0.495\textwidth]{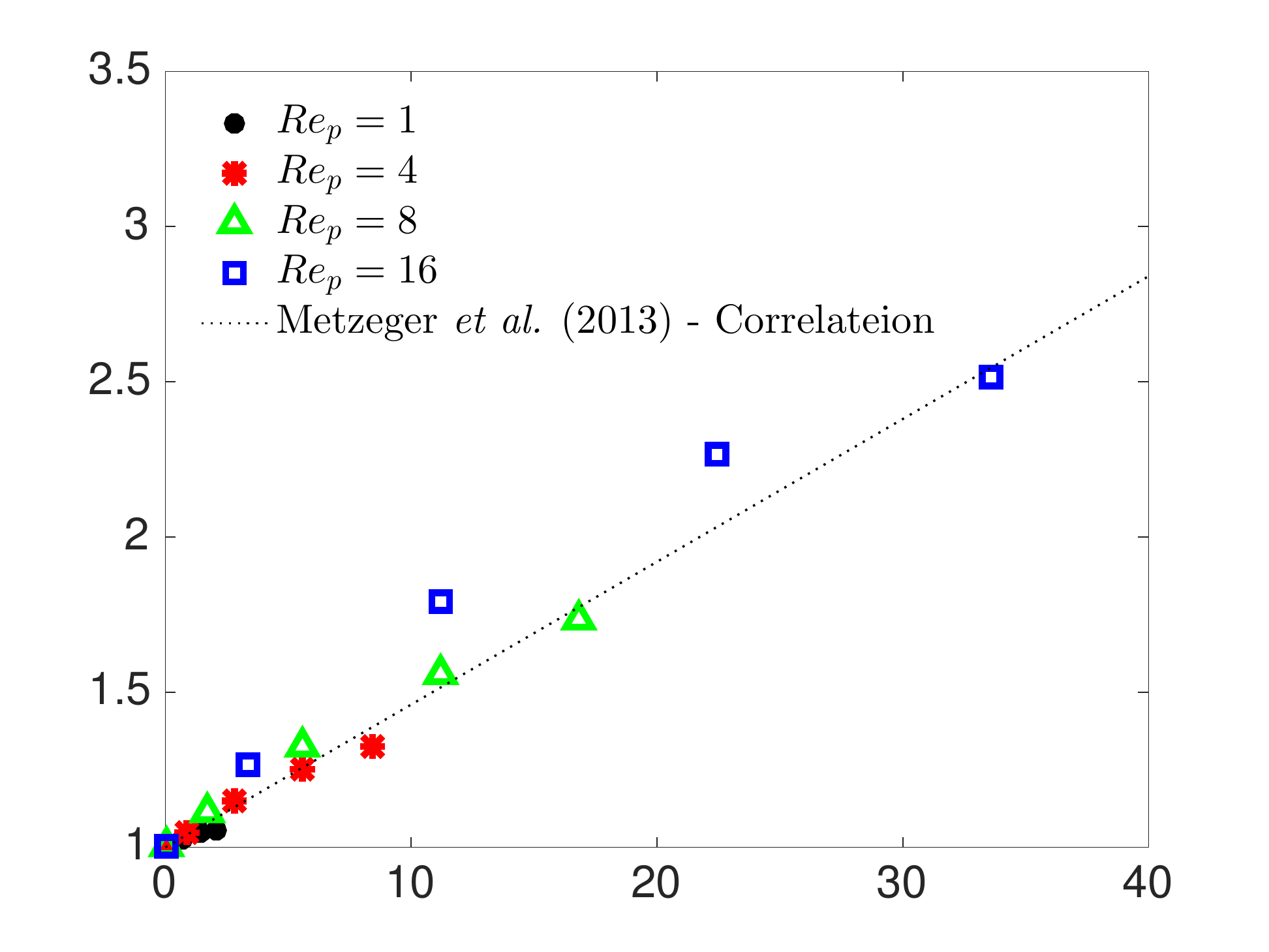}  
\put(-197,122){$(c)$}   
   \put(-100,-3){$\phi Pe$} 
   \put(-185,70){\rotatebox{90}{$\alpha_r$}}
   \vspace{5pt}
  \caption{The effective thermal diffusivity of the plain Couette flow, normalized with that for the single phase flow, $\alpha_r$ versus $(a)$ volume fraction percentage $\phi$,  $(b)$ particle Reynolds number $Re_p$ and $(c)$ $\phi Pe$. The linear correlation, suggested by \cite{Metzger2013} is depicted in $(a)$ and $(c)$ by black dotted lines.}
\label{fig:Re_Heat}
\end{figure}
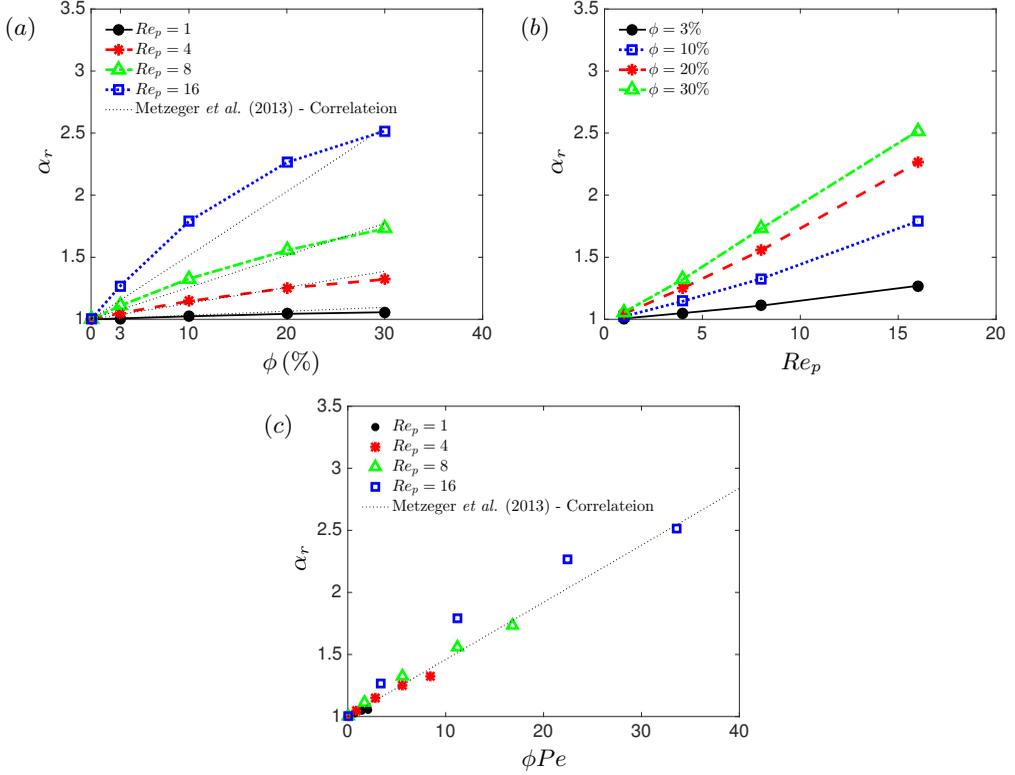
 
 The effective thermal diffusivity of the suspension, normalized with that of the single phase flow, $\alpha_r$,  is reported in figure~\ref{fig:Re_Heat}a for all cases under investigation. The correlation, suggested by \cite{Metzger2013} is also depicted in this figure. This correlation shows a linear variation of the effective thermal diffusivity of the suspension with the particle volume fraction in the Stokes regime. The data  are observed to follow the relation proposed in \cite{Metzger2013} at the two lowest $Re_p$ and then deviate as $Re_p$ increases. 
 At high $Re_p$, the slope of the curves is significantly higher than what predicted for vanishing inertia, and the points do not follow a straight line. 
 We also note that the sudden decrease of the effective thermal diffusivity observed at high volume fractions in  \cite{Metzger2013} seems to appear earlier in the inertial regime (cf.\ the curve at $Re_p=16$).
 
Figure~\ref{fig:Re_Heat}b depicts the same data as in panel (a), the normalized effective thermal diffusivity, now displayed versus $Re_p$ for the different volume fractions considered. 
The results show that the normalized effective diffusivity $\alpha_r$ varies almost linearly with $Re_p$ at fixed volume fraction in the range of $Re_p$ studied here. It should be noted that the thermal diffusivity inside the particles is the same as that of the fluid for the results in this section, hence only the wall-normal particle motions
 can enhance the mean diffusion. Depicting the results versus $\phi Pe$ as in figure~\ref{fig:Re_Heat}c, where $Pe$ is the Peclet number defined as $Pr Re_p$, demonstrates that the linear trend suggested by \cite{Metzger2013} can approximate the effective thermal diffusivity well roughly up to $Re_p =8$.    

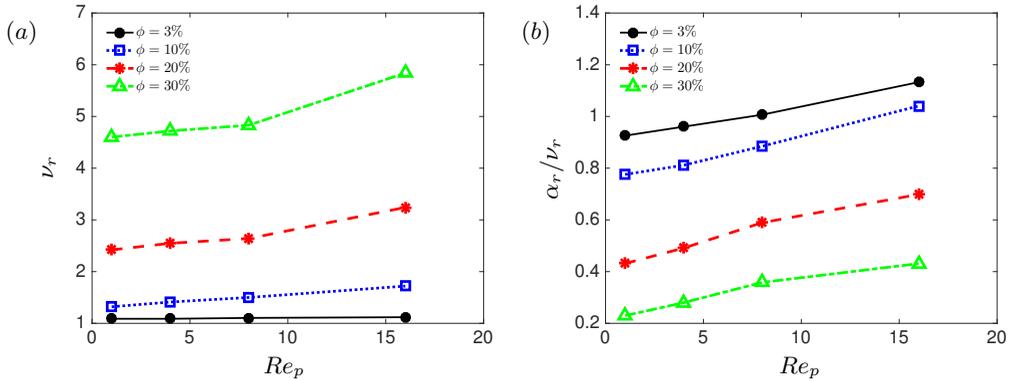
\begin{figure}
  \centering
   \includegraphics[width=0.495\textwidth]{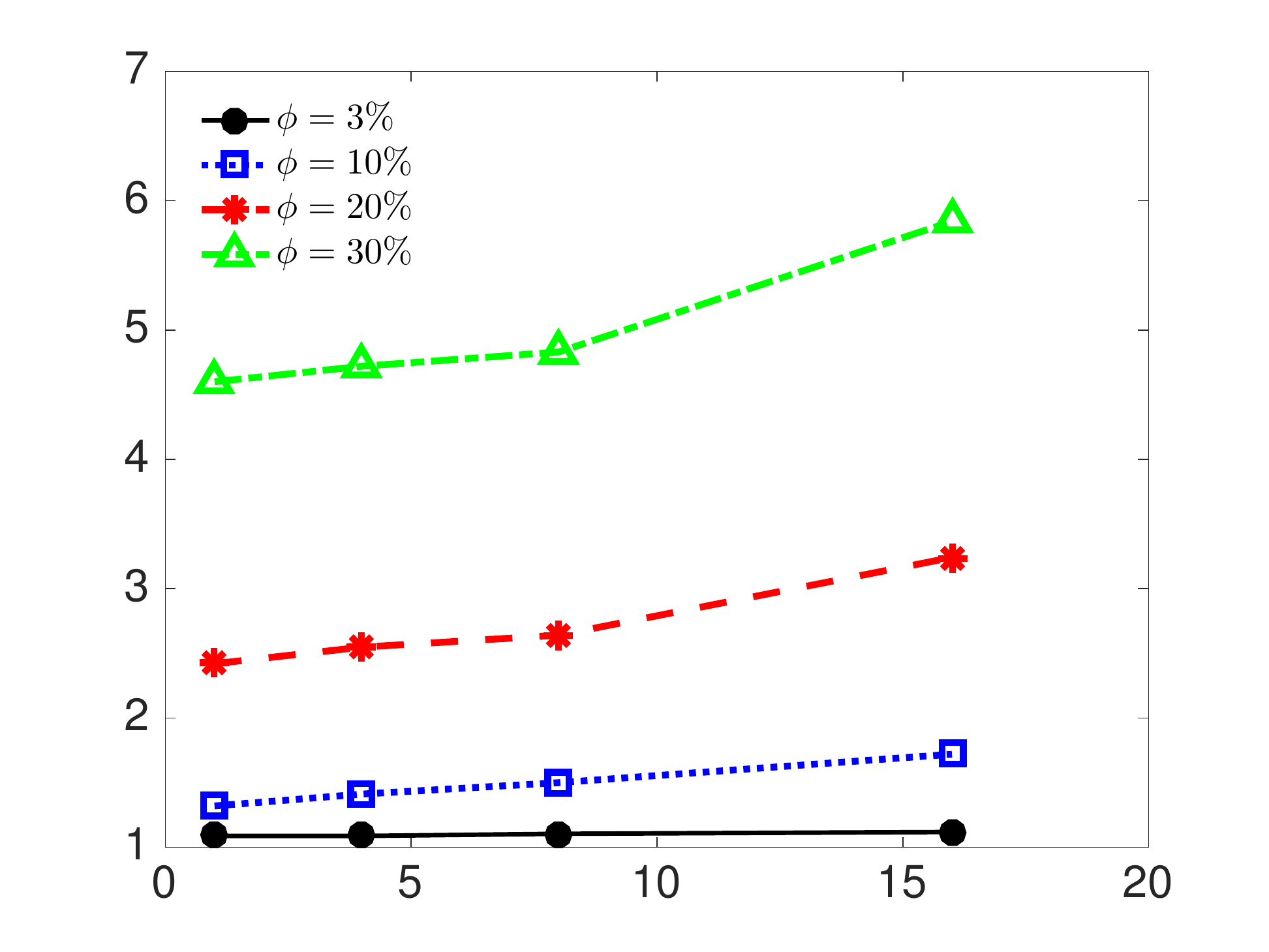}
   \includegraphics[width=0.495\textwidth]{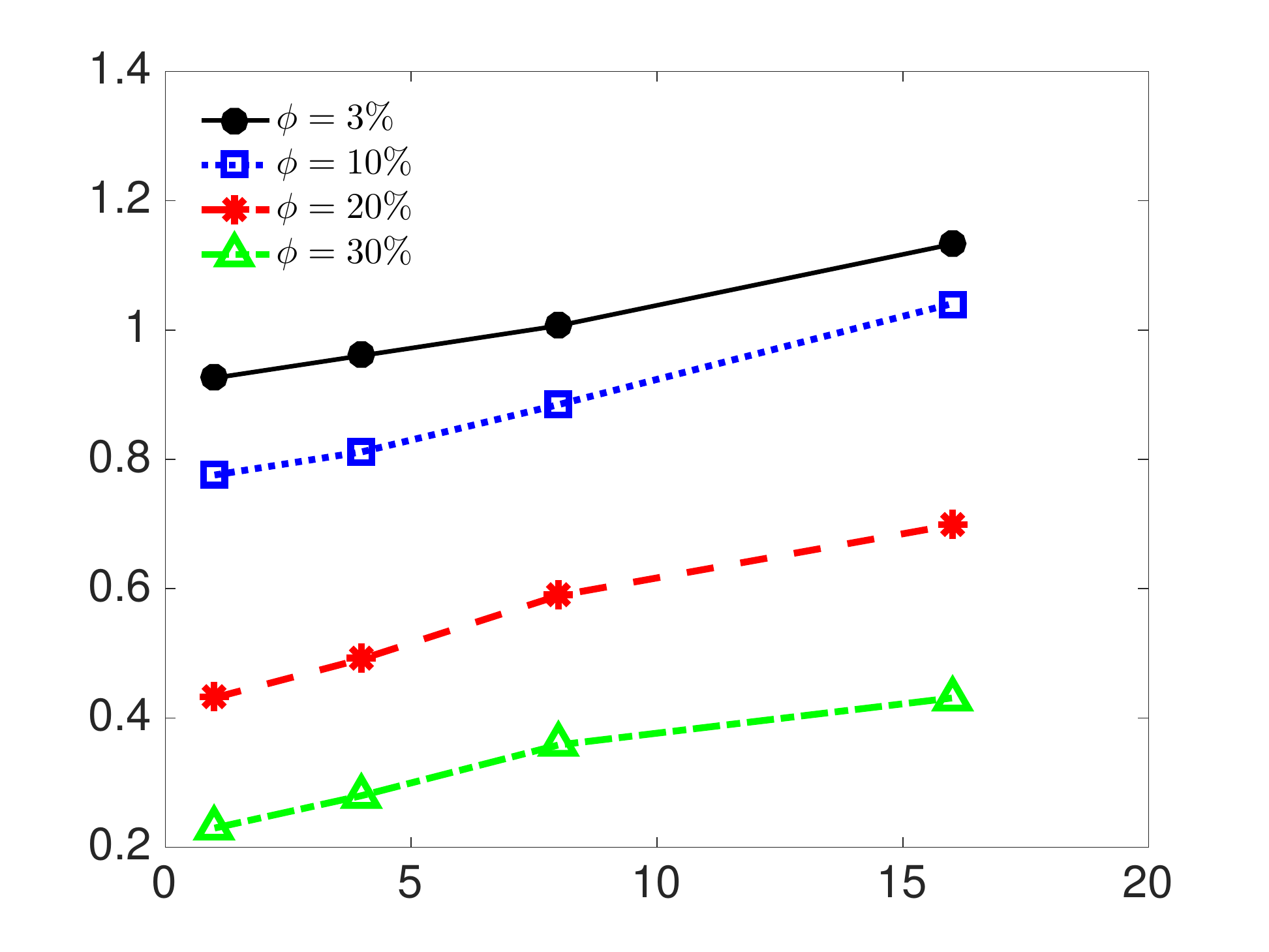}
 \put(-391,122){$(a)$}
\put(-197,122){$(b)$}    
\put(-294,-3){$Re_p$} 
   \put(-100,-3){$Re_p$} 
   \put(-190,62){\rotatebox{90}{$\alpha_r / \nu_r$}}   
   \put(-377,70){\rotatebox{90}{$\nu_r$}} \\   
  \caption{$(a)$ Effective viscosity of the suspension, normalized by the viscosity of the fluid and $(b)$ the ratio between the increase of the heat transfer $\alpha_r$ and the increase of the suspension effective viscosity $\nu_r$.}
\label{fig:viscosity}
\end{figure}

One aspect that should be considered when aiming to increase the heat transfer by employing particulate flows is that the increase of the effective thermal diffusivity usually comes at the price of an increase in the effective viscosity of the flow. This means a higher friction at the walls and  higher external power needed to drive the flow. 
To quantitatively address this issue, the effective viscosity of the suspension, normalized by the viscosity of the single phase flow, $\nu_r \equiv\nu_e/\nu_f$, is computed for all cases and depicted in figure~\ref{fig:viscosity}a versus $Re_p$. 
The data confirm an increase of the suspension viscosity with $\phi$ and with fluid and particle inertia: the former, at low $Re_p$, follows closely classical empirical fits, like Eiler's fit,  whereas the latter, also denoted as inertial shear-thickening, is discussed and explained in \cite{Picano2013} among others.

The ratio between $\alpha_r$ and $\nu_r$ can be used to measure the global efficiency of the heat transfer increase, like an effective Prandtl number of the suspension, similarly to the turbulent Prandtl used to quantify mixing in turbulent flows with respect to laminar flows. 
This ratio is depicted versus $Re_p$ in figure~\ref{fig:viscosity}b. The ratio is observed to increase with $Re_p$ and decrease with the volume fraction $\phi$.   
The increase of heat transfer obtained adding particles is always lower than the increase of the suspension viscosity for $\Gamma=1$ and $\phi>3 \%$ except for the case with $\phi=10 \%$ and $Re_p =16$; inertial effects, on the other hand, are more pronounced on the energy transfer than on the momentum transfer.        

\begin{figure}
  \centering
   \includegraphics[width=0.495\textwidth]{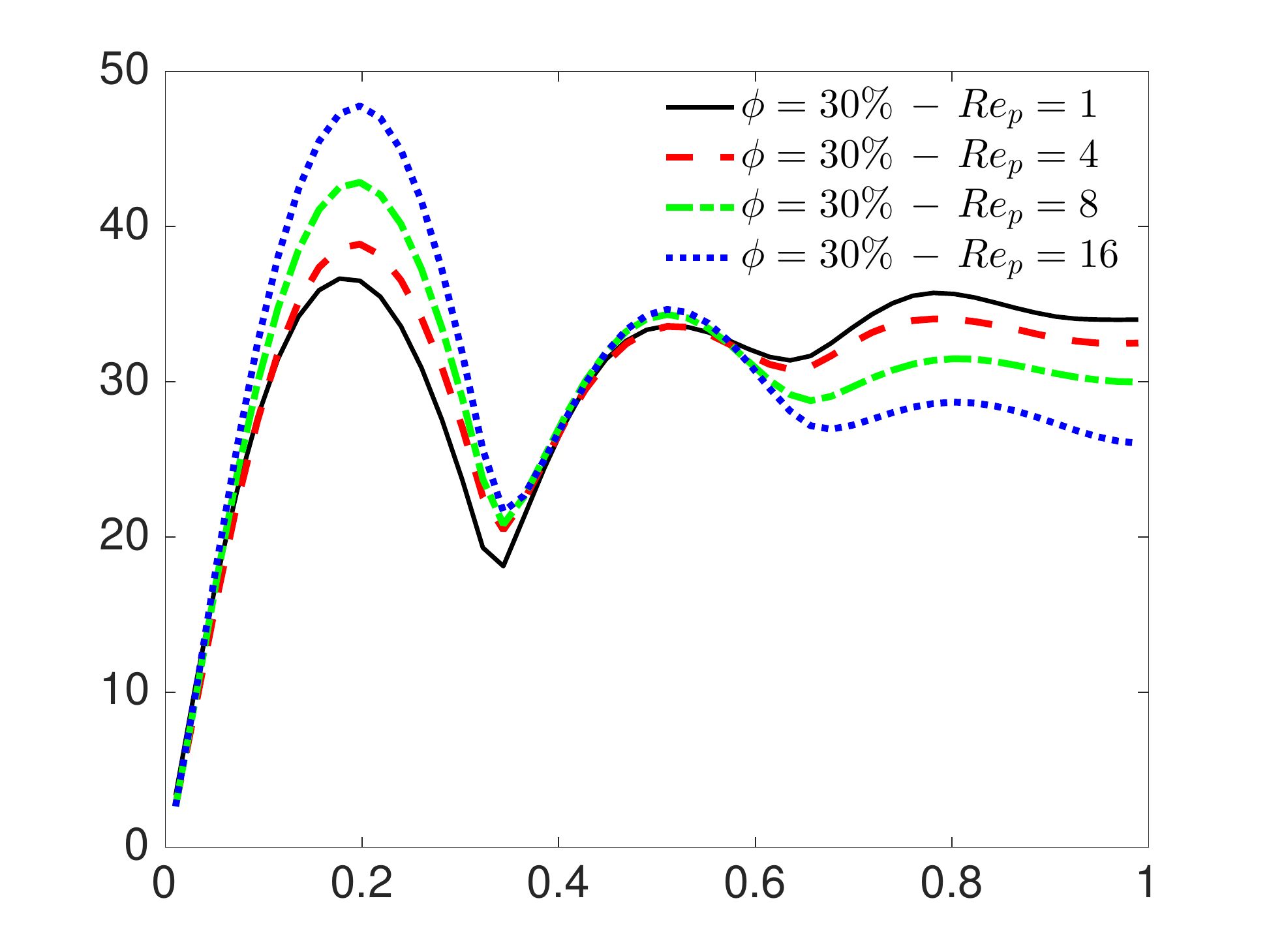}
   \includegraphics[width=0.495\textwidth]{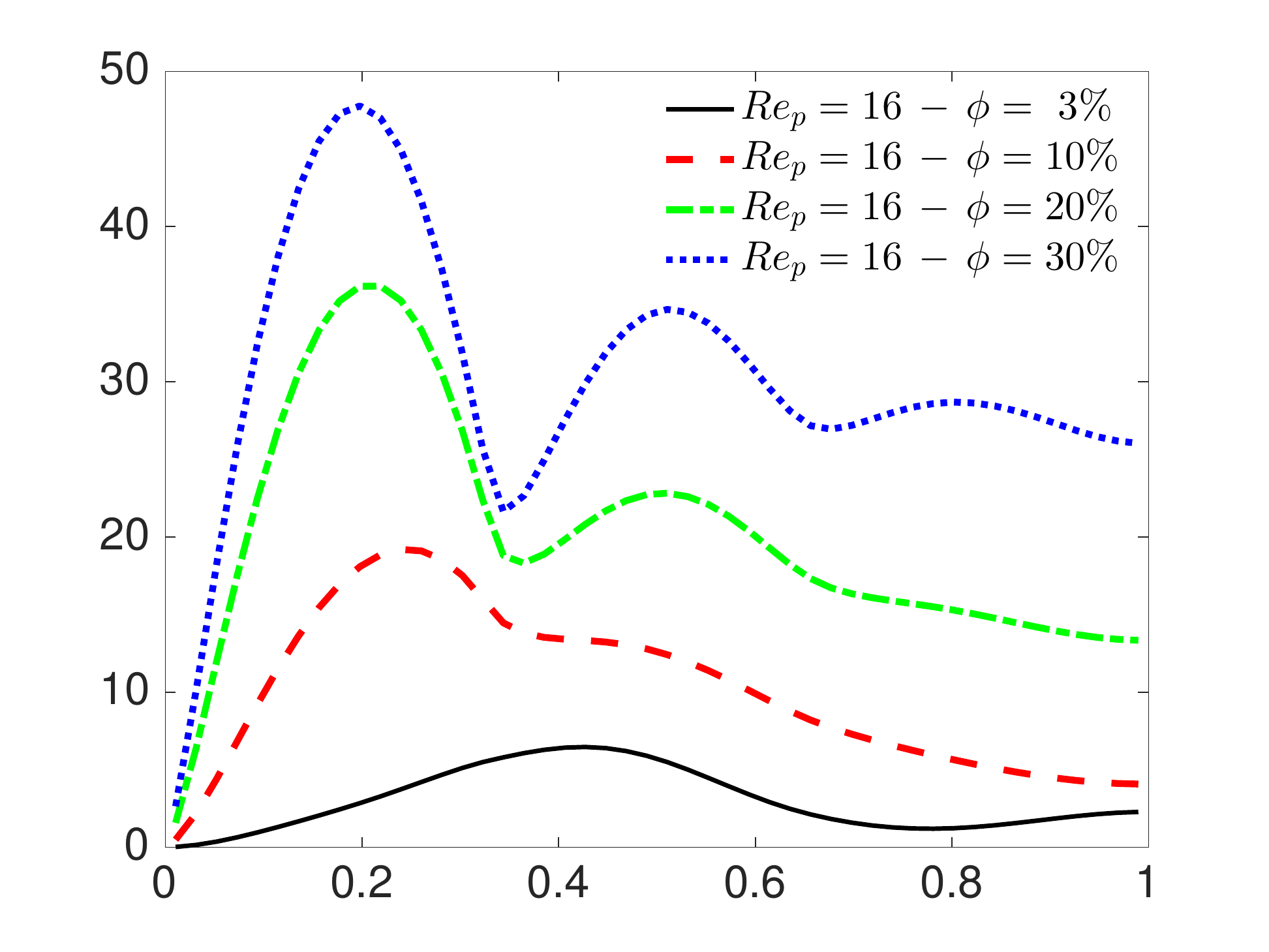}
 \put(-391,122){$(a)$}
\put(-197,122){$(b)$}    
\put(-293,-3){$y/h$} 
   \put(-100,-3){$y/h$} 
   \put(-190,66){\rotatebox{90}{$\Phi (y)$}}   
   \put(-384,66){\rotatebox{90}{$\Phi (y)$}} \\   
  \caption{Local volume fraction $\Phi (y)$ versus the normalized distance to the wall $y/h$ for $(a)$ the different particle Reynolds number under investigation at $\phi = 30 \%$ and $(b)$ different volume fractions at $Re_p = 16$.}
\label{fig:phi}
\end{figure}

\begin{figure}
  \centering
   \includegraphics[width=0.495\textwidth]{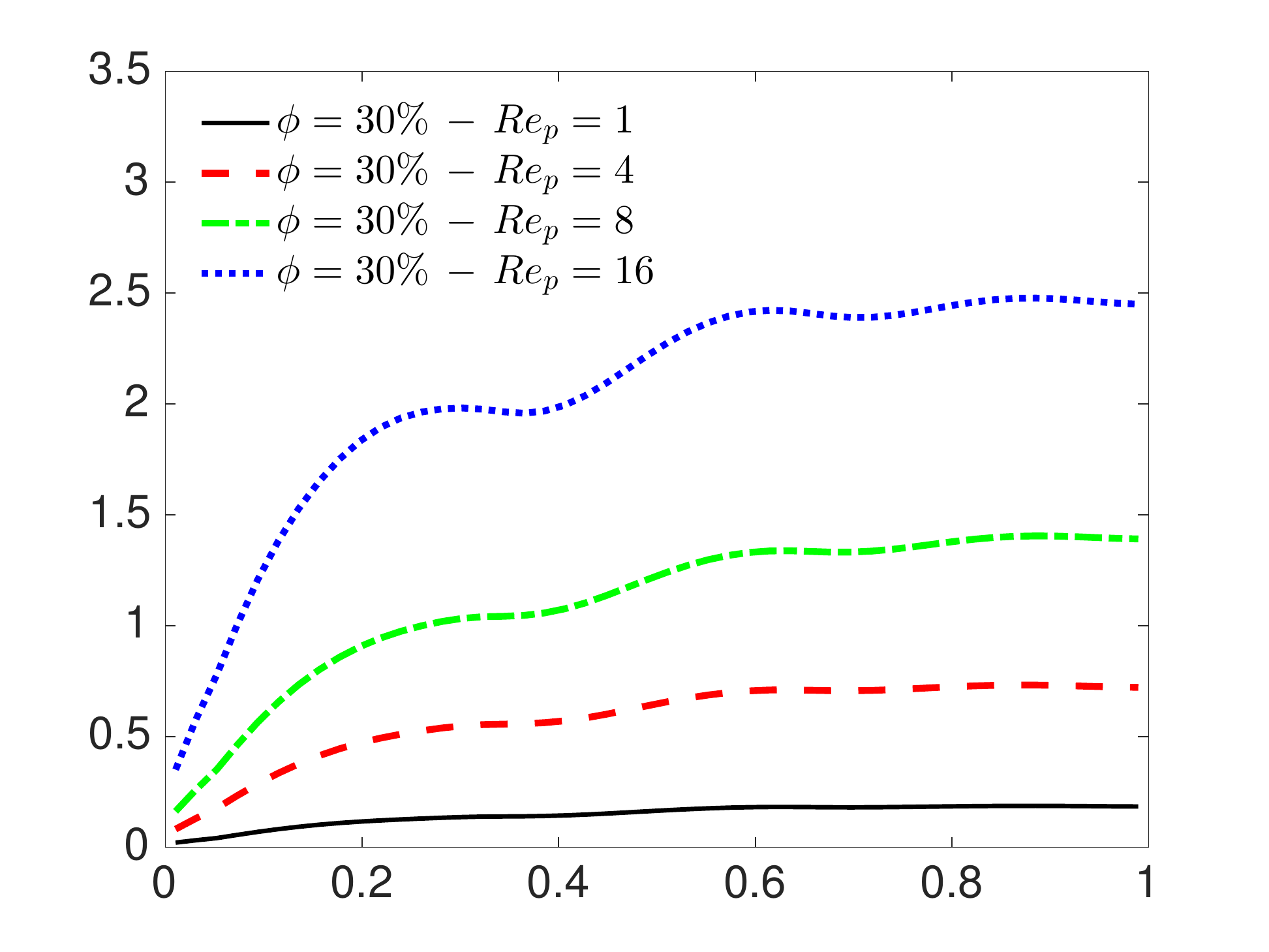}
   \includegraphics[width=0.495\textwidth]{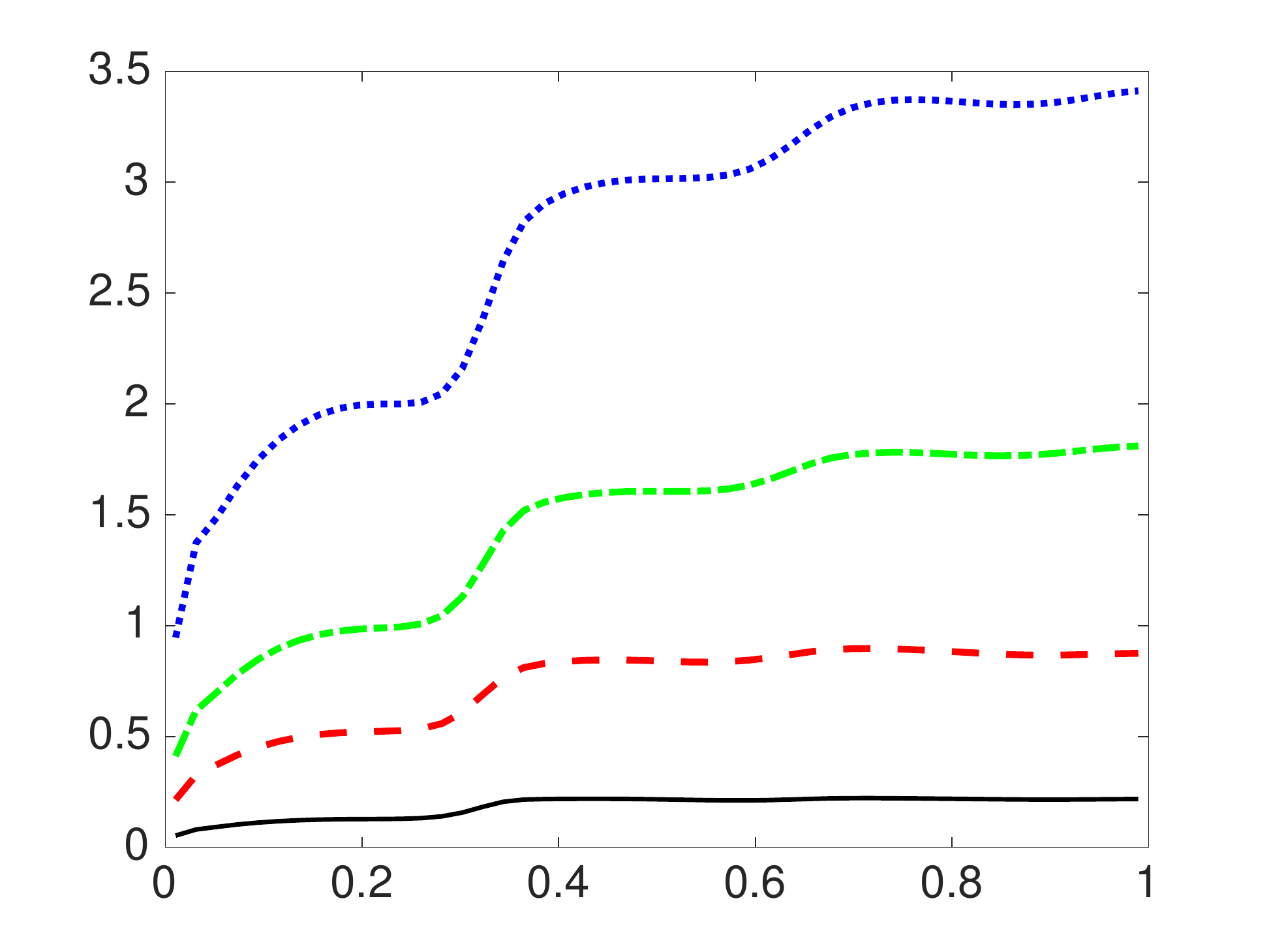}
 \put(-391,122){$(a)$}
\put(-197,122){$(b)$}    
\put(-293,-3){$y/h$} 
   \put(-100,-3){$y/h$} 
   \put(-190,62){\rotatebox{90}{$v^{\prime}_p D / \nu$}}   
   \put(-384,62){\rotatebox{90}{$v^{\prime}_f D / \nu$}} \\  
  \centering
   \includegraphics[width=0.495\textwidth]{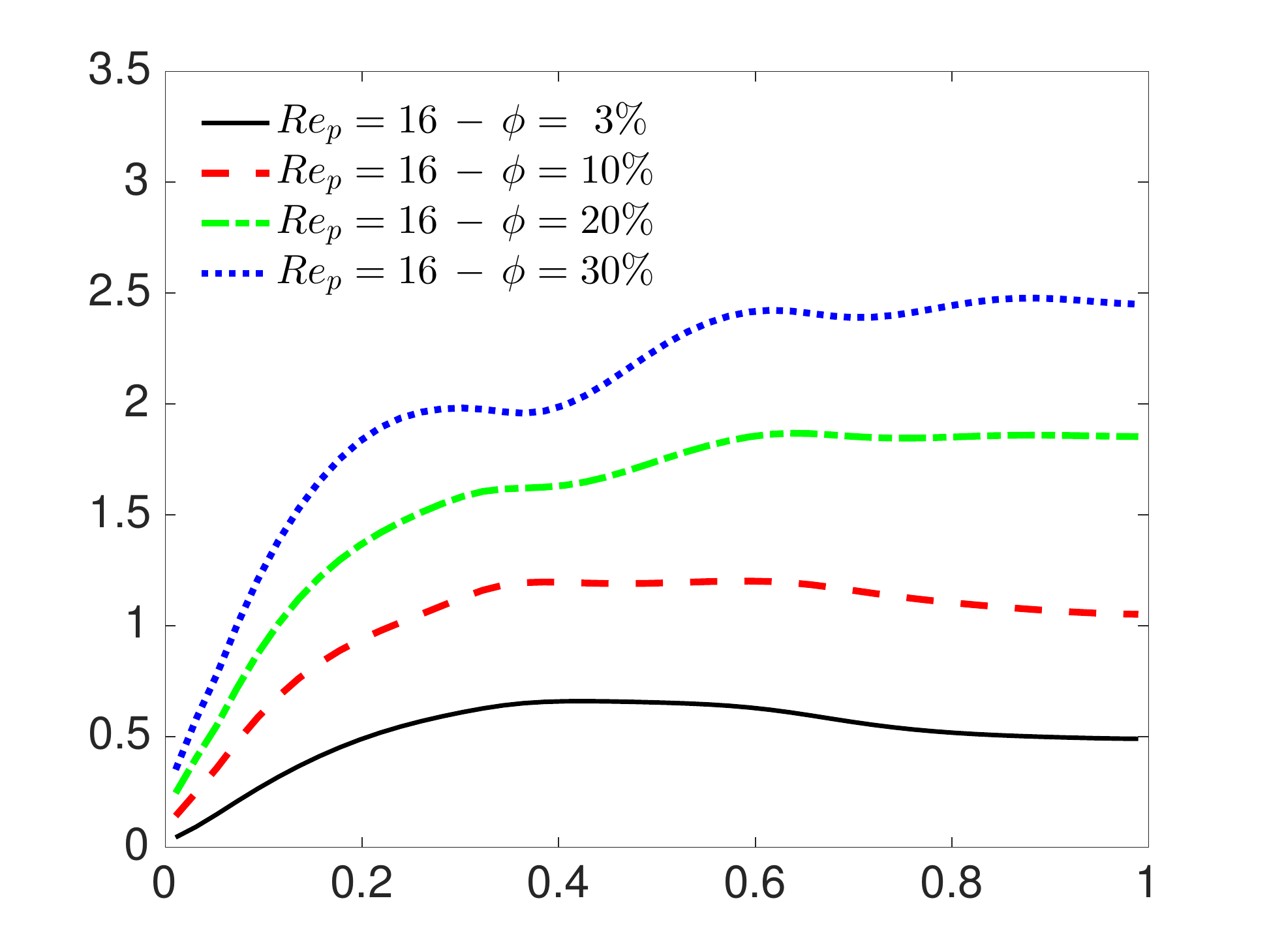}
   \includegraphics[width=0.495\textwidth]{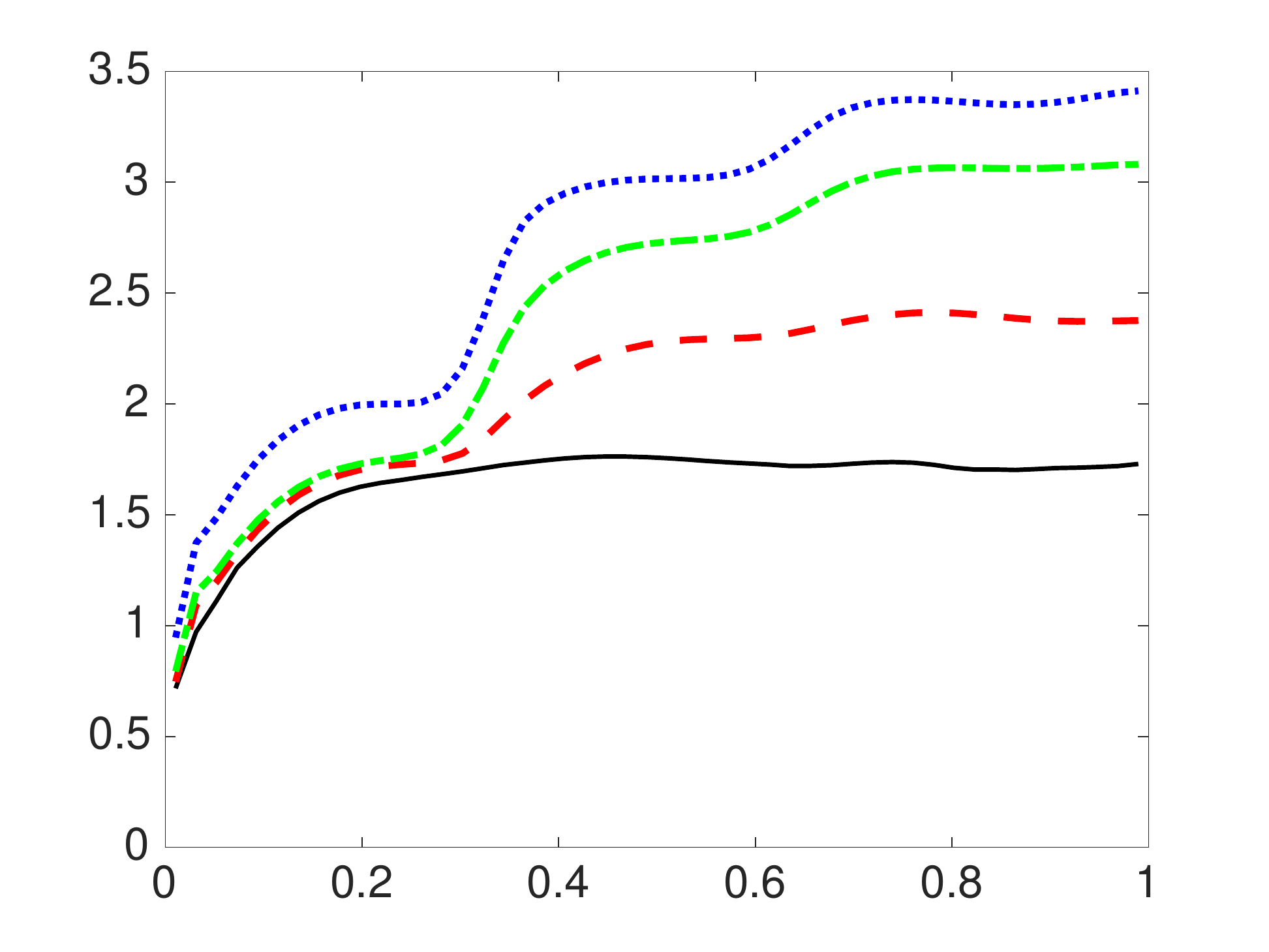}
 \put(-391,122){$(c)$}
\put(-197,122){$(d)$}    
\put(-293,-3){$y/h$} 
   \put(-100,-3){$y/h$} 
   \put(-190,62){\rotatebox{90}{$v^{\prime}_p D / \nu $}}   
   \put(-384,62){\rotatebox{90}{$v^{\prime}_f  D / \nu$}} \\      
  \caption{Root-mean-square velocity fluctuations, normalized by the diffusive velocity scale $\nu/D$, for the fluid phase ($v^{\prime}_f D / \nu$) and the particles ($v^{\prime}_p D / \nu$) versus the distance to the wall $y/h$: $(a)$ fluid phase for the different particle Reynolds number under investigation at $\phi = 30 \%$ ; $(b)$ particles -  same cases as in $(a)$ ; $(c)$ fluid phase - different volume fractions at $Re_p = 16$ ; $(d)$  particles -  same cases as in $(c)$ }
\label{fig:rms_velocity}
\end{figure}

\begin{figure}
  \centering
   \includegraphics[width=0.495\textwidth]{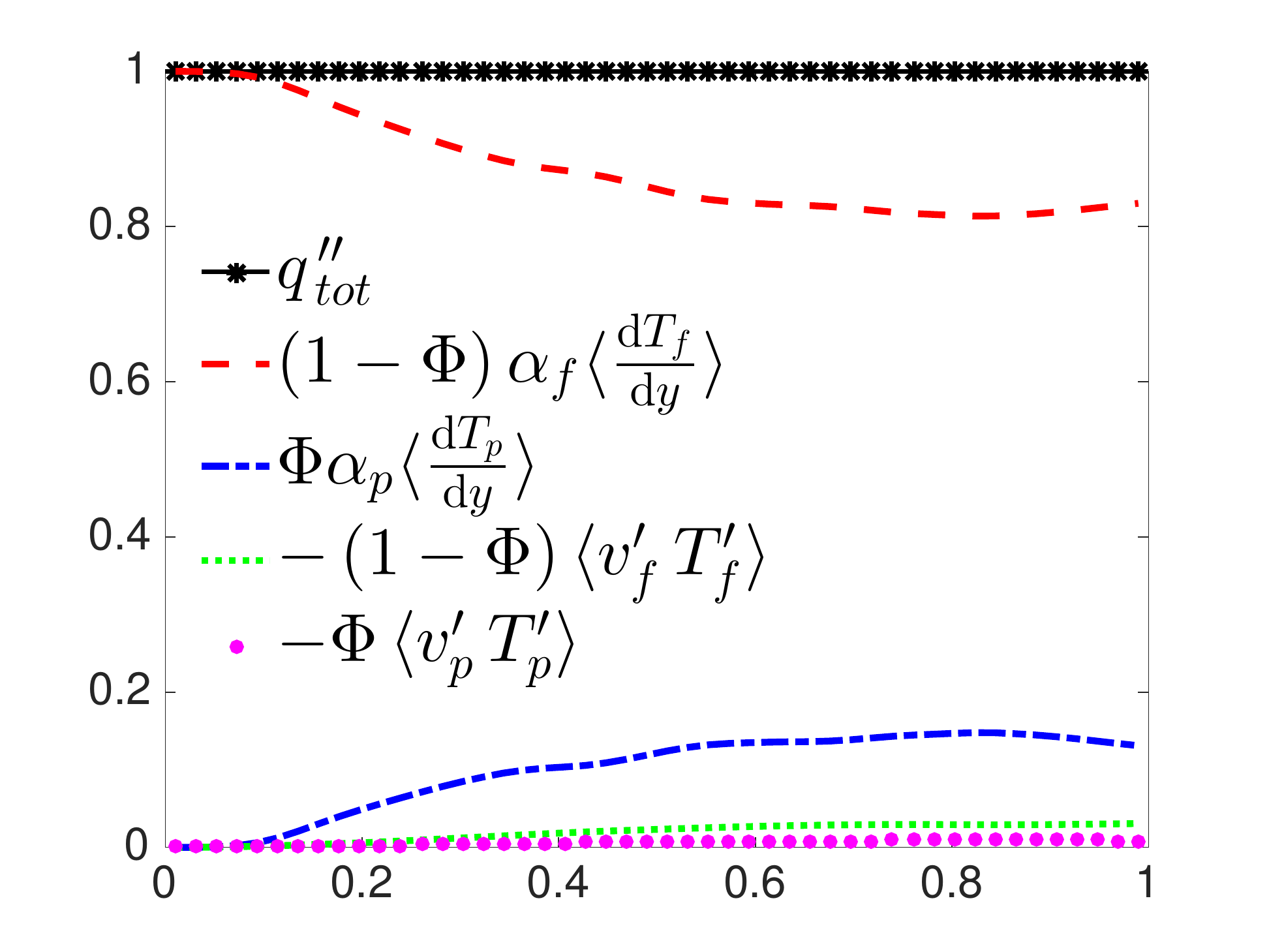}  
   \includegraphics[width=0.495\textwidth]{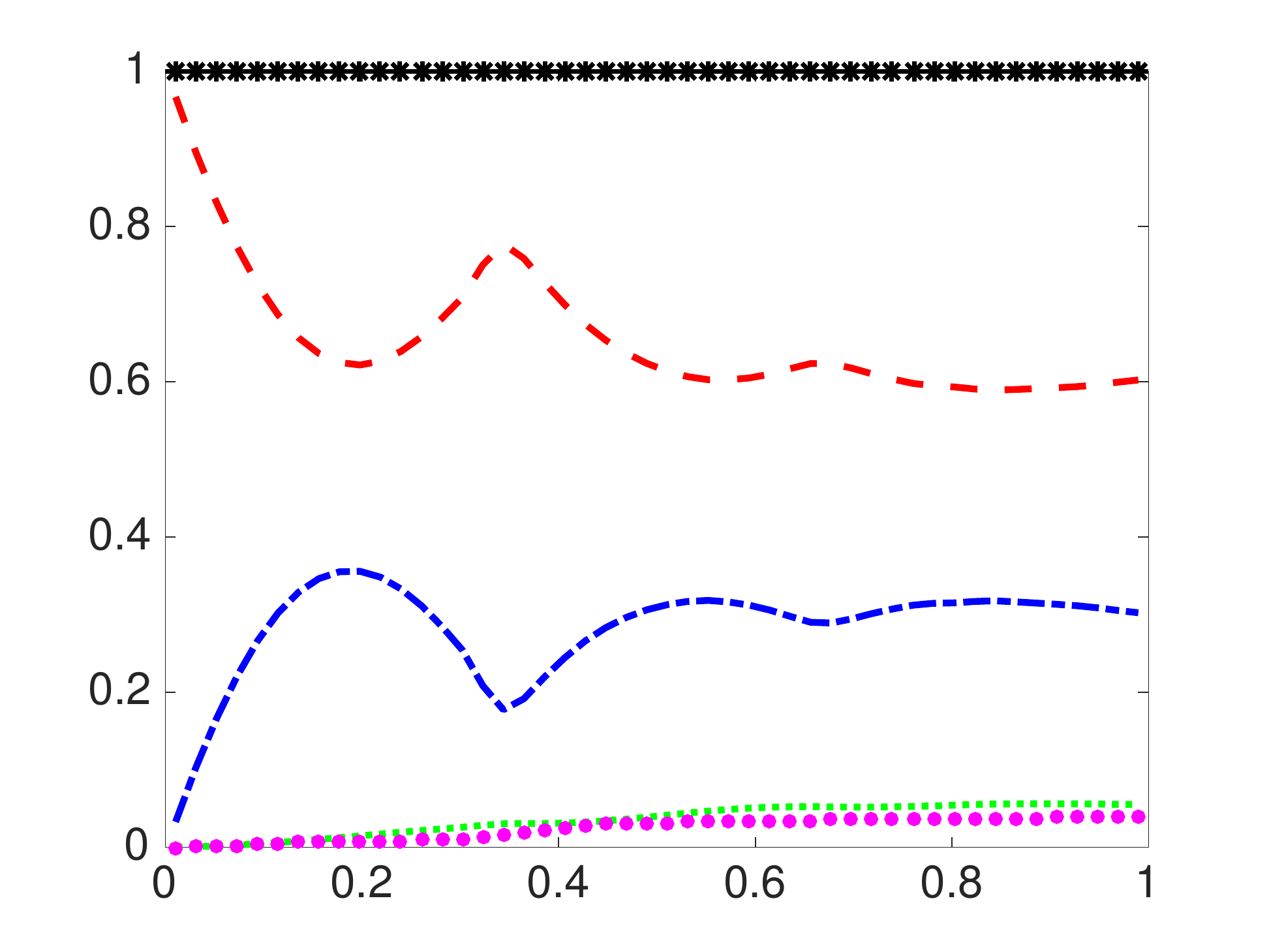}  
   \put(-391,122){$(a)$}
   \put(-197,122){$(e)$}    
   \put(-293,-3){$y/h$} 
   \put(-100,-3){$y/h$}    
   \put(-190,58){\rotatebox{90}{$q^{\,\prime \prime}_{\,\,i} / q^{\,\prime \prime}_{\,\,tot}$}}   
   \put(-384,58){\rotatebox{90}{$q^{\,\prime \prime}_{\,\,i} / q^{\,\prime \prime}_{\,\,tot}$}}\\
   \includegraphics[width=0.495\textwidth]{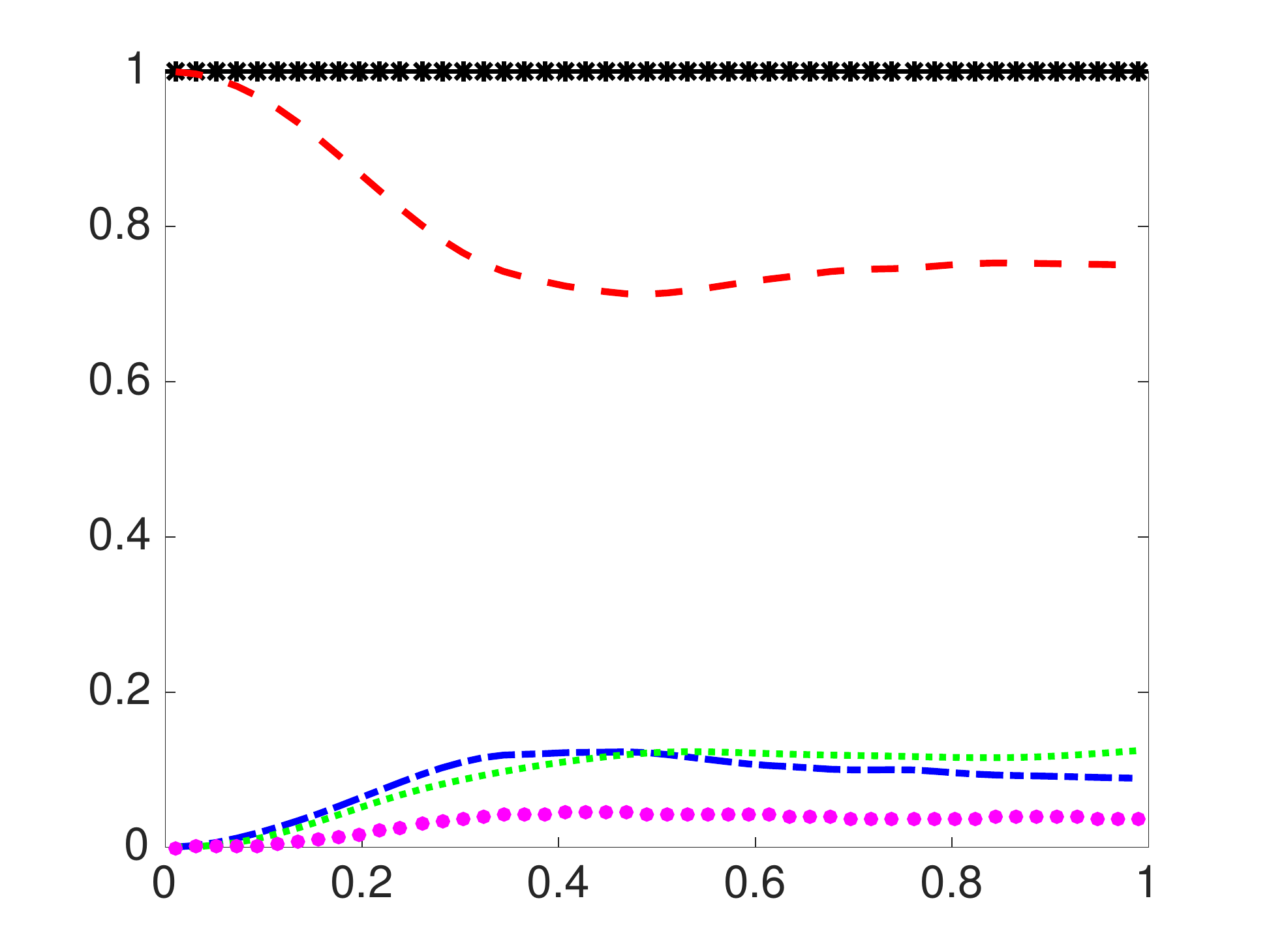} 
   \includegraphics[width=0.495\textwidth]{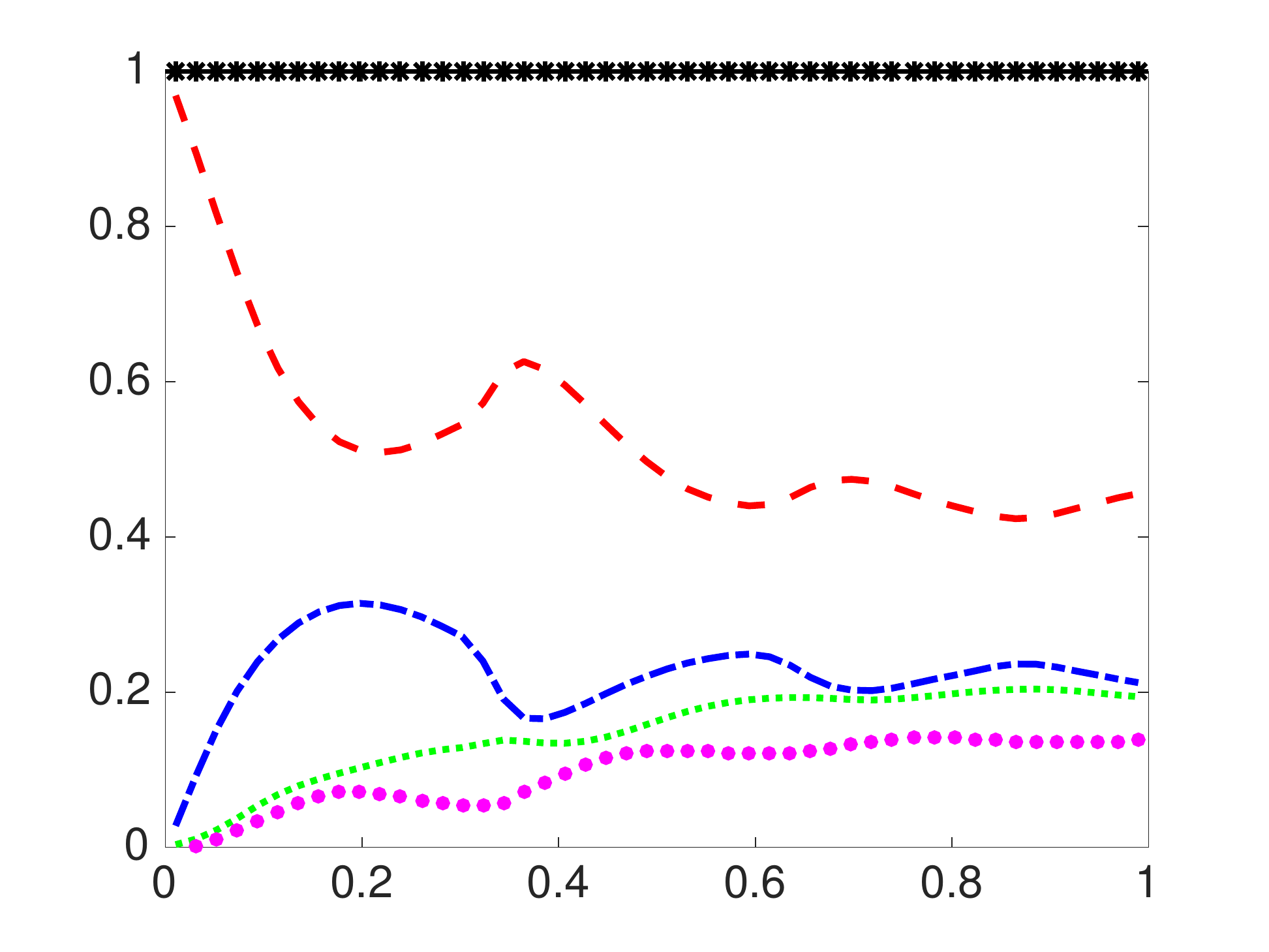} 
   \put(-391,122){$(b)$}
   \put(-197,122){$(f)$}    
   \put(-293,-3){$y/h$} 
   \put(-100,-3){$y/h$} 
   \put(-190,58){\rotatebox{90}{$q^{\,\prime \prime}_{\,\,i} / q^{\,\prime \prime}_{\,\,tot}$}}    
   \put(-384,58){\rotatebox{90}{$q^{\,\prime \prime}_{\,\,i} / q^{\,\prime \prime}_{\,\,tot}$}}\\      
   \includegraphics[width=0.495\textwidth]{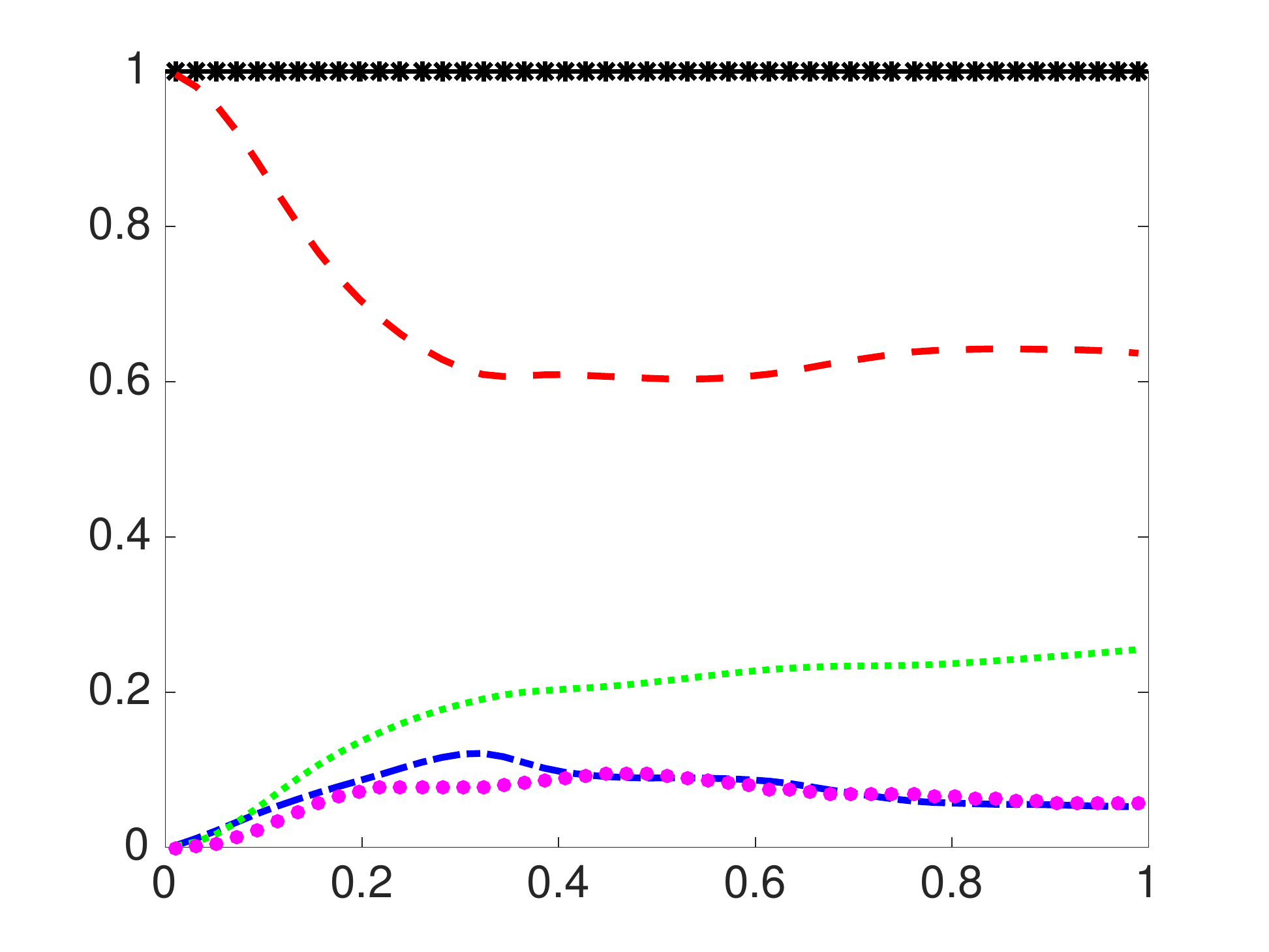} 
   \includegraphics[width=0.495\textwidth]{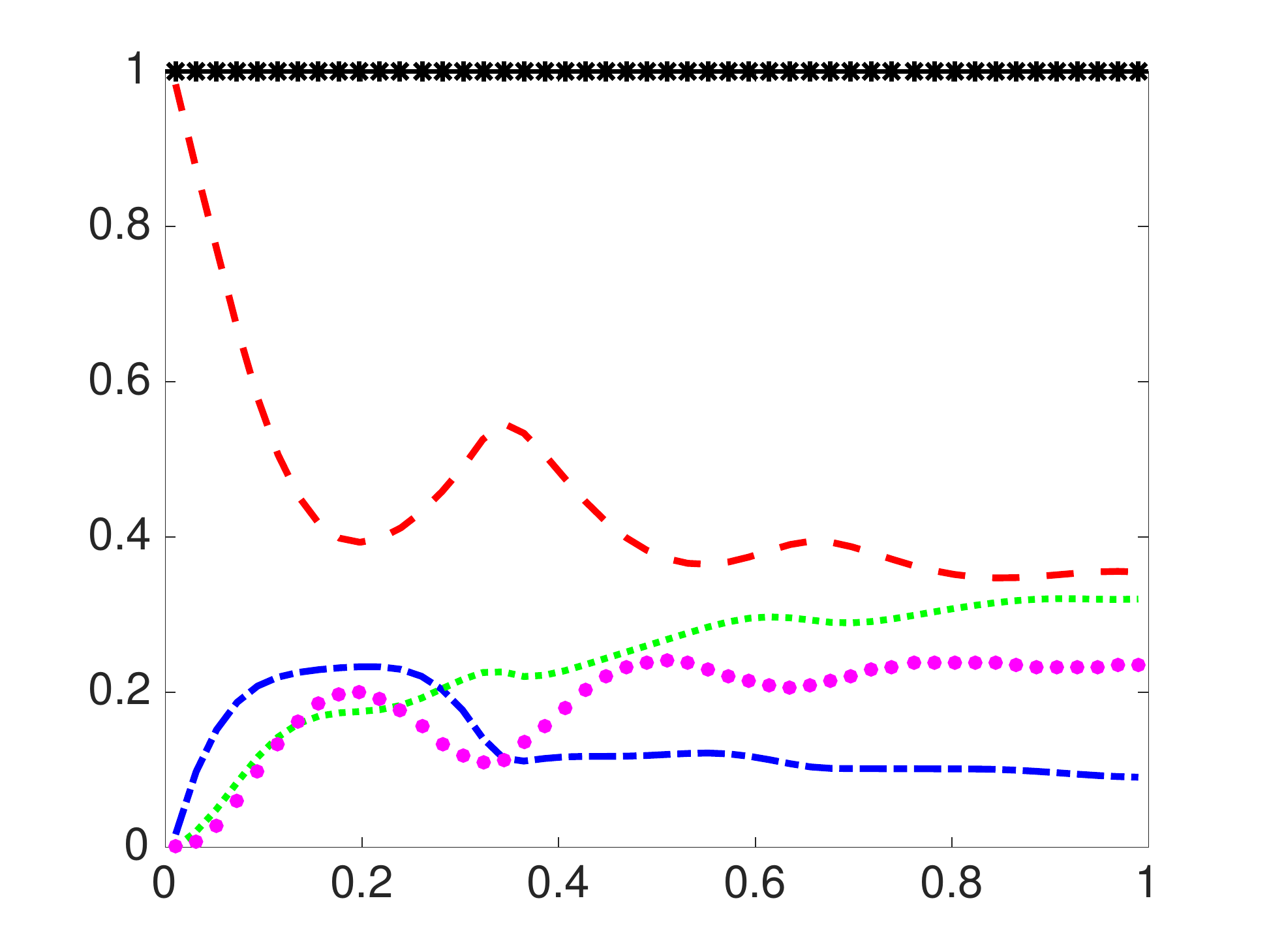}  
   \put(-391,122){$(c)$}
   \put(-197,122){$(g)$}    
   \put(-293,-3){$y/h$} 
   \put(-100,-3){$y/h$} 
   \put(-190,58){\rotatebox{90}{$q^{\,\prime \prime}_{\,\,i} / q^{\,\prime \prime}_{\,\,tot}$}}   
   \put(-384,58){\rotatebox{90}{$q^{\,\prime \prime}_{\,\,i} / q^{\,\prime \prime}_{\,\,tot}$}} \\   
   \includegraphics[width=0.495\textwidth]{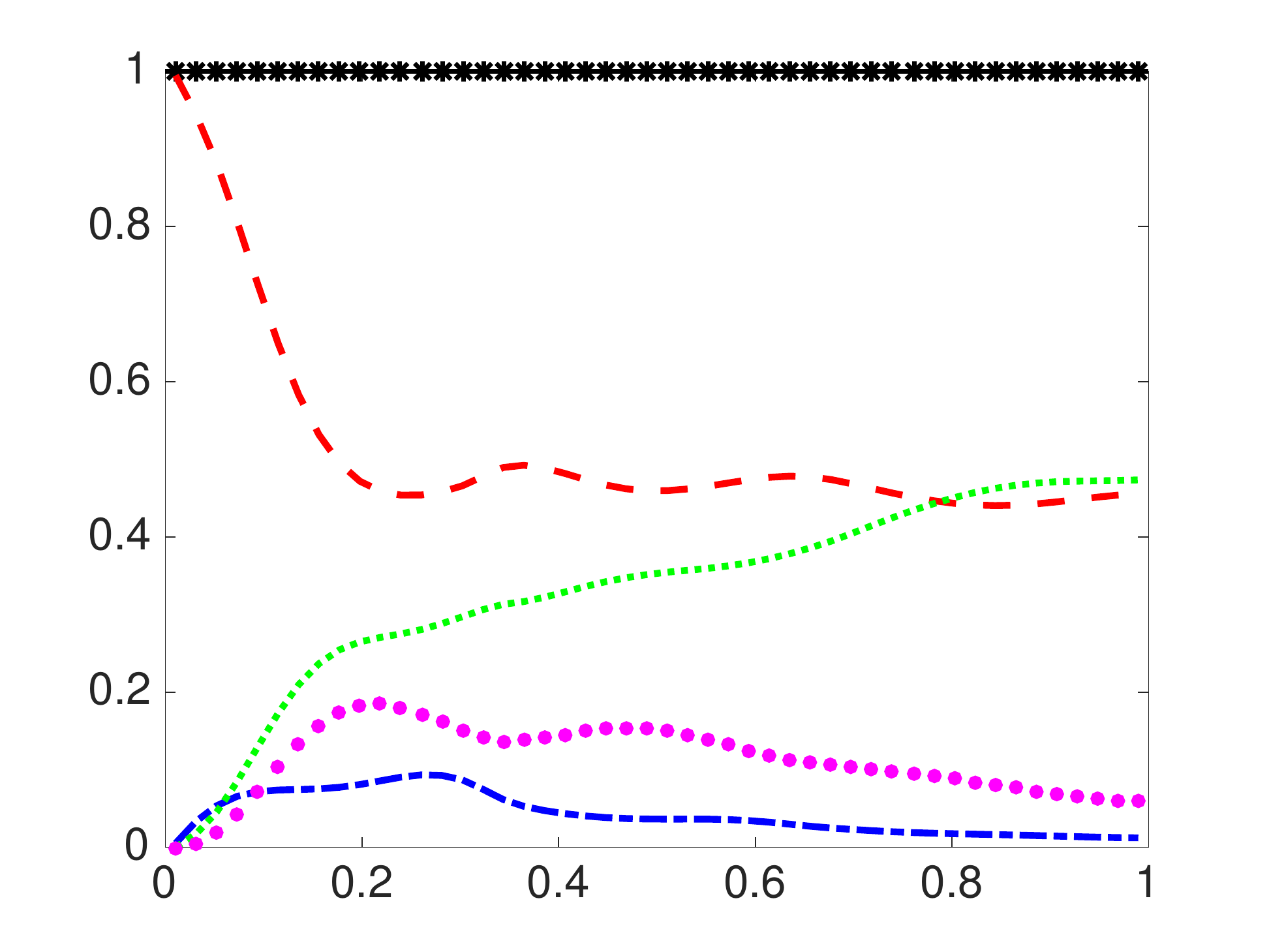} 
   \includegraphics[width=0.495\textwidth]{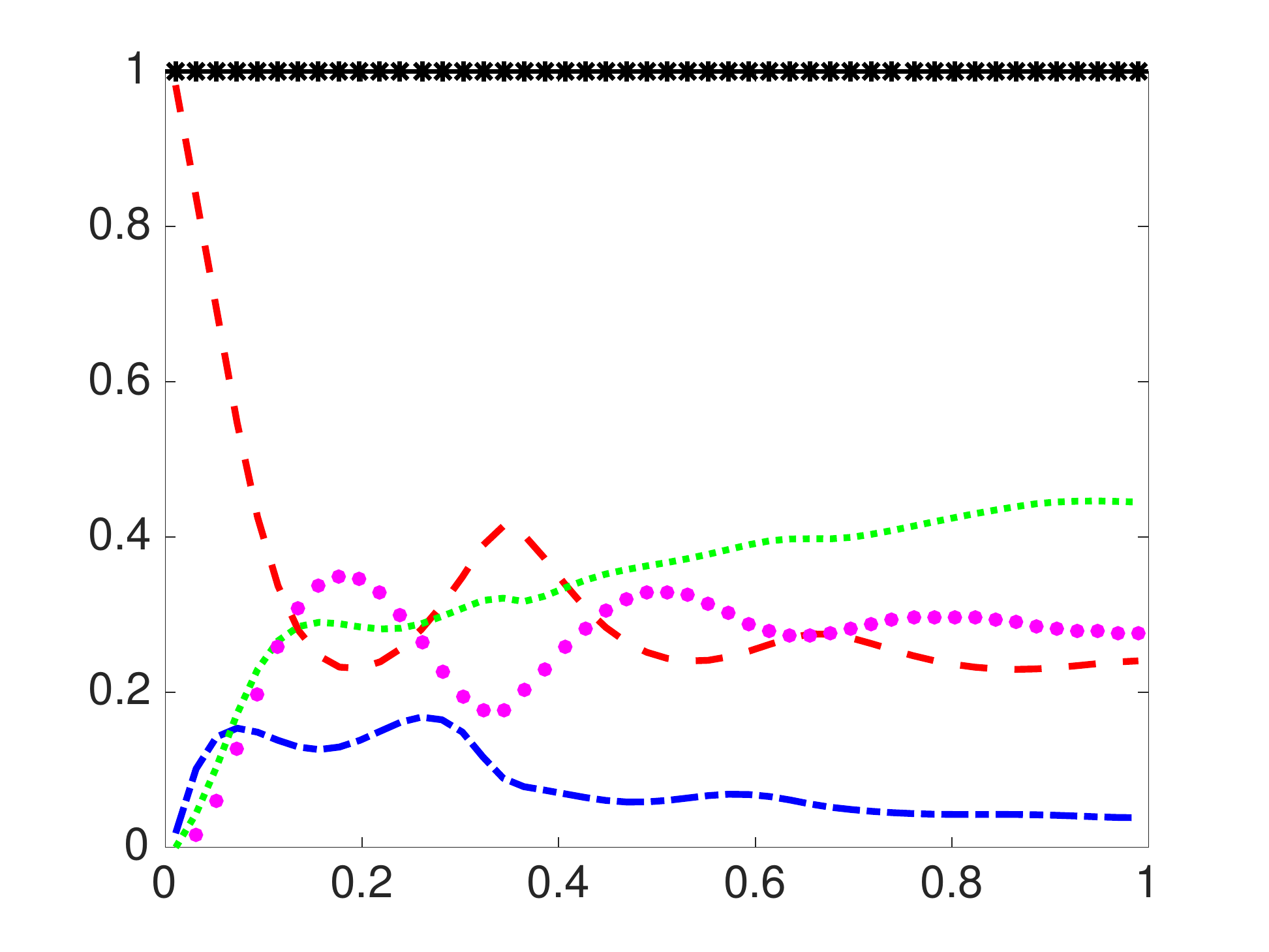} 
   \put(-391,122){$(d)$}
   \put(-197,122){$(h)$}    
   \put(-293,-3){$y/h$} 
   \put(-100,-3){$y/h$} 
   \put(-190,58){\rotatebox{90}{$q^{\,\prime \prime}_{\,\,i} / q^{\,\prime \prime}_{\,\,tot}$}}    
   \put(-384,58){\rotatebox{90}{$q^{\,\prime \prime}_{\,\,i} / q^{\,\prime \prime}_{\,\,tot}$}}\\   
  \caption{Heat flux budget for the volume fractions $\phi = 10\%$: $(a)$ $Re_p=1$; $(b)$ $Re_p=4$; $(c)$ $Re_p=8$; $(d)$ $Re_p=16$ and $\phi = 30\%$: $(e)$ $Re_p=1$; $(f)$ $Re_p=4$; $(g)$ $Re_p=8$; $(h)$ $Re_p=16$.}
  \label{fig:budget_gama1}
\end{figure}

To understand the mechanisms behind the heat transfer enhancement, statistics of the fluid and particle phase are given in figures~\ref{fig:phi} and \ref{fig:rms_velocity}. The local volume fraction $\Phi(y)$ is depicted for the different cases under investigation in figure~\ref{fig:phi} where a tendency for layering is observed. This tendency increases with particle Reynolds number (figure~\ref{fig:phi}a) and total volume fraction (figure~\ref{fig:phi}b). The tendency to form layers due to confinement from the wall is consistent with the findings of e.g.\ \cite{Fornari2016}.  A clear migration towards the wall is also observable for particles with higher inertia (higher $Re_p$) as the first peak significantly increases in figure~\ref{fig:phi}a. An explanation to this behaviour can be that the particles closer to the walls have larger velocity and momentum compared to the ones at the center of the domain, therefore those that reach the near-wall layer cannot leave easily as they are carried along by high momentum particles (and flow). At higher $Re_p$ viscous interactions among the particles are reduced and this increases the chance of trapping in already formed layers.     

The root-mean-square (RMS) of the wall-normal velocity fluctuations ($v^{\prime}$) are given in figure~\ref{fig:rms_velocity}, where the results are normalized by the diffusive velocity scale $\nu/D$. Figure~\ref{fig:rms_velocity}a and ~\ref{fig:rms_velocity}b depict the RMS of the wall-normal velocity for the fluid and the particles, showing the effect of $Re_p$ on the fluctuations at $\phi=30\%$. 
It can be observed that the maxima of $v^{\prime}_f$ and $v^{\prime}_p$ are smaller than the diffusive velocity scale $\nu/D$ for $Re_p \le 4$, suggesting that diffusion plays an important role in the heat transfer between the walls. 
For $Re_p > 4$ however, $v^{\prime}$ exceeds the diffusive velocity scale $\nu/D$, except for a thin region close to the wall. The consequences of layering on $v^{\prime}$ are observed especially in the particle fluctuations, which increase significantly from layer to layer and remain constant within each layer. 

The effect of volume fraction on the velocity fluctuations at $Re_p=16$ is shown in figure~\ref{fig:rms_velocity}c and \ref{fig:rms_velocity}d. The results show a saturation of  $v^{\prime}_p D/\nu$ as it slightly increases from $\phi = 20\%$ to $\phi = 30\%$; nonetheless $v^{\prime}_f$ continues to  increase with the volume fraction due to the presence of a larger number of the particles in the mentioned layers.

To further understand  the details of the transport mechanisms in a particle-laden plane Couette flow, it is useful to separate the different contributions to the total heat transfer.  We therefore phase and ensemble average the energy equation, as shown in appendix \ref{appA}, and write the wall-normal heat flux in the following form
\begin{eqnarray}
q^{\,\prime \prime}_{\,\,tot}  \,=\, -\Phi \, \langle v^{\prime}_p \, T^{\prime}_p  \rangle \,-\, \left( 1 - \Phi \right)  \langle v^{\prime}_f \, T^{\prime}_f  \rangle \,+\,  \Phi \,\alpha_p \, \langle \frac{\mathrm{d} T_p}{\mathrm{d} y} \rangle \,+\,    \left( 1 - \Phi \right) \alpha_f \,   \langle \frac{\mathrm{d} T_f}{\mathrm{d} y} \rangle   .
\label{eq:heatFlux} 
\end{eqnarray}

The total heat flux, on average constant across the channel, is divided into four terms: i) convection by the particle velocity fluctuations; ii) convection by the fluid; iii)  molecular diffusion in the solid phase (solid conduction) and iv) molecular diffusion in the fluid phase (fluid conduction). 

\begin{figure}
  \centering
   \includegraphics[width=0.495\textwidth]{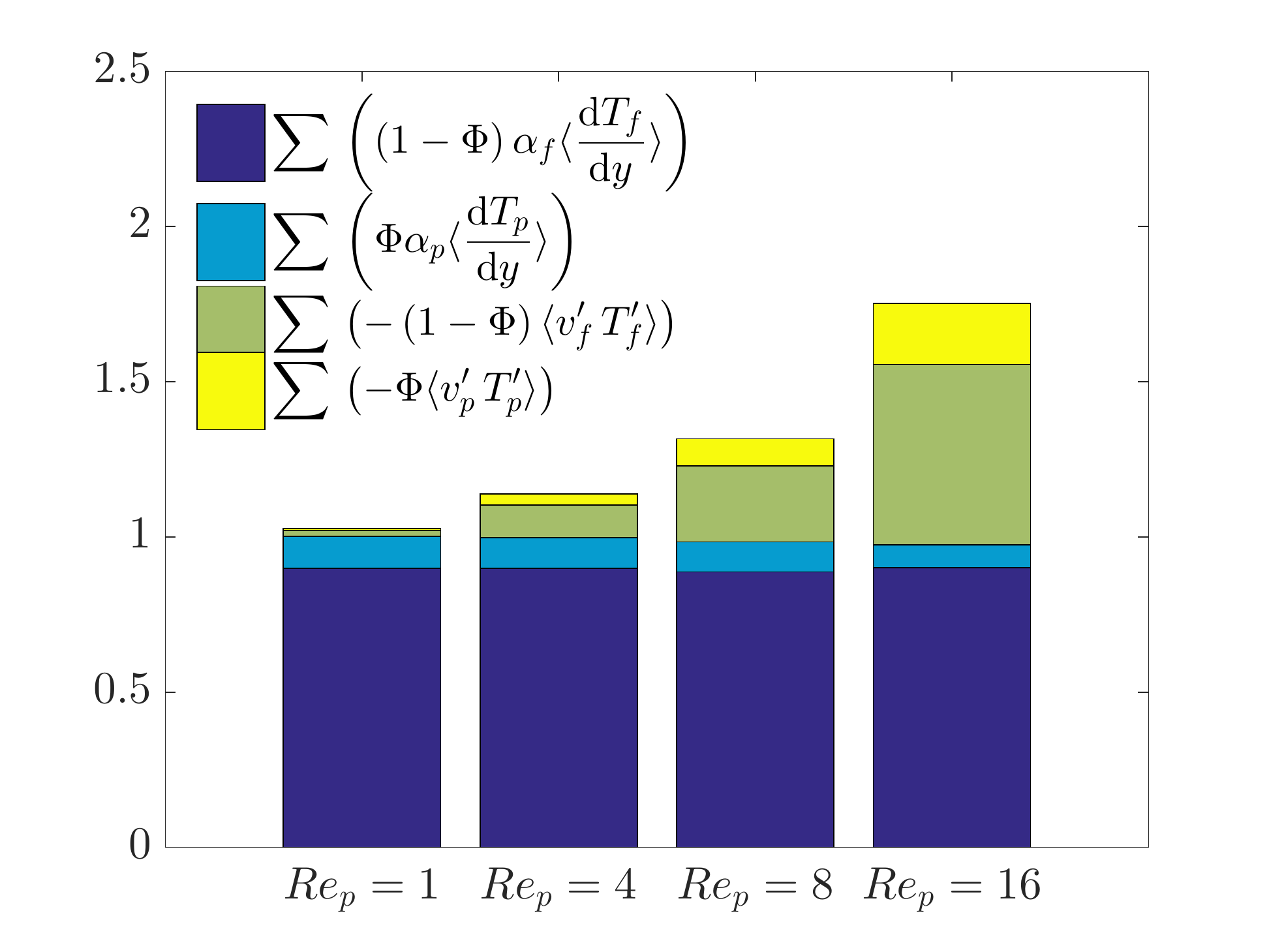}
   \includegraphics[width=0.495\textwidth]{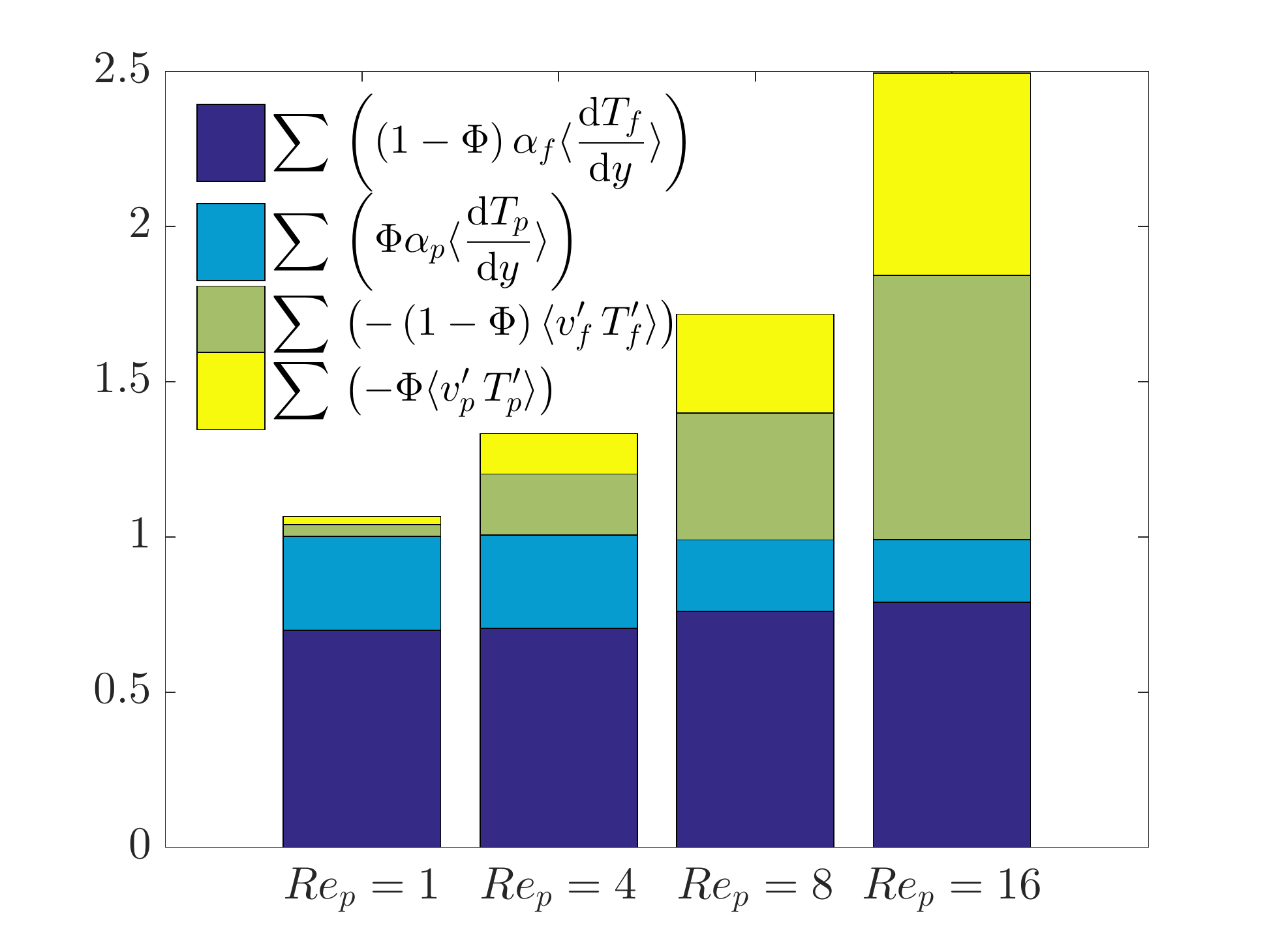}
   \put(-391,122){$(a)$}
   \put(-197,122){$(b)$}    
   \put(-294,-3){$Re_p$} 
   \put(-100,-3){$Re_p$} 
   \put(-190,44){\rotatebox{90}{{$\Sigma \,q^{\,\prime \prime}_{\,\,i} \, / \, \Sigma \, q^{\,\prime \prime}_{\,\,tot_{\phi=0\%}}$}}}  
   \put(-382,44){\rotatebox{90}{{$\Sigma \,q^{\,\prime \prime}_{\,\,i} \, / \, \Sigma \, q^{\,\prime \prime}_{\,\,tot_{\phi=0\%}}$}}}  \\   
  \caption{Wall-normal integral of the heat fluxes transferred by different mechanisms, normalized by total heat flux in single phase flow for: $(a)$ $\phi=10\%$ and $(b)$ $\phi=30\%$.}
\label{fig:share}
\end{figure}
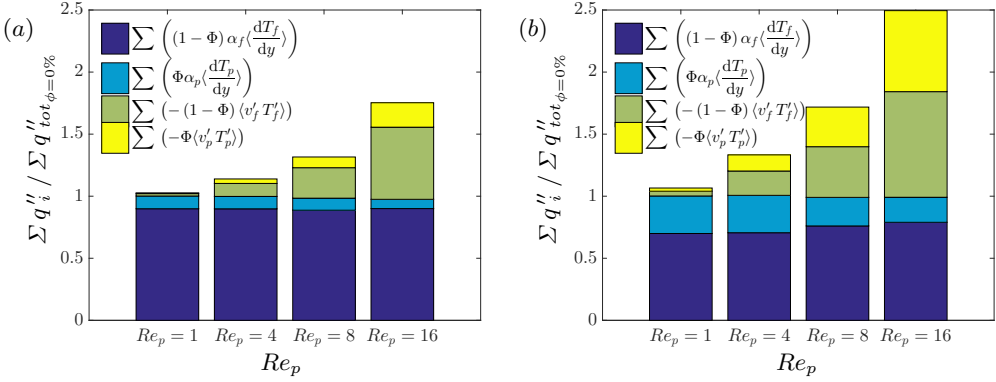

Figure \ref{fig:budget_gama1} shows the wall-normal profiles, from the wall to the centreline, of these four different terms for volume fractions $\phi=10\%$ and $30\%$ at the different particle Reynolds numbers under investigation. 
The results at $Re_p= 1$ (panel $(a)$ and $(e)$), which can be considered as representative of the Stokes regime, reveal that the heat flux is almost completely due to the heat molecular diffusion in the fluid and solid phase and the role of the fluctuations is negligible in both phases. Interestingly, when $Re_p$ increases, the contribution from the term $-\left( 1 - \Phi \right)  \langle v^{\prime}_f \, T^{\prime}_f  \rangle$,  the correlation between the fluctuations in the temperature and wall-normal fluid velocity, increases. At $Re_p=16$, this term gives the largest contribution at the centreline. The heat transfer due to conduction in the solid phase, $\Phi \,\alpha_p \, \langle \frac{\mathrm{d} T_p}{\mathrm{d} y} \rangle$,  is found to reduce when increasing $Re_p$, when the particle velocity fluctuations ($-\Phi \, \langle v^{\prime}_p \, T^{\prime}_p  \rangle$) transfer heat by the motion of the particles across layers.  

The integral of the contribution to the total heat flux of each term is depicted in figure~\ref{fig:share}, where the results are normalized by the total heat flux in the absence of particles. 
The figure shows that  the heat flux transferred by the velocity fluctuations significantly increase with $Re_p$ and this explains the increase of the effective suspension diffusivity shown above;  the the total amount of diffusive heat flux (in the fluid and solid phase) saturates. Interestingly, the heat flux transferred by conduction in the solid phase reduces as $Re_p$ increases for $\phi=10\%$ (figure~\ref{fig:share}a) and $\phi=30\%$ (figure~\ref{fig:share}b). 
To explain this behaviour we refer to figure~\ref{fig:rms_velocity}b, where we report the particle wall-normal velocity fluctuations normalized by the diffusive velocity scale. 
As discussed above,  the order of magnitude of the velocity fluctuations is significantly larger than the diffusive velocity scale for the highest $Re_p$ (note that the heat diffusion velocity is smaller than $\nu/D$ by a factor of the Prandtl number, $Pr$). This difference in velocity scales, particle fluctuations and heat diffusion,
explains the reduction of heat transferred by conduction in the solid; in other words, heat diffusion in the solid phase 
is slow compared to the time it takes the particle to move to a new position. 
Nevertheless, heat transfer by convective processes in both phases accounts for more than half of the total except for the case at $\phi=30\%$ and $Re_p=16$. An increase in the thermal diffusivity of the solid phase can reduce the difference between the velocity scales, increasing the heat transfer.

\subsection{Different thermal diffusivity of the particle}

In this section we present results obtained varying the particle thermal diffusivity and examine the implications on the  heat transfer in the same Couette geometry. 
We first consider  a solid thermal diffusivity higher ($ \Gamma = \alpha_p / \alpha_f = 10$), lower ($\Gamma = 0.1$) and equal to the fluid thermal diffusivity at $Re_p =0.5$. The effect of inertia  in the presence of particles with different thermal diffusivity is investigated by performing simulations at $Re_p =16$ for $\Gamma = 0.1$ and $10$. These flow cases are studied at volume fractions $\phi =10 , \, 20$ and $30\%$ with constant Prandtl number $Pr = 7$.

\begin{figure}
  \centering
   \includegraphics[width=0.495\textwidth]{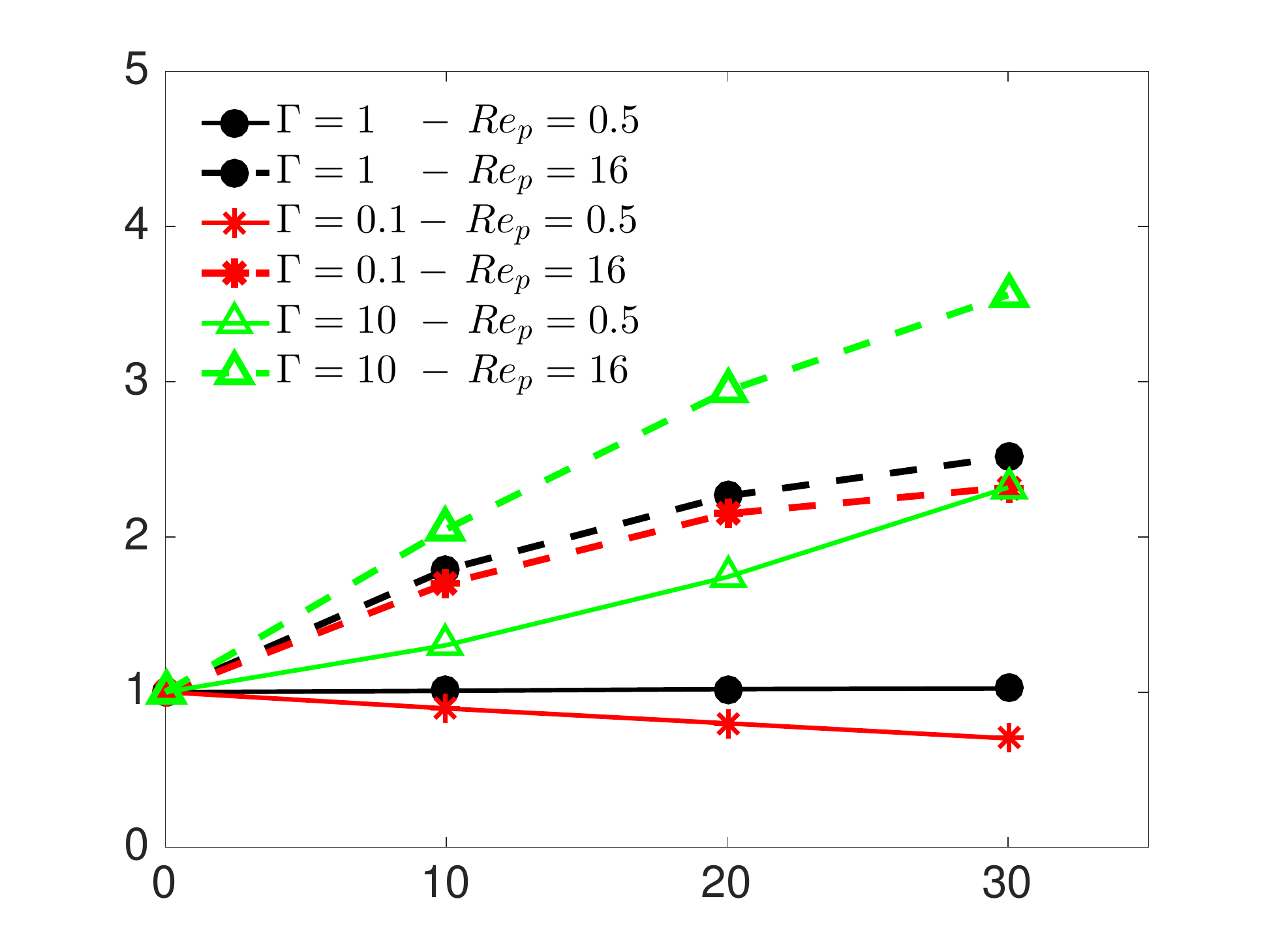}
   \includegraphics[width=0.495\textwidth]{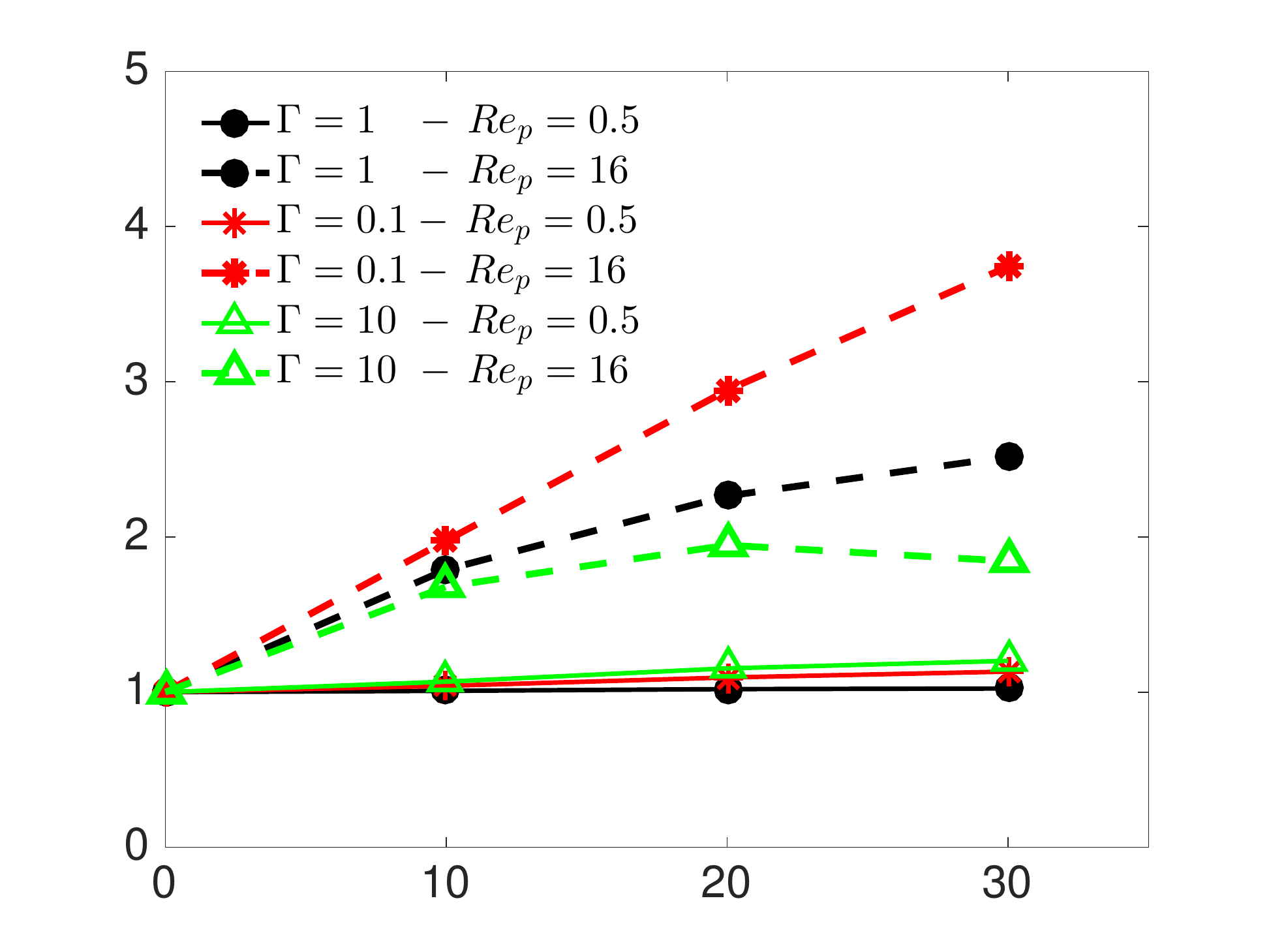}
   \put(-391,122){$(a)$}
   \put(-197,122){$(b)$}    
   \put(-300,-3){$\phi \,(\%)$} 
   \put(-107,-3){$\phi \,(\%)$} 
   \put(-187,60){\rotatebox{90}{{$\alpha_e \, / \,  \alpha_{ec}$}}}  
   \put(-378,70){\rotatebox{90}{{$\alpha_r$}}}  \\   
  \caption{$(a)$ The effective thermal diffusivity of the suspension, normalized with that of the single phase flow, $\alpha_r$ versus the particle volume fraction $\phi$ and $(b)$ same data normalized with the effective thermal conductivity of a composite ($\alpha_{ec}$) with the same volume fraction, estimated according to the Lewis-Nielsen model \citep{Nielsen1974}.}
\label{fig:gama}
\end{figure}
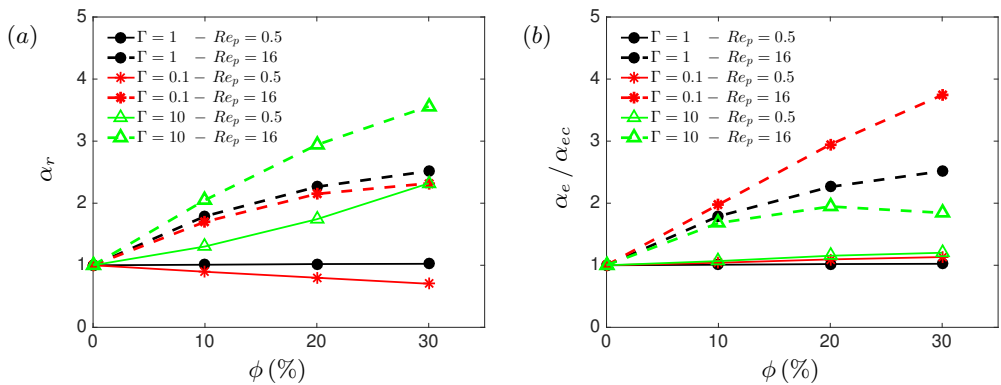

Figure~\ref{fig:gama}a depicts the effective diffusivity of the suspension, $\alpha_r$, versus the volume fraction of the solid phase for the different values of the thermal diffusivity ratio $\Gamma$ investigated. As expected, at small particle Reynolds number $Re_p =0.5$ $\alpha_r$  increases or decreases with the particle volume fraction when the thermal diffusivity of the solid particles is higher or lower than that of the fluid. Interestingly, at $Re_p=16$, the inertial effect overcomes the lower diffusivity, see case $\Gamma =0.1$, resulting in a considerable global  increase of the heat transfer across the flow. 
The results for the case at $Re_p =16$ and $\Gamma =1$ are again reported in the figure for a better comparison. When inertial effects become important the difference between the results with  $\Gamma =1$ and $\Gamma =0.1$ is negligible. 

In figure~\ref{fig:gama}a the effective thermal diffusivity is normalized with that of the single phase flow. This normalization might not be the most relevant when the thermal diffusivity inside the particles is different than that of the surrounding fluid due to the change in the average thermal diffusivity of the suspension. To highlight how the motion of the particles can enhance the heat transfer of a suspension, it may therefore be better to normalize the effective thermal diffusivity of a sheared suspension to that of a suspension at rest. 
\cite{Pietrak2015} recently reviewed the models for effective thermal conductivity of composite materials. The first analytical expression for the effective conductivity of a heterogenic medium was suggested by \cite{Maxwell1904} in his pioneering work on electricity and magnetism.  Maxwell's formula was found to be valid in the range of $\phi <  25\%$ \citep{Pietrak2015}. Later, \cite{Nielsen1974} suggested an empirical model, now frequently used in the literature, which provides relatively good results up to $\phi < 40\%$ and covers a wide range of particle shapes. The effective thermal conductivity of a composite ($\alpha_{ec}$) according to the Lewis-Nielsen model is given by
\begin{eqnarray}
\alpha_{ec}  = \frac{1\,+\,a\,b\,\phi}{1-\,a\,b\,\psi} \,\,\,\,;\,\,\,\, b = \frac{\alpha_p / \alpha_f - 1}{\alpha_p / \alpha_f + a} \,\,\,\,;\,\,\,\, \psi = 1 \,+\, \frac{1\,-\,\phi_m}{\phi_m^2} \,\phi\, , 
\label{eq:alpha_ec} 
\end{eqnarray} 
where $\phi_m$ is the maximum filler volume fraction and $a$ is the shape coefficient for the filler particles. $\phi_m$ for random packing of spherical particles is takes as $0.637$, while $a=1.5$ for spherical particles \citep{Nielsen1974}. 

The effective thermal diffusivities extracted form the simulations are depicted in figure~\ref{fig:gama}b normalized with the effective conductivity of the suspension at rest ($\alpha_{ec}$). 
The data show that the largest relative increase of the heat transfer when particles move occurs for
 $\Gamma = 0.1$ and $Re_p=16$. 
 This can be explained by the fact that $\alpha_{ec}$ is miminum in this case while inertial effects are large at $Re_p=16$. Interestingly, a peak is observed at $\phi=20\%$ for the case with $\Gamma=10$ and $Re_p=16$, indicating that  $\alpha_{ec}$ grows faster than inertial effects at $\phi=30\%$. 
The trend is different when inertia is negligible, $Re_p=0.5$. In this case, $\alpha_{ec}$, calculated by Lewis-Nielsen model, matches the value of $\alpha_e$ from the simulations. The small deviation observed can be related to the layering of particles discussed above, as in the model of Lewis-Nielsen particles are assumed ot be randomly distributed.
This effect is more pronounced for the case with $\Gamma=10$.

\begin{figure}
  \centering
   \includegraphics[width=0.495\textwidth]{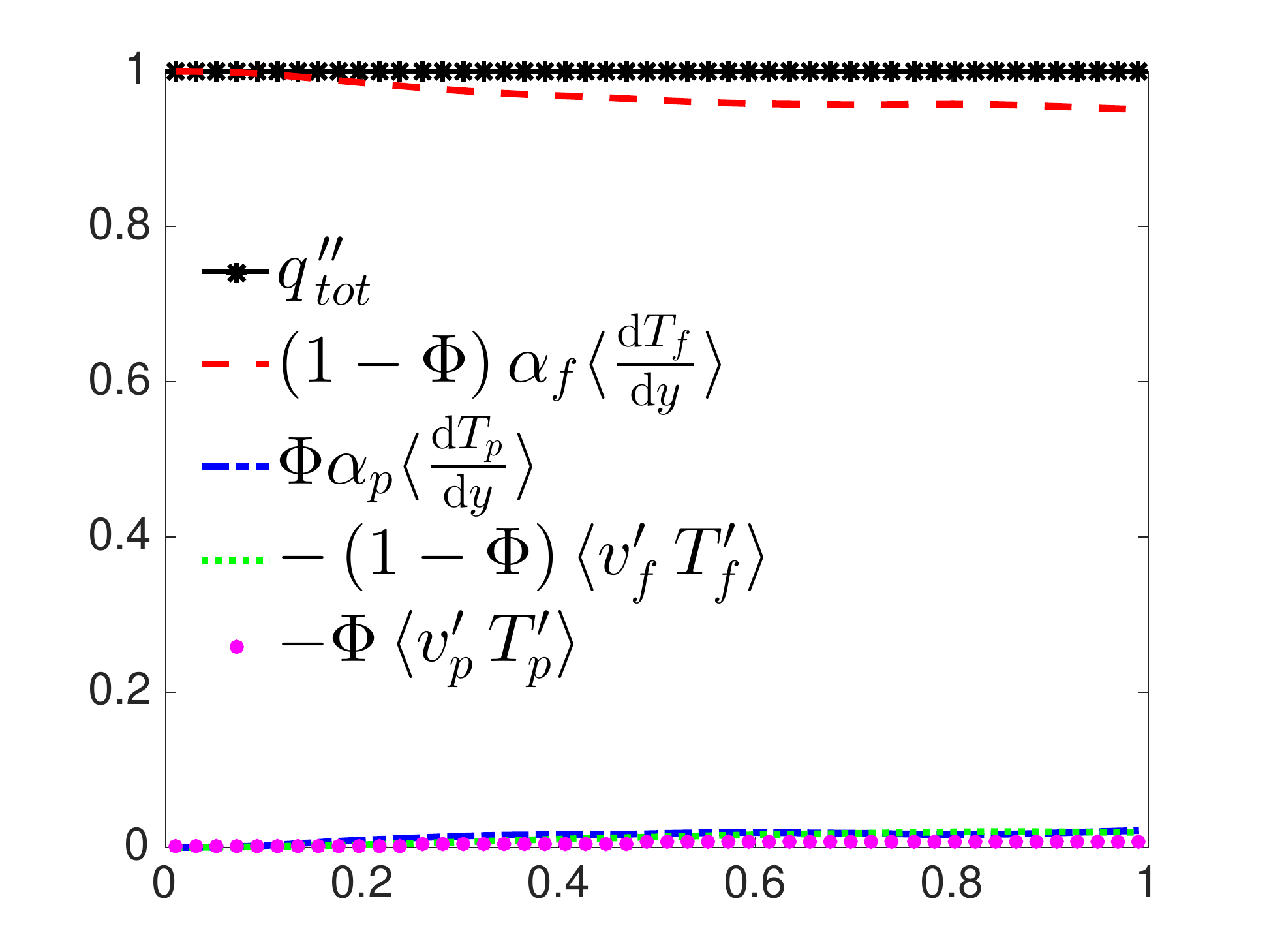}  
   \includegraphics[width=0.495\textwidth]{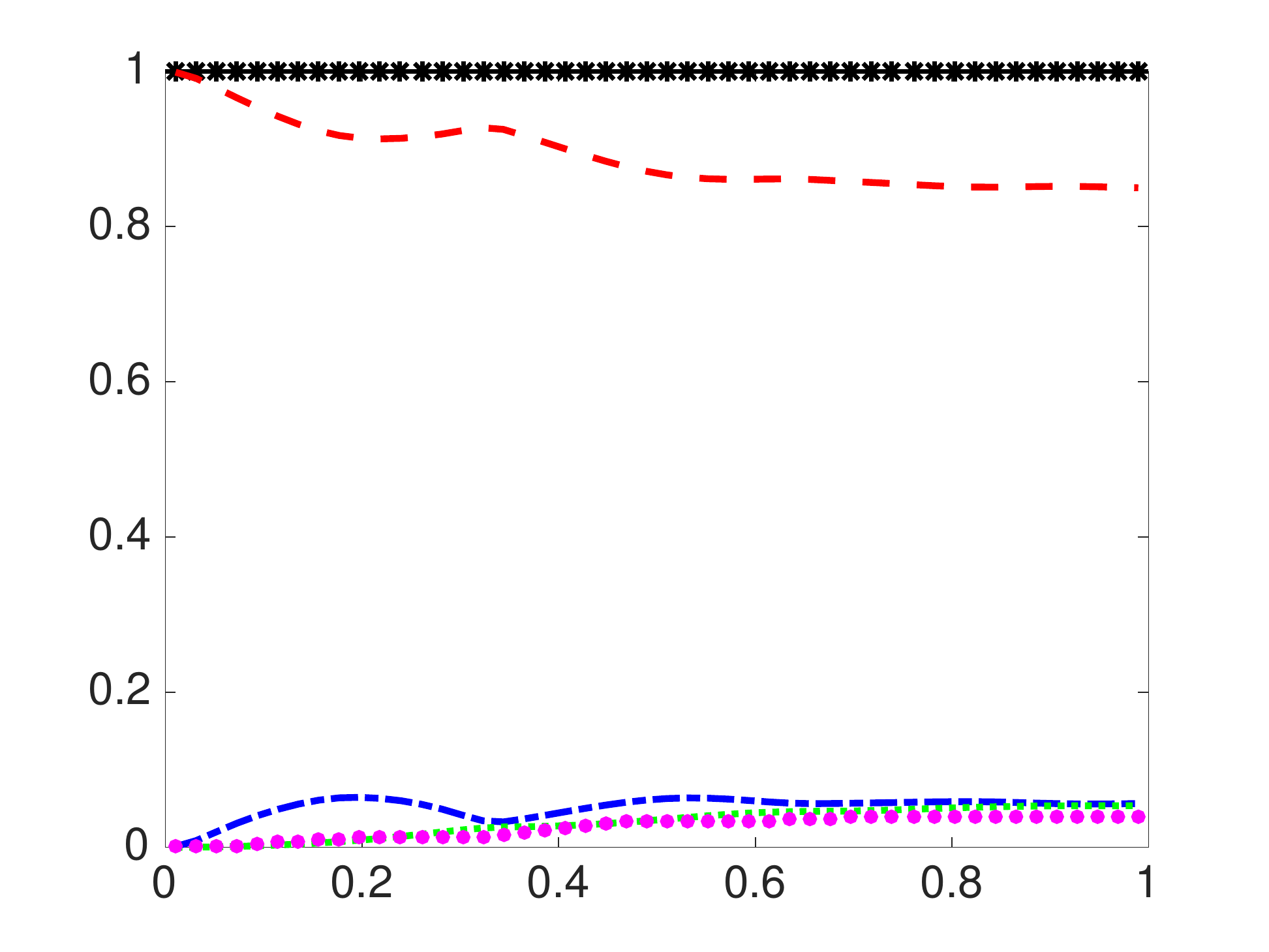}  
   \put(-391,122){$(a)$}
   \put(-197,122){$(e)$}    
   \put(-293,-3){$y/h$} 
   \put(-100,-3){$y/h$}    
   \put(-190,58){\rotatebox{90}{$q^{\,\prime \prime}_{\,\,i} / q^{\,\prime \prime}_{\,\,tot}$}}   
   \put(-384,58){\rotatebox{90}{$q^{\,\prime \prime}_{\,\,i} / q^{\,\prime \prime}_{\,\,tot}$}}\\
   \includegraphics[width=0.495\textwidth]{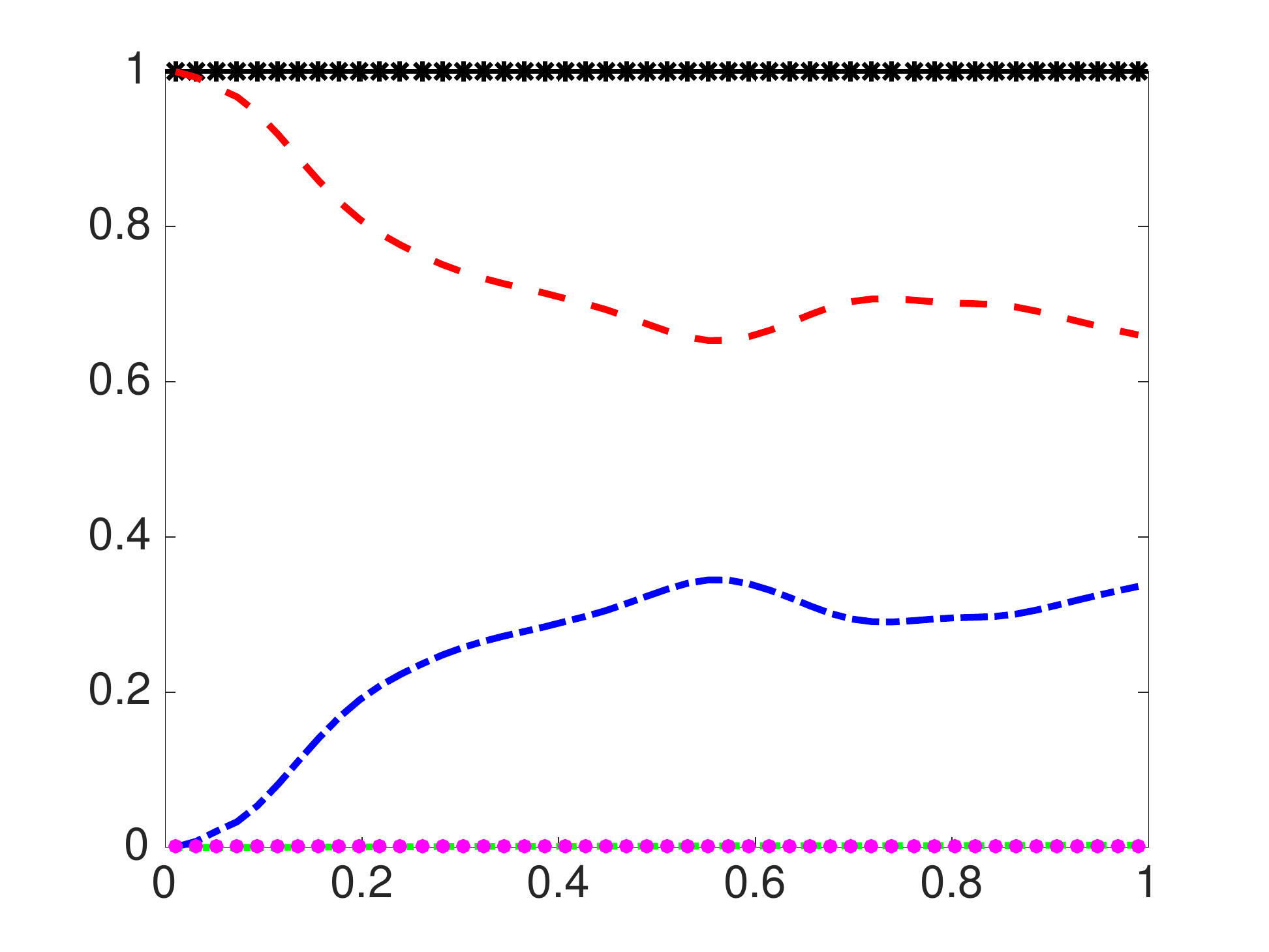} 
   \includegraphics[width=0.495\textwidth]{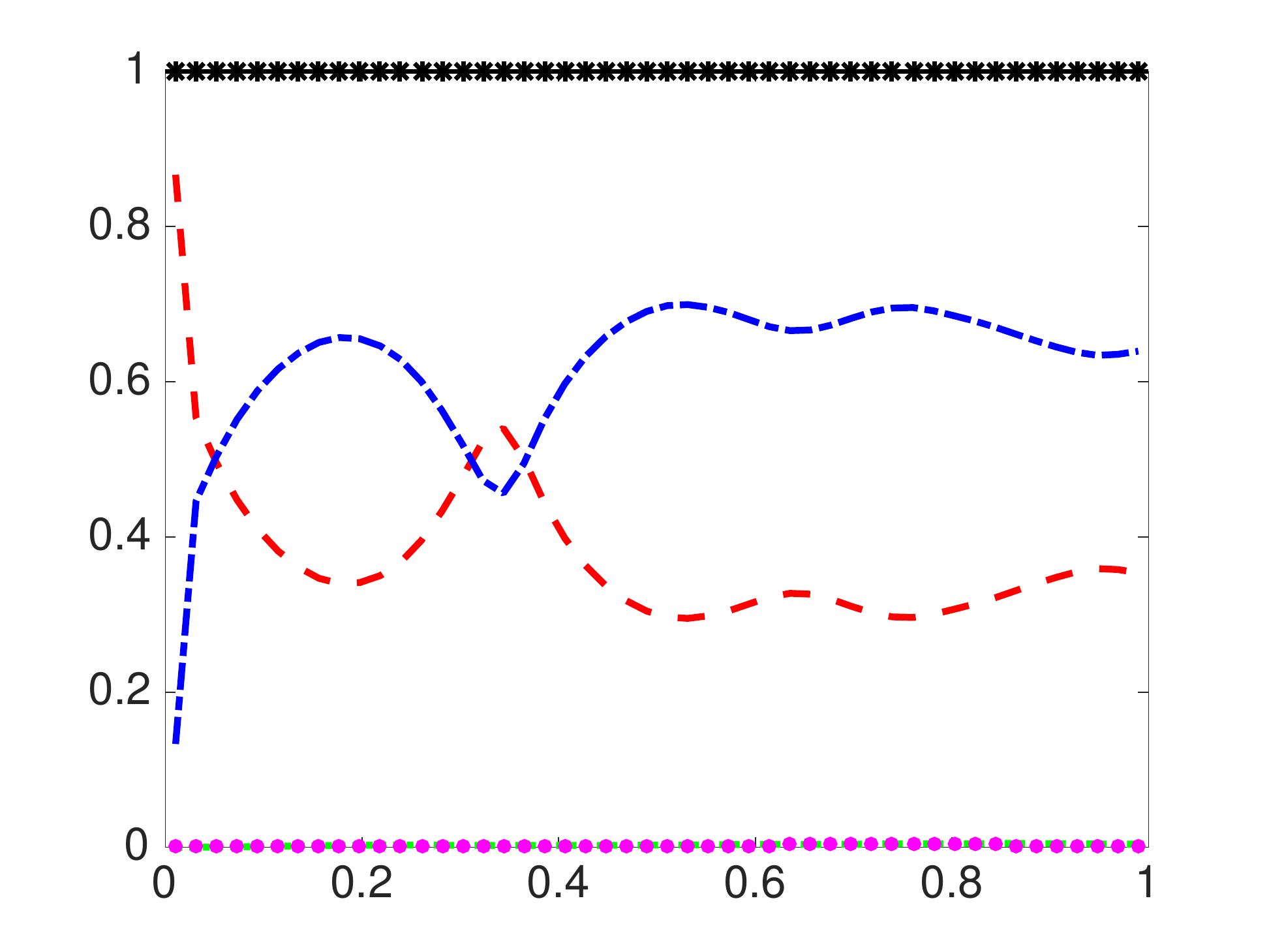} 
   \put(-391,122){$(b)$}
   \put(-197,122){$(f)$}    
   \put(-293,-3){$y/h$} 
   \put(-100,-3){$y/h$} 
   \put(-190,58){\rotatebox{90}{$q^{\,\prime \prime}_{\,\,i} / q^{\,\prime \prime}_{\,\,tot}$}}    
   \put(-384,58){\rotatebox{90}{$q^{\,\prime \prime}_{\,\,i} / q^{\,\prime \prime}_{\,\,tot}$}}\\      
   \includegraphics[width=0.495\textwidth]{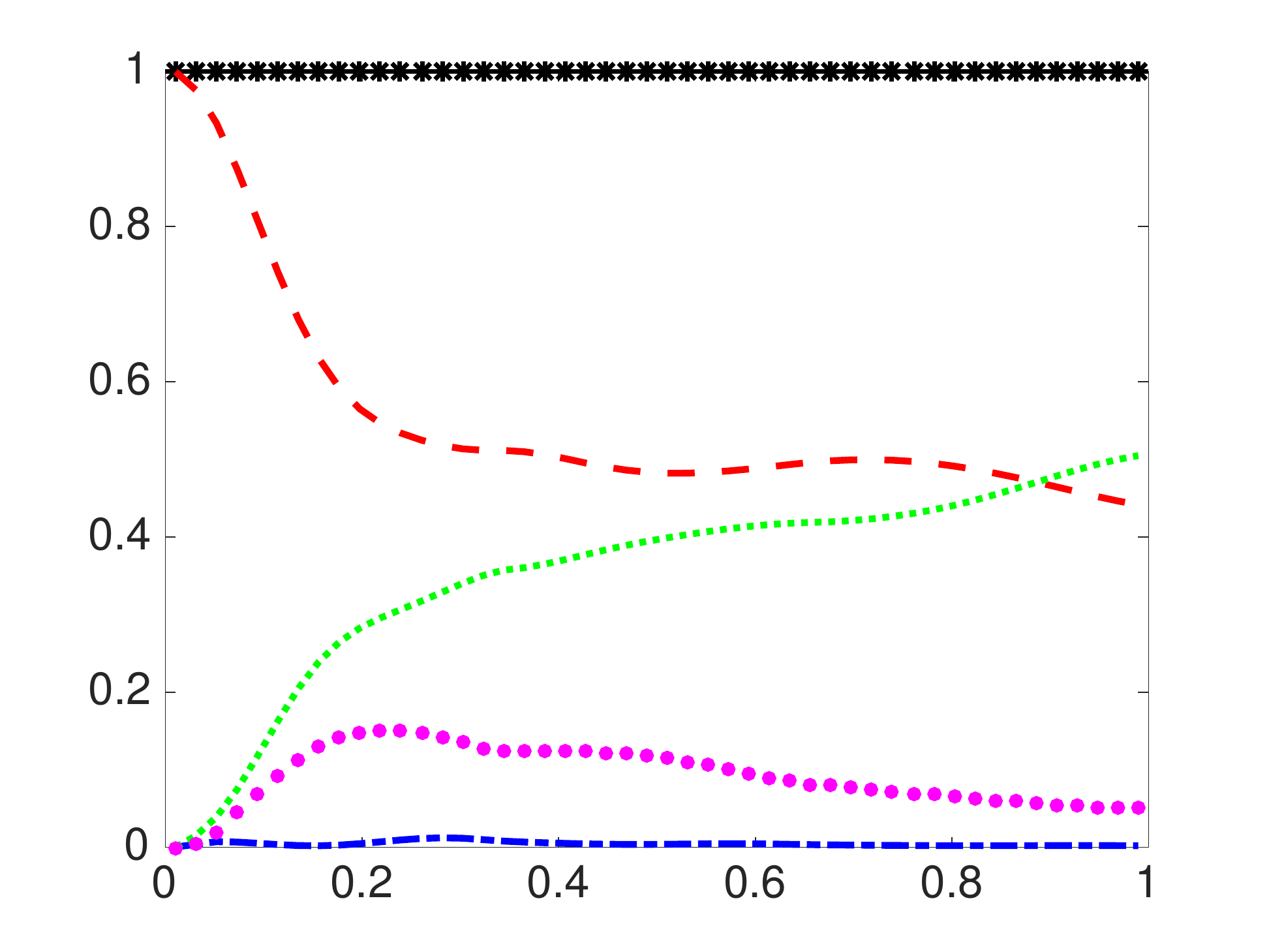} 
   \includegraphics[width=0.495\textwidth]{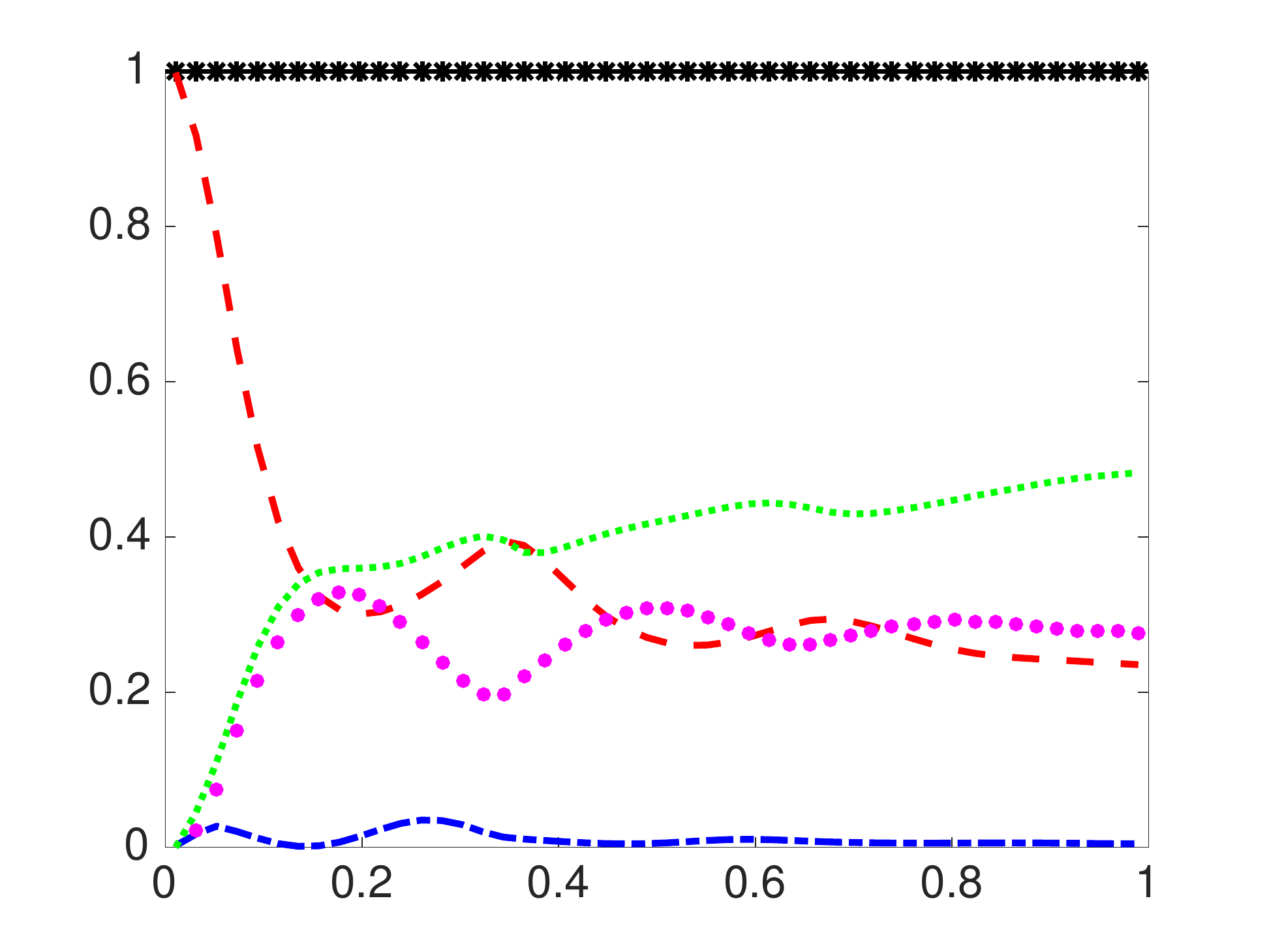}  
   \put(-391,122){$(c)$}
   \put(-197,122){$(g)$}    
   \put(-293,-3){$y/h$} 
   \put(-100,-3){$y/h$} 
   \put(-190,58){\rotatebox{90}{$q^{\,\prime \prime}_{\,\,i} / q^{\,\prime \prime}_{\,\,tot}$}}   
   \put(-384,58){\rotatebox{90}{$q^{\,\prime \prime}_{\,\,i} / q^{\,\prime \prime}_{\,\,tot}$}} \\   
   \includegraphics[width=0.495\textwidth]{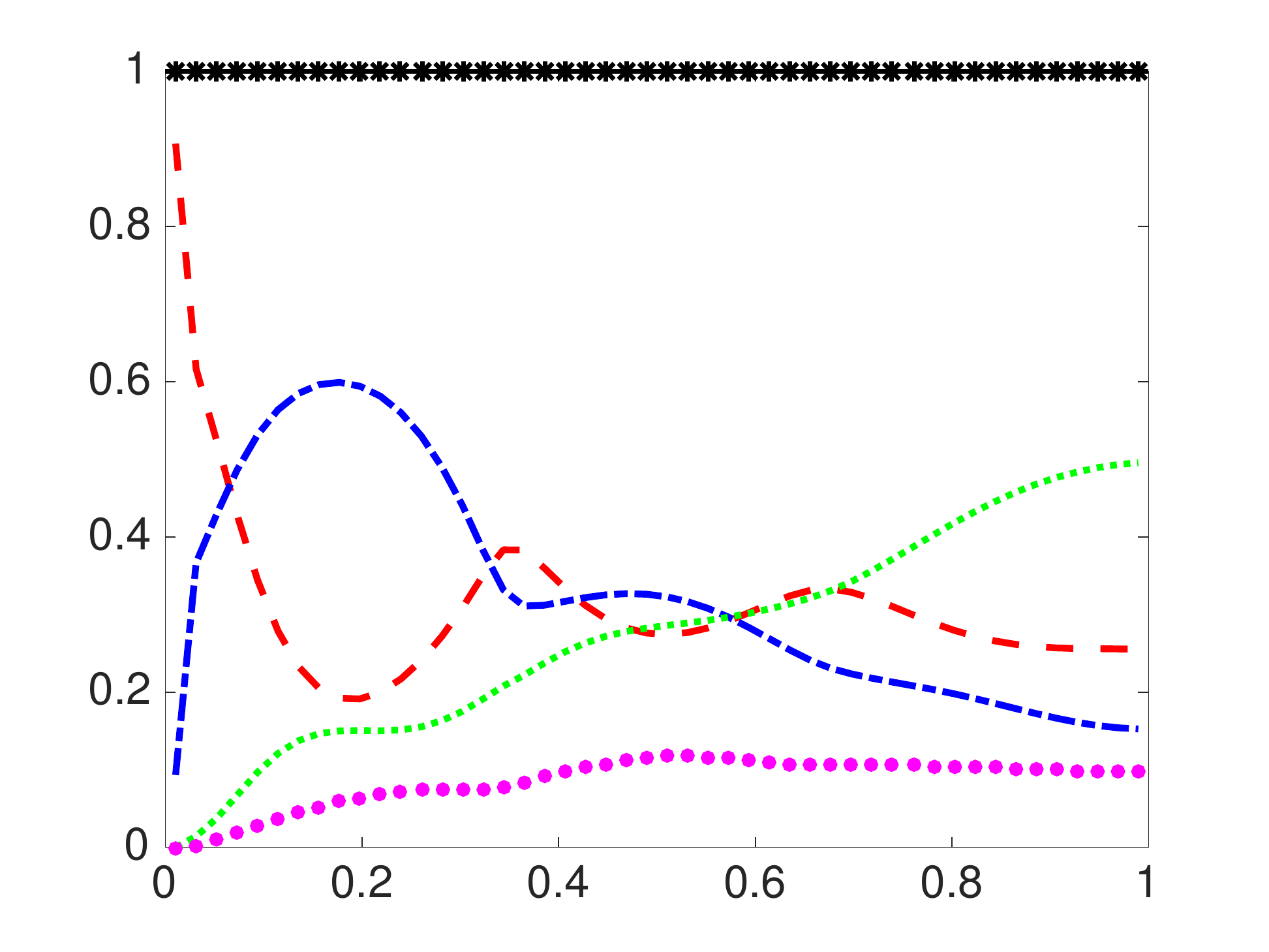} 
   \includegraphics[width=0.495\textwidth]{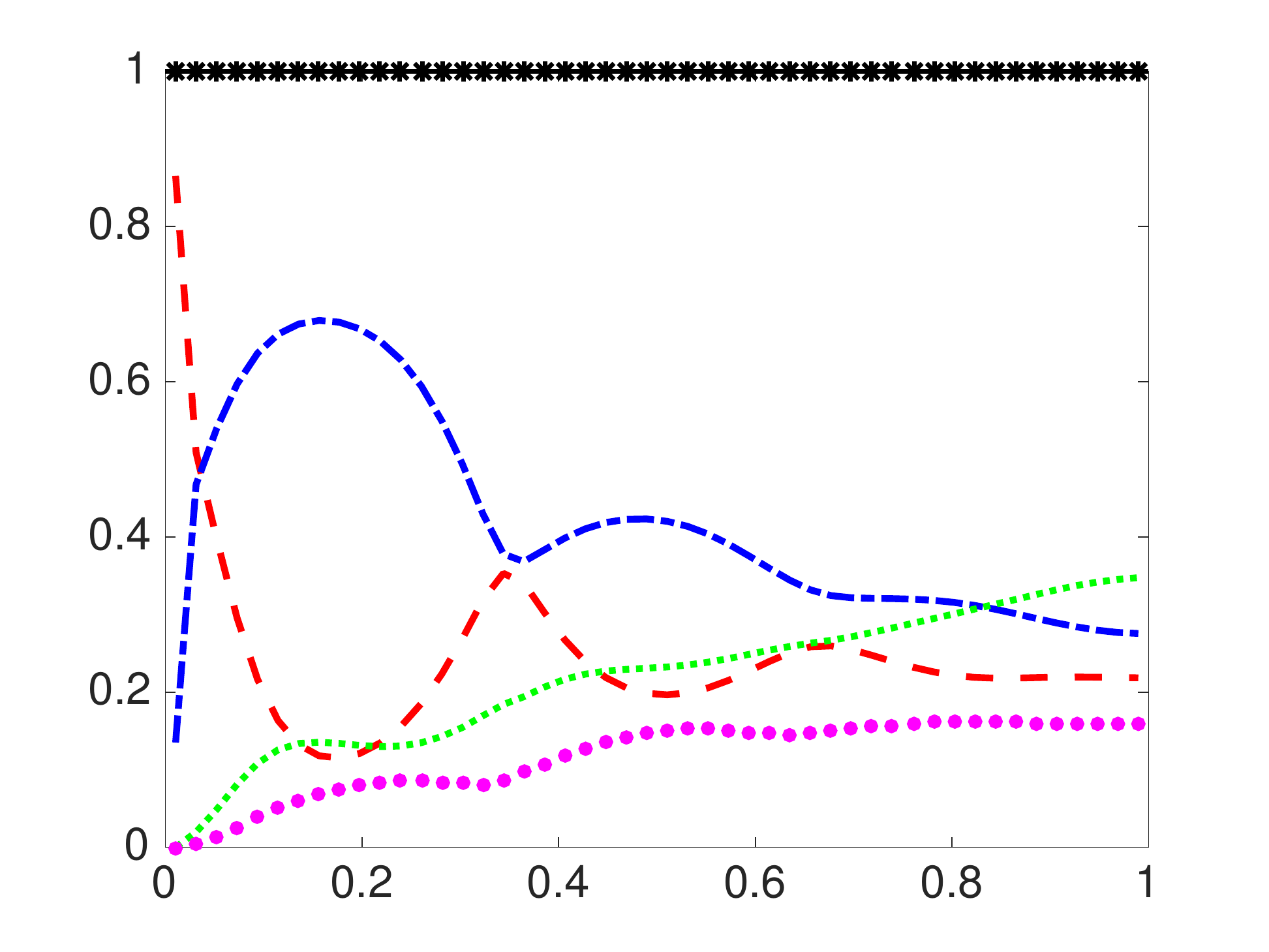} 
   \put(-391,122){$(d)$}
   \put(-197,122){$(h)$}    
   \put(-293,-3){$y/h$} 
   \put(-100,-3){$y/h$} 
   \put(-190,58){\rotatebox{90}{$q^{\,\prime \prime}_{\,\,i} / q^{\,\prime \prime}_{\,\,tot}$}}    
   \put(-384,58){\rotatebox{90}{$q^{\,\prime \prime}_{\,\,i} / q^{\,\prime \prime}_{\,\,tot}$}}\\   
  \caption{Heat flux budget for the volume fractions $\phi = 10\%$: $(a)$ $Re_p=0.5 , \Gamma=0.1$; $(b)$ $Re_p=0.5 , \Gamma=10$; $(c)$ $Re_p=16 , \Gamma=0.1$; $(d)$ $Re_p=16 , \Gamma=10$ and $\phi = 30\%$: $(e)$ $Re_p=0.5 , \Gamma=0.1$; $(f)$ $Re_p=0.5 , \Gamma=10$; $(g)$ $Re_p=16 , \Gamma=0.1$; $(h)$ $Re_p=16 , \Gamma=10$.}
  \label{fig:budget_gama_not1}
\end{figure}

\begin{figure}
  \centering
   \includegraphics[width=0.495\textwidth]{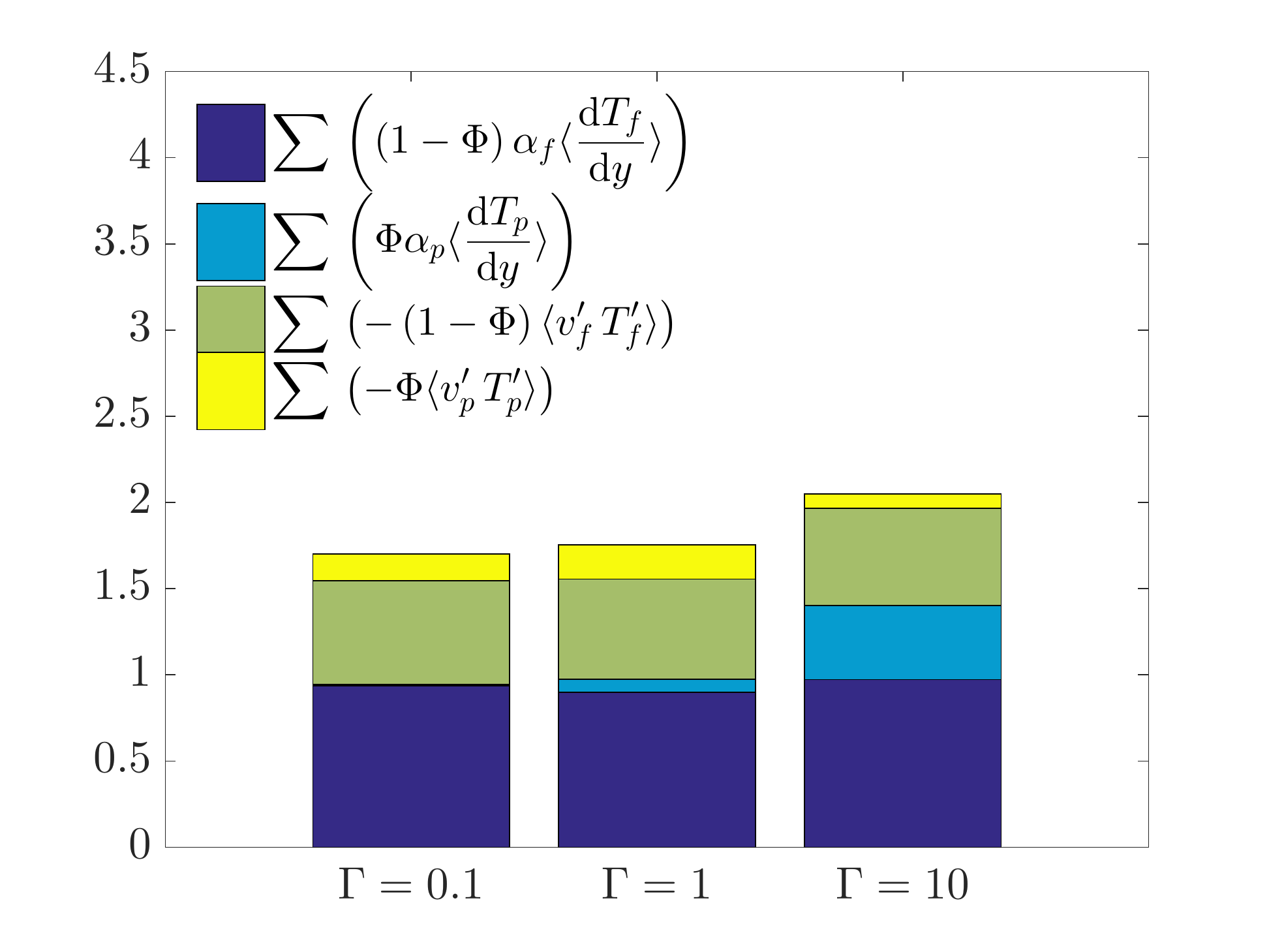}
   \includegraphics[width=0.495\textwidth]{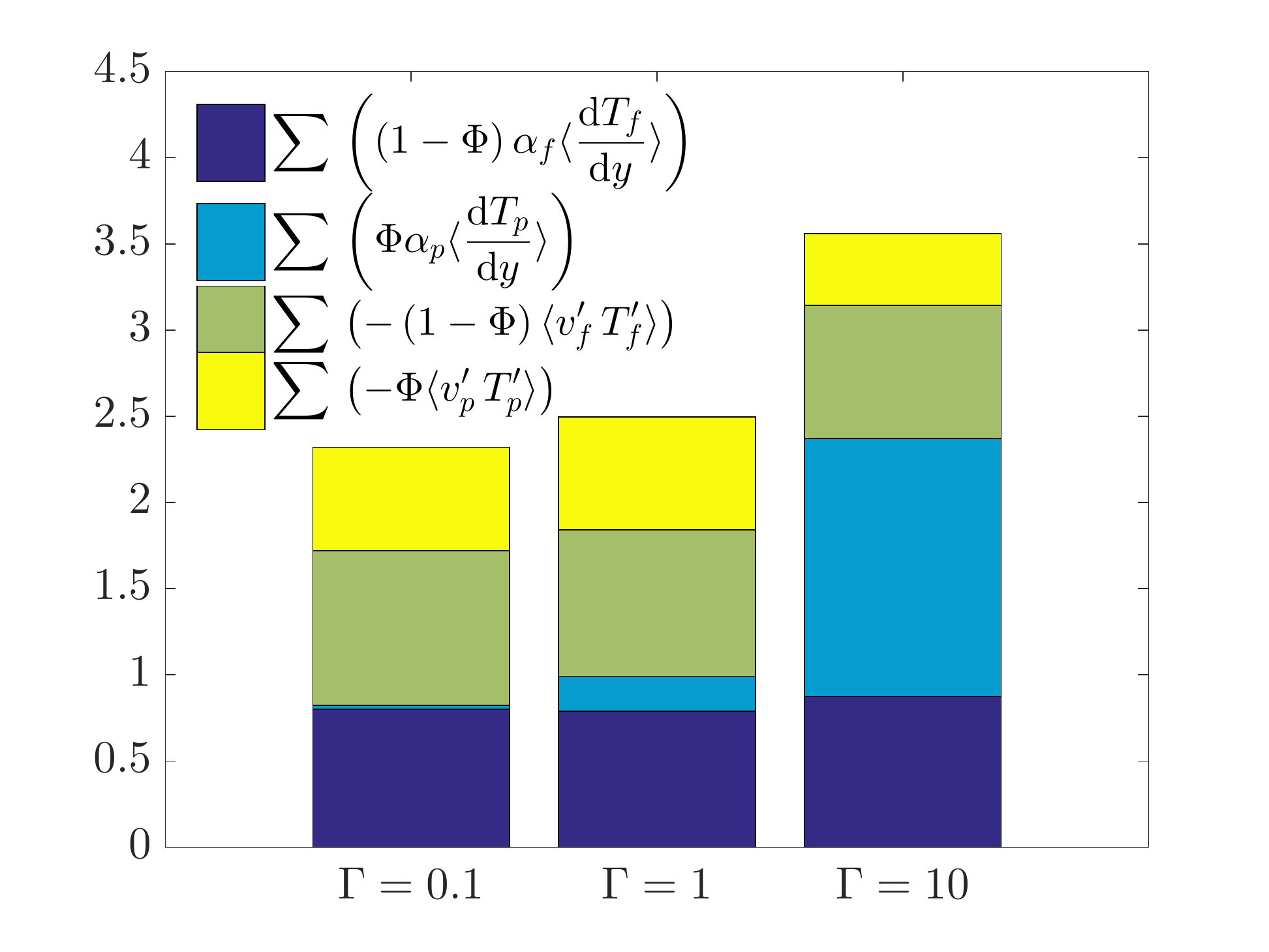}
   \put(-391,122){$(a)$}
   \put(-197,122){$(b)$}    
   \put(-289,-3){$\Gamma$} 
   \put(-96,-3){$\Gamma$} 
   \put(-190,44){\rotatebox{90}{{$\Sigma \,q^{\,\prime \prime}_{\,\,i} \, / \, \Sigma \, q^{\,\prime \prime}_{\,\,tot_{\phi=0\%}}$}}}  
   \put(-382,44){\rotatebox{90}{{$\Sigma \,q^{\,\prime \prime}_{\,\,i} \, / \, \Sigma \, q^{\,\prime \prime}_{\,\,tot_{\phi=0\%}}$}}}  \\   
  \caption{Wall-normal integral of the different terms defining the average heat flux across the channel, normalized by the total heat flux in single phase flow for: $(a)$ $\phi=10\%$ and $(b)$ $\phi=30\%$.}
\label{fig:Gamma_share}
\end{figure}

The analysis of the averaged heat flux across the gap width between the planes, see equation (\ref{eq:heatFlux}), is repeated here for the cases with different $\Gamma$.
The results are presented in figure~\ref{fig:Gamma_share} for $\phi = 10\%$ ($(a)$ to $(d)$) and $\phi = 30\%$ ($(e)$ to $(h)$) at $\Gamma = 0.1$ , $10$ and $Re_p=0.5$ , $16$.  
As mentioned before, all terms in equation (\ref{eq:heatFlux}) are normalized with the total wall-normal heat flux.

The data show that for the case at $Re_p= 0.5$ and $\Gamma = 0.1$, when the flow around the particles is in the Stokes regime, 
the molecular diffusion in the fluid is the dominant contribution to the overall heat flux. The contribution from the diffusion in the solid phase is small in this case owing to the low thermal diffusivity of the particles;  the terms related to the velocity fluctuations play a very minor role because of the low particle Reynolds number and low level of fluctuations. Indeed, the fluid and particles fluctuations are almost negligible when $Re_p= 0.5$. When $\Gamma = 10$, the diffusion in the solid is larger than that in the fluid only when the volume fraction $\phi=30\%$, except for a small region close to the wall where the local solid volume fraction tends to zero.

Finally, we consider inertial effects, the last two rows of  figure~\ref{fig:Gamma_share} pertaining the results at $Re_p= 16$.
When the particle diffusivity is lower than the fluid one, $\Gamma = 0.1$, the fraction of heat diffusing through the solid particles is negligible and the fluctuations in the fluid and solid particles play a major role in the heat transfer process. Already at $\phi=10\%$, the heat transfer due to the fluctuations in the fluid velocity is comparable to the heat diffusing in the fluid towards the centreline of the domain. At higher particle volume fractions, the term $-\left( 1 - \Phi \right)  \langle v^{\prime}_f \, T^{\prime}_f  \rangle$  dominates over the other transport mechanisms except for a region close to the wall where the velocity and temperature fluctuations vanish. 
At the highest volume fraction investigated, $\phi=30\%$, the wall-normal heat transfer due to the combined effect of temperature and particle velocity fluctuations 
is almost equal to the heat diffusion in the fluid.
For the cases with $\Gamma = 10$, however, all four terms play an important role and none of them can be a priori discarded.
At the lowest volume fraction, $\phi=10\%$, the heat diffusion and the transport related to the fluid velocity fluctuations, $\left( 1 - \Phi \right) \alpha_f  \langle \frac{\mathrm{d} T_f}{\mathrm{d} y} \rangle$ and $-\left( 1 - \Phi \right)  \langle v^{\prime}_f \, T^{\prime}_f  \rangle$,  are dominant especially in the centre of the channel.
On the contrary, the diffusion in the solid phase, $  \Phi \, \alpha_p \, \langle \frac{\mathrm{d} T_p}{\mathrm{d} y} \rangle$, dominates  at higher volume fractions, except for the small region close to the wall where the local volume fraction approaches zero.

The global contribution of each term to the  steady state heat flux
is depicted in figure~\ref{fig:Gamma_share}, where the results are normalized by the total heat flux in the absence of particles. Interestingly, the heat flux due to the particle velocity fluctuations reduces when $\Gamma=10$. Indeed, as we discussed in the previous section, the ratio between the velocity and time scale of thermal diffusion inside the particles and transport with the particles (proportional to the particle velocity fluctuations) can affect the relative importance of these two transport mechanisms; e.g. for $\Gamma=10$ the velocity of thermal diffusion is significantly larger than for $\Gamma=1$, causing the particles to reach to the surrounding fluid temperature fast in comparison to the particle own motion.

The above results shed some light on the contribution of the solid-phase thermal diffusivity to the heat transfer in particle suspensions. We show that in a suspension of solid particles with a lower thermal diffusivity than the fluid the heat transfer through the suspension can become smaller than that in single-phase flow. However, as the inertia of the particle increases, the motion of particles and fluid becomes chaotic and the heat transfer can be enhanced even in the presence of the particles with lower diffusivity, $\Gamma < 1$. The results indicate that the thermal diffusivity of the solid phase is more important when the flow is in the Stokes regime and inertial effects negligible.       

\section{Final remarks}

We report results from interface-resolved direct numerical simulations (DNS) of plane Couette flow with rigid spherical particles. In this study, we focus on the heat transfer enhancement when varying particle Reynolds number, total volume fraction (number of particles) and the ratio between the particle and fluid thermal diffusivity. 
Simulations are performed using a numerical approach proposed in this study to address heat transfer in both phases. 
The numerical algorithm is based on an immersed boundary method (IBM) to resolve fluid-solid interactions with lubrication and contact models for the short-range particle-particle (particle-wall) interactions. A volume of fluid (VoF) model is used to solve the energy equation both inside and outside of the particles, enabling us to consider different thermal diffusivities in the two phases.

The results of the simulations show that the effective thermal diffusivity of the suspension increases linearly with the volume fraction of the particles at small particle Reynolds numbers in agreement with the experimental and numerical findings in \cite{Metzger2013}, which serve as validation for the present method. 
The results show that the heat-transfer increase with respect to the unladen case is larger at finite particle inertia. The effective suspension diffusivity also scales linearly  with $\phi$ at finite $Re_p$  and  low $\phi$ (with higher slope) but eventually saturates for $\phi> 20\%$. 
The empirical relation suggested by \cite{Metzger2013} is found to be a good approximation for $Re_p<8$. 

We also vary the ratio between particle diffusivity and show that the scaling derived in \cite{Metzger2013} over-predicts the heat transfer for $\Gamma=0.1$, low conductivity in the solid phase, and under-predicts the simulation data for $\Gamma=10$, lower conductivity the fluid. However, the data are found to scale reasonably good at vanishing inertia when rescaled with the conductivity of a composite estimated by Lewis-Nielsen model \citep{Nielsen1974}. Inertial effects trigger an increase of the heat transfer that is, in absolute value, more pronounced for $\Gamma=10$. However, when scaling the data with the average conductivity of a composite, the increase due to inertial effects is relatively more important at $\Gamma=0.1$. 

It should be noticed that the increase of the effective thermal diffusivity on addition of particles usually comes at the price of an increase in the effective viscosity of the flow, resulting in higher external power needed to drive the flow. We show in this study that for particle volume fractions lower than $10\%$ 
inertial effects ($Re_p =16$)  are more pronounced on the energy transfer than on the momentum transfer; in other words the enhancement in the effective thermal diffusivity of the suspension is more than the increase of the effective viscosity, which we denote as an efficient heat transfer enhancement. 

To better understand the heat-transfer process, 
we ensemble-average the energy equation and  obtain 4 different contributions making up the total heat transfer: 
i) transport associated to the particle motion; ii) convection by fluid velocity; iii)  molecular diffusion in the solid phase (solid conduction) and iv) molecular diffusion in the fluid phase (fluid conduction). The analysis shows that the increase of the effective conductivity observed at finite inertia can be associated to an increase of the transport associated to fluid and particle velocity.
Interestingly, the total contribution of the solid conduction term reduces when increasing the particle Reynolds number. This can be explained by the ratio between the time scale of molecular diffusion in the solid and of the transport by particle motions. As particles move faster, conduction inside the solid becomes negligible.
Given the small contribution of the solid conduction term at high particle Reynolds numbers, we expect that decreasing the particle thermal diffusivity  does not have a large influence on the total heat transfer. Indeed, the results of the simulations with different thermal diffusivities match the mentioned expectation as the effective thermal diffusivity at $\Gamma=0.1$ is close to that at $\Gamma=1$.
On the other hand, a particle thermal diffusivity higher than that of the fluid significantly increases the effective heat transfer. 

We investigate in this study the effect of particle Reynolds number, total volume fraction (number of particles) and the thermal diffusivity of the particles on the effective thermal diffusivity of the suspension, however the effect of the particles shape on the heat transfer still remains unexplored, which needs to be addressed in the near future.

\section*{Acknowledgments}
L. B. and M. N. acknowledge financial support by
European Research Council, Grant No. ERC-2013-CoG-
616186, TRITOS, whereas
O. A. acknowledges the financial support of Shiraz University for the granted sabbatical leave. 
Computer time has been provided by
SNIC (the Swedish National Infrastructure for Computing).

\appendix 
\section{Different contributions to the total heat transfer}\label{appA}

In this section we examine the heat flux in suspension mixtures by phase ensemble averaging the energy equation using the framework developed and employed in \cite{Marchioro1999,Zhang2010,Picano2015},  and find the different contributions to the total heat transfer reported in the main text. 
We define the phase indicator $\xi$, whose value varies between $0$ and $1$ based on the solid fraction in the considered volume. Defining the phase-ensemble average $\langle \, \, \rangle$ as the ensemble average (implicitly) conditioned to the considered phase, the local volume fraction  is defined as
\begin{eqnarray}
\Phi  = \langle \, \xi \, \rangle \, . 
\label{eq:xi_avg} 
\end{eqnarray} 
   
A generic observable of the combined phase, $O_c$, can be constructed in terms of $O_p$ and $O_f$ (the same observable inside the particles and in the fluid phase), using  the phase indicator $\xi$. The phase-ensemble average of $O_c$ is   
\begin{eqnarray}
\langle \, O_c \, \rangle  \,= \, \langle \, \xi\, O_p \,+\, (1\,-\,\xi)\, O_f \, \rangle \,  \,= \, \Phi\, \langle \, O_p \, \rangle   \,+\,    (1\,-\,\Phi)\,\langle \, O_f \, \rangle \, , 
\label{eq:O_avg} 
\end{eqnarray} 
where we use the subscript inside the brackets to indicate the phase conditioning.

The differential heat equation, 
\begin{eqnarray}
\rho \,C_p \, \frac{\mathrm{D} T}{\mathrm{D} t}  \,=\, \nabla \cdot \left( k \, \nabla T \right) \, ,
\label{eq:heateq} 
\end{eqnarray} 
can be re-written in terms of both phases as
\begin{eqnarray}
 \xi {\left(\rho C_p \right)}_p   \frac{\mathrm{D}  T_p}{\mathrm{D}  t}  +  \left(1-\xi \right) {\left(\rho C_p \right)}_f   \frac{\mathrm{D} T_f}{\mathrm{D} t}   \,=\, \nabla \cdot \left[  \xi k_p \, \nabla T_p  +  \left(1-\xi \right) k_f \, \nabla T_f  \right]  \, .
\label{eq:heateq2} 
\end{eqnarray} 

Phase-ensemble averaging equation~\ref{eq:heateq2} and using equations~(\ref{eq:xi_avg}) and (\ref{eq:O_avg}) result in 
\begin{eqnarray}
 \Phi {\left(\rho C_p \right)}_p   \left[ \langle \frac{\partial T_p}{\partial t} \rangle +  \langle \textbf{u}_p \cdot \nabla T_p \rangle \right]  \,\, &+& \,\,  \left(1-\Phi \right) {\left(\rho C_p \right)}_f   \left[ \langle \frac{\partial T_f}{\partial t} \rangle +  \langle \textbf{u}_f \cdot \nabla T_f \rangle \right]  \, \nonumber\\[5pt] & = & \,\,\, \nabla \cdot \left[  \Phi k_p \, \langle \nabla T_p \rangle +  \left(1-\Phi \right) k_f \, \langle\nabla T_f \rangle \right]  \, .
\label{eq:heateq3} 
\end{eqnarray} 

Next we decompose temperature and velocity into the average and fluctuating part ($\textbf{u} = \textbf{U} + \textbf{u}^{\prime}$ and  $T = \overline{T} + T^{\prime}$)
and consider the mean of these quantities. With the above decomposition, equation~\ref{eq:heateq3} can be rewritten as:   
\begin{eqnarray}
 \Phi {\left(\rho C_p \right)}_p   \left[ \textbf{U}_p \cdot \nabla \overline{T}_p +  \langle \textbf{u}^{\prime}_p \cdot \nabla T^{\prime}_p \rangle \right]  \,\, &+& \,\,  \left(1-\Phi \right) {\left(\rho C_p \right)}_f   \left[ \textbf{U}_f \cdot \nabla \overline{T}_f +  \langle \textbf{u}^{\prime}_f \cdot \nabla T^{\prime}_f \rangle \right]  \, \nonumber\\[5pt] & = & \,\,\, \nabla \cdot \left[  \Phi k_p \, \langle \nabla T_p \rangle +  \left(1-\Phi \right) k_f \, \langle\nabla T_f \rangle \right]  \, .
\label{eq:heateq4} 
\end{eqnarray}    
   
Assuming  ${\left(\rho C_p \right)}_p = {\left(\rho C_p \right)}_f$ and exploiting the symmetries in the two homogeneous directions, the projection of equation~\ref{eq:heateq4} in the inhomogeneous wall-normal direction $y$ gives
\begin{eqnarray}
\frac{\, \mathrm{d} \,}{\mathrm{d} y } \left[  - \Phi \langle v^{\prime}_p T^{\prime}_p \rangle -  \left(1-\Phi \right) \langle v^{\prime}_f T^{\prime}_f \rangle   + \Phi \alpha_p \langle \frac{\mathrm{d} T_p}{\mathrm{d} y } \rangle +  \left(1-\Phi \right) \alpha_f \langle \frac{\mathrm{d} T_f}{\mathrm{d} y } \rangle  \right]  \, \, = \,\, 0 \, ,
\label{eq:heateq5} 
\end{eqnarray}    
where $\alpha_p = k_p/{(\rho C_p )}_p$ and $\alpha_f = k_f/{(\rho C_p )}_f$.

Finally, integrating equation~(\ref{eq:heateq5}) in the wall-normal direction results in the following equation      
for the total heat flux $q^{\,\prime \prime}_{\,\,tot}$
\begin{eqnarray}
q^{\,\prime \prime}_{\,\,tot}  \, \, = \,\, -\Phi \langle v^{\prime}_p T^{\prime}_p \rangle -  \left(1-\Phi \right) \langle v^{\prime}_f T^{\prime}_f \rangle      +\Phi \alpha_p \langle \frac{\mathrm{d} T_p}{\mathrm{d} y } \rangle +  \left(1-\Phi \right) \alpha_f \langle \frac{\mathrm{d} T_f}{\mathrm{d} y } \rangle    \, .
\label{eq:heateq6} 
\end{eqnarray}

\bibliographystyle{jfm}
\bibliography{Heat}

\end{document}